\documentclass[11pt,a4paper]{article}
\pdfoutput=1                     % to allow pdflatex compilation in JCAP
\usepackage{jheppub}
\usepackage{float}
\usepackage{subfig}
\usepackage[table]{xcolor}
\usepackage{multirow}
\usepackage{feynmp}
\usepackage{color}
\usepackage{mathtools}          % Fixes and extends (and also loads) amsmath.
\usepackage{bbm}                % Blackboard math variants of CM fonts. (For \mathbbm{1}.)
\usepackage{bm}                 % Bolded math symbols (e.g. bolded greek alphabet).

\usepackage{tikz}
\usepackage{stackengine}
\usepackage{scalerel} 

\usepackage[titletoc]{appendix}

\usepackage{caption}

\usepackage{hyperref}
\usepackage{cleveref}          % better cross-refs using~\cref (load after hyperref!)
   \crefname{equation}{}{}
   \crefname{align}{}{}
   \crefname{figure}{}{}       % example:~\crefformat{figure}{#figure~#1#3}
   \crefname{table}{}{}
   \crefname{section}{}{}      % example:~\crefformat{section}{#2section~#1#3}
   \crefname{appendix}{}{}
   \crefname{footnote}{}{}
   
   \crefalias{subequation}{equation}

\newcommand{\bardelta}{\mathord{\stackinset{c}{}{c}{0.45ex}{\rule{0.35em}{0.08ex}}{$\delta$}}}

\newcommand{\luin}[1]{{ \mkern-3mu \raisebox{-0.5pt} {$\scaleobj{0.60}{{(#1)}}$} }}
\newcommand{\duin}[1]{{ \mkern 0mu \raisebox{-0.5pt} {$\scaleobj{0.60}{{(#1)}}$} }}

\newcommand{\lin}[1]{{  \raisebox{1pt}{$\scaleobj{0.65}{#1}$} }}
\newcommand{\linb}[1]{{ \raisebox{0pt}{$\scaleobj{0.60}{#1}$} }}

\newcommand{\rchi}{\raisebox{1pt}{$\chi$}}

\newcommand{\oline}[1]{\overline{#1}}
\newcommand{\widebar}[1]{\mkern 1.3mu\overline{\mkern-1.3mu#1\mkern-1.3mu}\mkern 1.3mu}

\newcommand{\sqmup}{ \mu^2_{\varphi} }
\newcommand{\sqmuc}{ \mu^2_{\chi} }

\newcommand{\sqmupF}[1]{ \mu^2_\lin{\varphi,\!#1} }
\newcommand{\sqmucF}[1]{ \mu^2_\lin{\rchi,\!#1}   }

\newcommand{\ssqmupF}[1]{{ \sqmupF{#1} \!+\! \delta^\duin{#1}_{\mu\varphi}}}
\newcommand{\ssqmucF}[1]{{ \sqmucF{#1} \!+\! \delta^\duin{#1}_{\mu\chi}}}

\newcommand{\lpp}{\lambda_{\varphi\varphi}}
\newcommand{\lcc}{\lambda_{\chi\chi}}
\newcommand{\lpc}{\lambda_{\varphi\chi}}

\newcommand{\sqmpp}{m^2_{\varphi\varphi}}
\newcommand{\sqmcc}{m^2_{\chi\chi}}
\newcommand{\sqmpc}{m^2_{\varphi\chi}}
\newcommand{\hsqmpp}{\hat m^2_{\varphi\varphi}}
\newcommand{\hsqmcc}{\hat m^2_{\chi\chi}}
\newcommand{\hsqmpc}{\hat m^2_{\varphi\chi}}

\newcommand{\DeltaEps}{\Delta_\epsilon}
\newcommand{\DeltaFO}{\Delta_{\rm FO}}
\newcommand{\hDeltaFO}{\hat\Delta_{\rm FO}}
\newcommand{\DeltaF}{\Delta^\lin{\!\rm F}}

\newcommand{\sfrac}[2]{{\textstyle{\frac{#1}{#2}}}} % small fraction (for display math)

\newcommand{\laaidx}[1] {{ \lambda_\lin{\alpha\alpha}^\luin{#1} }}
\newcommand{\labidx}[1] {{ \lambda_\lin{\alpha\beta}^\luin{#1} }}
\newcommand{\lppidx}[1] {{ \lambda_\lin{\varphi\varphi}^\luin{#1} }}
\newcommand{\lccidx}[1] {{    \lambda_\lin{\chi\chi}^\luin{#1}    }}
\newcommand{\lpcidx}[1] {{    \lambda_\lin{\varphi\chi}^\luin{#1} }}

\newcommand{\blpcidx}[1]{{\bar\lambda_\lin{\varphi\chi}^\luin{#1} }}

\newcommand{\dlaaidx}[1] {{ \delta_\lin{\lambda\alpha\alpha}^\duin{#1} }}
\newcommand{\dlabidx}[1] {{ \delta_\lin{\lambda\alpha\beta}^\duin{#1} }}
\newcommand{\dlppidx}[1] {{ \delta_\lin{\lambda\varphi\varphi}^\duin{#1} }}
\newcommand{\dlccidx}[1] {{    \delta_\lin{\lambda\chi\chi}^\duin{#1}    }}
\newcommand{\dlpcidx}[1] {{    \delta_\lin{\lambda\varphi\chi}^\duin{#1} }}

\newcommand{\dblpcidx}[1]{{\bar\delta_\lin{\lambda\varphi\chi}^\duin{#1} }}

\newcommand{\slaa}[1] {{ \laaidx{#1}  \!+\! \dlaaidx{#1}  }}
\newcommand{\slab}[1] {{ \labidx{#1}  \!+\! \dlabidx{#1}  }}

\newcommand{\slpc}[1] {{ \lpcidx{#1}  \!+\! \dlpcidx{#1}  }}

\newcommand{\sblpc}[1]{{ \blpcidx{#1} \!+\! \dblpcidx{#1} }}

\newcommand{\klaa}[1] {{ ( \laaidx{#1} \!+\! \dlaaidx{#1})  }}
\newcommand{\klab}[1] {{ ( \labidx{#1} \!+\! \dlabidx{#1})  }}
\newcommand{\klpp}[1] {{ ( \lppidx{#1} \!+\! \dlppidx{#1})  }}
\newcommand{\klcc}[1] {{ ( \lccidx{#1} \!+\! \dlccidx{#1})  }}
\newcommand{\klpc}[1] {{ ( \lpcidx{#1} \!+\! \dlpcidx{#1})  }}

\newcommand{\kblpc}[1]{{ (\blpcidx{#1} \!+\! \dblpcidx{#1}) }}

\newcommand{\scRB}{{\raisebox{0pt}{\scaleobj{0.65}{\rm B}}}}
\newcommand{\varphiB}{\varphi_\scRB}
\newcommand{\chiB}{\chi_\scRB}

\def\lsim{\mathrel{\raise.3ex\hbox{$<$\kern-.75em\lower1ex\hbox{$\sim$}}}}
\def\gsim{\mathrel{\raise.3ex\hbox{$>$\kern-.75em\lower1ex\hbox{$\sim$}}}}

\title{Consistent Thermal Resummation and Phase Transitions with 2PI Methods}

\author[a,b,c,d]{Amitayus Banik}
\author[e,f]{and Kimmo Kainulainen}

\affiliation[a]{Department of Physics, National Tsing Hua University, Hsinchu 300, Taiwan}
\affiliation[b]{Center for Theory and Computation, National Tsing Hua University, Hsinchu 300, Taiwan}
\affiliation[c]{Department of Physics, Chungbuk National University, Cheongju, Chungbuk 28644, Korea}
\affiliation[d]{Research Institute for Nanoscale Science and Technology,
Chungbuk National University,\\
Cheongju, Chungbuk 28644, Korea}
\affiliation[e]{Department of Physics, University of Jyv\"{a}skyl\"{a}, \\
PO Box 35 (YFL), FI-40014 Jyv\"{a}skyl\"{a}, Finland}
\affiliation[f]{Helsinki Institute of Physics, University of Helsinki, \\
PO Box 64, FI-00014 Helsinki, Finland}

\emailAdd{abanik@cbnu.ac.kr}
\emailAdd{kimmo.kainulainen@jyu.fi}

\abstract{We apply the two-particle irreducible (2PI) formalism as a framework for a consistent thermal resummation in studies of cosmological phase transitions. Considering a model with two mixing real scalar fields, we work within the Hartree approximation and renormalize the 2PI effective action, while introducing a connection to physical parameters. This yields the Hartree-resummed finite-temperature effective potential, which is valid for all temperatures and avoids the limitations of conventional methods based on the high-temperature approximation. With this potential, we study one- and two-step transitions within the model, and compare our results with those obtained using resummation schemes widely employed in the literature. Finally, we evaluate the gravitational wave spectrum generated from first-order phase transitions, demonstrating the impact of the choice of resummation scheme on the predicted spectrum.}


\begin{document}
\begin{flushright}
\end{flushright}
\maketitle
%
%%%%%%%%%%%%%%%%%%%%%%%%%%%%%%%%%%%%%%%%%%%%%%%%%%%%%%%%%%%%%%%%%%%%%%%%%%%%%%%%%%%%%%%%%%%%%%%%%%%%%%%%%%
%%%%%%%%%%%%%%%%%%%%%%%%%%%%%%%%%%%%%%%%%%%%%%%%%%%%%%%%%%%%%%%%%%%%%%%%%%%%%%%%%%%%%%%%%%%%%%%%%%%%%%%%%%

%%%%%%%%%%%%%%%%%%%%%%%%%%%%%%%%%%%%%%%%%%%%%%%%%%%%%%%%%%%%%%%%%%%%%%%%%%%%%%%%%%%%%%%%%%%%%%%%%%%%%%%%%%
%%%%%%%%%%%%%%%%%%%%%%%%%%%%%%%%%%%%%%%%%%%%%%%%%%%%%%%%%%%%%%%%%%%%%%%%%%%%%%%%%%%%%%%%%%%%%%%%%%%%%%%%%%
%
\section{Introduction}
\label{sec:intro}
%
%%%%%%%%%%%%%%%%%%%%%%%%%%%%%%%%%%%%%%%%%%%%%%%%%%%%%%%%%%%%%%%%%%%%%%%%%%%%%%%%%%%%%%%%%%%%%%%%%%%%%%%%%%
%%%%%%%%%%%%%%%%%%%%%%%%%%%%%%%%%%%%%%%%%%%%%%%%%%%%%%%%%%%%%%%%%%%%%%%%%%%%%%%%%%%%%%%%%%%%%%%%%%%%%%%%%%

First-order phase transitions (PTs) in cosmology have long been studied in the context of electroweak baryogenesis~\cite{Kuzmin:1985mm,Morrissey:2012db}, providing the necessary out-of-equilibrium conditions~\cite{Sakharov:1967dj} to generate the matter-antimatter asymmetry of the Universe. In recent years, cosmological first-order PTs have gained traction due to their possibility of also sourcing stochastic gravitational wave (GW) backgrounds~\cite{Hogan:1986dsh}, which may be detectable at future observatories such as the Laser Interferometer Space Antenna (LISA)~\cite{LISA:2017pwj}, the Cosmic Explorer (CE)~\cite{Reitze:2019iox}, the Big Bang Observer (BBO)~\cite{Crowder:2005nr, Corbin:2005ny, Harry:2006fi}, and the International Pulsar Timing Array (IPTA)~\cite{Antoniadis:2022pcn, IPTA:2023ero}.  The detection of these GWs would provide valuable insight into the physics of the early Universe and would serve as an indicator of physics beyond the Standard Model (SM). Predicting the GW signal of a cosmological PT requires a good understanding of its thermodynamics~\cite{Caprini:2019egz, Athron:2023xlk, Croon:2024mde}. This can be captured by the scalar effective potential at finite temperature $V_{\rm eff}$, evaluated in some approximation in thermal perturbation theory~\cite{Dolan:1973qd, Weinberg:1974hy,Kirzhnits:1976ts}. Four-dimensional lattice simulations offer an alternative, fully non-perturbative approach, but challenges in incorporating chiral fermions on a lattice~\cite{Luscher:2000hn}, and the high computational resources required, limit their applicability. 

Perturbative evaluations of the finite temperature effective potential are notoriously sensitive to infrared (IR) corrections from light bosonic fields~\cite{Linde:1980ts, Gross:1980br}, whose couplings receive enhancements of the form $\kappa \to \kappa \,(T/m)$. These effective couplings may become large for $T\gg m$ even for small $|\kappa|$, as indeed often happens in cosmological PTs~\cite{Laine:2016hma,Laine:2000kv}. The infrared problem can be tackled by {\em resummations} that give rise to thermal masses, and restore the convergence of the perturbative series at finite temperature~\cite{Andersen:2004fp}. Commonly used lowest-order schemes include the ring~\cite{Carrington:1991hz,Arnold:1992rz} and the Parwani~\cite{Parwani:1991gq} approximations. In the former, only the lightest zero Matsubara modes\footnote{The notion of Matsubara modes refers to the imaginary time formalism of thermal field theory. For a review, see Ref.~\cite{Laine:2016hma}.}, which primarily contribute to the IR problem, are resummed, while in the latter corrections are applied to all Matsubara modes. Both schemes inherently rely on the high-temperature limit $T \gg m$ to identify thermal corrections and yield simple expressions that are relatively straightforward to implement. The ring-improved potential breaks down for $m/T \gsim 1$, failing to display the correct Boltzmann suppression with heavy fields. The Parwani potential is Boltzmann-suppressed, but as it does not include thermal corrections in a self-consistent manner, the validity of the scheme is not obvious for $m/T \sim 1$.

Alternatively, one can use the dimensional reduction method, by matching the original four-dimensional theory to a three-dimensional (3D) effective field theory~\cite{Farakos:1994kx,Kajantie:1995dw,Braaten:1995cm}. One can then study the effective 3D theory on the  lattice~\cite{Kajantie:1995kf}, or derive further improved actions by use of a 3D perturbation theory~\cite{Ekstedt:2024etx}. These two approaches agree well~\cite{Gould:2023ovu,Ekstedt:2024etx} within the dimensionally reduced setup and in many cases provide the most accurate results available. However, these methods also rely on the validity of the high-temperature limit through the dimensional reduction step, which involves integrating out the non-zero Matsubara modes under the assumption that $m \ll T$. 

There are scenarios where the high-temperature approximation breaks down, especially involving strong PTs~\cite{Laine:2017hdk,Kainulainen:2019kyp,Niemi:2024axp,Ekstedt:2024etx,Chala:2024xll,Chala:2025aiz,Chala:2025oul,Bernardo:2025vkz,Kierkla:2026bnm}. A self-consistent resummation scheme, valid across a wide temperature range, and extendable to higher orders in the perturbative expansion, is then clearly needed. This is an intricate task that involves setting up a renormalization procedure for ultraviolet divergences that avoids spurious effects or ambiguities at finite temperature, and several studies~\cite{Boyd:1993tz,Funakubo:2012qc,Kneur:2015uha,Curtin:2016urg,Curtin:2022ovx,Funakubo:2023cyv,Funakubo:2023eic,Bahl:2024ykv,Bittar:2025lcr,Navarrete:2025yxy,Klose:2026pux} have already studied the issue. In this work, we solve the problem by employing a self-consistent resummation framework based on the two-particle irreducible (2PI) formalism~\cite{Jackiw:1974cv,Cornwall:1974vz}.

The 2PI action involves a two-point function, which is self-consistently solved from the 2PI equations of motion to a given truncation of the 2PI loop expansion, which each correspond to an infinite resummation of ordinary one-particle irreducible (1PI) diagrams. In the lowest order (Hartree) approximation, these equations reduce to a \textit{gap equation} for the renormalized masses. The inherent resummation feature of the 2PI formalism then allows treating the high- and low-temperature corrections to thermal masses on equal footing, avoiding a reliance on the high-temperature limit~\cite{Amelino-Camelia:1992qfe,Amelino-Camelia:1993rvt,Boyd:1993tz,Kainulainen:2021eki}. On the other hand, the 2PI renormalization procedure is more complex, as orders in perturbation theory get mixed. Yet the 2PI action can be renormalized using local counterterms~\cite{Berges:2005hc} as has been explicitly demonstrated for a $\lambda \varphi^4$-theory in~\cite{vanHees:2001ik,VanHees:2001pf,Blaizot:2003br, Blaizot:2003an, Carrington:2014lba, Carrington:2017lry,Kainulainen:2021eki} and for $\text{O}(N)$ symmetric theories in~\cite{Patkos:2008ik, Patkos:2008sg, Pilaftsis:2013xna, Pilaftsis:2015cka, Brown:2016vaj,Pilaftsis:2017enx}.

The 2PI resummation has been previously applied to study quark-gluon plasma~\cite{Blaizot:2000fc,Andersen:2004re,Berges:2004hn}, demonstrating improved convergence over the standard resummed perturbation theory~\cite{Braaten:1989mz,Blaizot:1999ap}. The 2PI effective action has also been used to derive thermodynamic quantities in scalar theories~\cite{Carrington:2016zlc} and in the study of symmetry breaking~\cite{Amelino-Camelia:1993rvt,Pilaftsis:2013xna,Pilaftsis:2017enx}. However, direct applications to cosmological PTs remain limited. Moreover, while phenomenological studies would require renormalization in terms of observable parameters, this is usually not carried through (see however~\cite{Kainulainen:2021eki,Banik:2023nqm}). In this work, we study the cosmological PT carefully in a simple model comprising two scalar fields, deriving its 2PI effective potential at finite temperature, and imposing renormalization in terms of physical parameters. An additional novel aspect of this work is the renormalization of the 2PI action involving mixing of scalar fields, which, to our knowledge, has not been studied previously in the literature. The methods demonstrated here carry over to the study of more realistic models such as the singlet extension of the SM.

The rest of this paper is organized as follows: in Sec.~\ref{sec:ht_approx}, we review the 2PI effective action techniques and introduce the Hartree truncation for our toy model. Then, in Sec.~\ref{sec:renorm}, we systematically renormalize the 2PI action; in particular, we obtain the renormalized two-point functions and field equations. With the renormalization procedure established, we obtain the vacuum scale-invariant Hartree effective potential for two scalar fields in Sec.~\ref{sec:eff_pot}. Importantly, we demonstrate the procedure to connect the parameters in the Hartree effective potential to physical quantities. Next, in Sec.~\ref{sec:PT}, we discuss finite-temperature corrections to this effective potential, brought about through the thermally corrected masses, which are consistently resummed within the 2PI formalism. We also introduce the Parwani and ring approximations commonly used in the phenomenological studies of PTs, and end this section with a brief review of PT dynamics. Section~\ref{sec:numerical_results} is devoted to our numerical results obtained for benchmark points featuring one- and two-step transitions and the predicted GW signals corresponding to these scenarios. We conclude in Sec.~\ref{sec:conclusion}. Finally, appendices~\cref{app:Z2_wave_function_ren_factors}--\cref{app:mass-derivatives} contain details of the computations, in particular, the calculation of the wave-function renormalization factors within the 2PI formalism and the proof of scale invariance of the Hartree effective potential.

%%%%%%%%%%%%%%%%%%%%%%%%%%%%%%%%%%%%%%%%%%%%%%%%%%%%%%%%%%%%%%%%%%%%%%%%%%%%%%%%%%%%%%%%%%%%%%%%%%%%%%%%%%
%%%%%%%%%%%%%%%%%%%%%%%%%%%%%%%%%%%%%%%%%%%%%%%%%%%%%%%%%%%%%%%%%%%%%%%%%%%%%%%%%%%%%%%%%%%%%%%%%%%%%%%%%%%%%%%
%
\section{The 2PI action in the Hartree Approximation}
\label{sec:ht_approx}
%
%%%%%%%%%%%%%%%%%%%%%%%%%%%%%%%%%%%%%%%%%%%%%%%%%%%%%%%%%%%%%%%%%%%%%%%%%%%%%%%%%%%%%%%%%%%%%%%%%%%%%%%%%%%%%%%
%%%%%%%%%%%%%%%%%%%%%%%%%%%%%%%%%%%%%%%%%%%%%%%%%%%%%%%%%%%%%%%%%%%%%%%%%%%%%%%%%%%%%%%%%%%%%%%%%%%%%%%%%%%%%%%

In this section, we introduce our setup. We work in the lowest-order perturbative or Hartree approximation for the 2PI effective action, from which we derive the equations of motion (EOMs) for the classical fields and for the classical two-point functions. We study a model with two interacting real scalar fields described by the Lagrangian,
\begin{align}
    \mathcal{S}[\phi_\scRB,\eta_\scRB] = \int\! {\rm d}^4x \,\Big( 
             &  \sfrac12(\partial_{\nu}\phi_\scRB)^2  
              + \sfrac12(\partial_{\nu}\eta_\scRB)^2 
            \nonumber \\ 
            + &\sfrac{1}{2}\sqmup \phi_\scRB^2 + \sfrac{1}{2}\sqmuc\eta_\scRB^2 
              - \sfrac{1}{4!}\lpp \phi_\scRB^4 - \sfrac{1}{4!} \lcc\eta_\scRB^4 
              - \sfrac{1}{4} \lpc \phi_\scRB^2\eta_\scRB^2 \Big),
    \label{eq:cl_action}
\end{align}
where we imposed a $\mathbb{Z}_2 \otimes \mathbb{Z}'_2$ symmetry, which prevents cubic terms in the potential. In view of the renormalization procedure, which we later discuss, the subscript ``B" will always indicate the bare fields. The generic expression for the 2PI effective action for our model is given by
\begin{align}
     \Gamma_{\rm{2PI}}[\varphiB,\chiB;\Delta_\scRB] = \mathcal{S}_{\rm cl}[\varphiB,\chiB] - \frac{i}{2}\text{Tr}\ln \Delta_\scRB +\frac{i}{2}\text{Tr} [\Delta^{-1}_\lin{\rm B 0} \Delta_\scRB] + \Gamma_2[\varphiB,\chiB;\Delta_\scRB]\,.
    \label{eq:Gamma_2PI}
\end{align}
Here $\varphiB$ and $\chiB$ are the classical fields (one-point functions) corresponding to the divisions $\phi_\scRB \equiv \varphi_{\rm B} + \hat\phi_\scRB$ and $\eta_\scRB \equiv \chi_{\rm B} + \hat\eta_\scRB$, where $\hat\phi_\scRB,\,\hat\eta_\scRB$ are the corresponding quantum fields, and $\Delta_\scRB$ is the matrix of the connected two-point functions (full propagators). The trace involves an integration over the complex Keldysh contour~\cite{Keldysh:1964ud}. In the real-time representation, this becomes the usual time integration and a summation over the Keldysh indices that indicate the real-time branches. For example, the classical action becomes:
\begin{equation}
 \mathcal{S}_{\rm cl} = \displaystyle \sum_{a = \pm} a\,\mathcal{S}[\varphiB^{a},\chiB^{a}]\,, 
\label{eq:Bare_classical_action}
\end{equation}
where the action within the sum has exactly the same functional form as~\cref{eq:cl_action}. Similarly, the inverse of the classical propagators becomes
\begin{align}
    i (\Delta^{ab}_{\scRB,0})^{-1}(x,y) = -\left[\Box_x 
    +\mathcal{M}^2(\varphiB^a,\chiB^a)\right]\,\delta^{(4)}(x-y)\delta^{ab},
    \label{eq:inv_prop}
\end{align}
where the mass matrix is given by\footnote{In order to simplify the notation, we always suppress the Keldysh indices when there is no danger of confusion. This applies to equations~\cref{eq:mass_matrix,eq:Sint,eq:Gamma2_HT} where we show only the functional dependencies on fields and propagators that actually have Keldysh indices as indicated in the main text. The same practice will also be followed later when we present the renormalized 2PI action and equations of motion. }
\begin{align}
    \mathcal{M}^2(\varphiB,\chiB) = 
    \begin{pmatrix}
    -\sqmup + \frac12\lpp\varphiB^2+ \frac12\lpc\chiB^2 & \lpc\varphiB\chiB\\[2mm]
    \lpc\varphiB\chiB &  -\sqmuc + \frac12\lcc\chiB^2+ \frac12\lpc\varphiB^2
    \label{eq:mass_matrix}
\end{pmatrix}\,.	
\end{align}

\begin{figure}[t!]
    \centering
    \includegraphics[width=0.8\columnwidth]{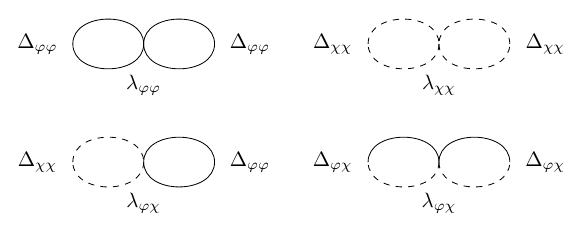}
    \caption{The diagrams obtained from $S_{\rm int}$ in Eq.~\cref{eq:Sint}, that correspond to the Hartree approximation in Eq.~\cref{eq:Gamma2_HT}.}
    \label{fig:ht_approx}
\end{figure}

Next, we consider the shifted action, which is not directly part of the action $\Gamma_{\rm 2PI}$, but defines the perturbative expansion for the interaction term $\Gamma_2$. It has a similar structure to the classical action~\cref{eq:Bare_classical_action} in Keldysh indices: $\mathcal{S}_{\rm int} = \sum_\pm a\mathcal{S}[\varphiB^a,\chiB^a;\hat\phi_\scRB^a,\hat\eta_\scRB^a]$, with 
\begin{align}
    \mathcal{S}[\varphiB,\chiB;\hat\phi_\scRB,\hat\eta_\scRB] =
     -\int_x &\Big( \; \sfrac{1}{6} \lpp \varphiB \hat\phi^3_\scRB 
                     + \sfrac{1}{6} \lcc \chiB    \hat\eta^3_\scRB 
     + \sfrac12\lpc\left( \varphiB \hat\phi_{\scRB} \hat\eta^2_\scRB
     + \chiB \hat\eta_\scRB \hat\phi_\scRB^2\right)\nonumber \\
   & + \sfrac{1}{4!}\lpp \hat\phi^4_\scRB  
     + \sfrac{1}{4!}\lcc \hat\eta^4_\scRB
     + \sfrac{1}{4} \lpc \hat\phi^2_\scRB \hat\eta^2_\scRB \Big)\,.
    \label{eq:Sint}
\end{align}
Here we have used $\int_x \equiv \int {\rm d}^4x$ for brevity. Interactions in~\cref{eq:Sint} generate $\Gamma_2$, which consists of all 2PI vacuum diagrams involving the full propagators. In this work, we will consider the Hartree approximation, which is the leading-order coupling constant truncation, corresponding to diagrams shown in Fig.~\cref{fig:ht_approx}. These loops are generated by the terms in the last line of~\cref{eq:Sint}. Explicitly, we find $\Gamma_2^{\rm HT} = \sum_\pm a\Gamma^{\rm HT}_{2}[\Delta^{\!\scRB aa}_{\alpha\beta}]$,
\begin{align}
    \Gamma_{2}^{\rm HT}[\Delta^{\!\scRB}_{\alpha\beta}] = 
    &-\frac{\lpp}{8}\int_x \,[\Delta^{\!\scRB}_{\varphi\varphi}(x,x)]^2
    -\frac{\lcc}{8}\int_x \; [\Delta^{\!\scRB}_{\chi\chi}(x,x)]^2
    \nonumber \\
    &-\frac{\lpc}{4}\int_x \Delta^{\!\scRB}_{\varphi\varphi}(x,x)\Delta^{\!\scRB}_{\chi\chi}(x,x)
    -\frac{\lpc}{4}\int_x \big(\Delta^{\!\scRB}_{\varphi\chi}(x,x) + \Delta^{\!\scRB}_{\chi\varphi}(x,x)\big)^2,
    \label{eq:Gamma2_HT}
\end{align}
where we have written the last term on the second line, featuring mixed propagators in a fully symmetric form.
One can, in fact, leave out the Keldysh indices from the local correlation functions, because all propagator functions $\Delta^{ab}_{\alpha\beta}$ have the same local limit. 

%=======================================================================================================
%=======================================================================================================
%=======================================================================================================
%=======================================================================================================

%%%%%%%%%%%%%%%%%%%%%%%%%%%%%%%%%%%%%%%%%%%%%%%%%%%%%%%%%%%%%%%%%%%%%%%%%%%%%%%%%%%%%%%%%%%%%%%%%%%%%%%%%%
\paragraph{Equations of motion.}
%%%%%%%%%%%%%%%%%%%%%%%%%%%%%%%%%%%%%%%%%%%%%%%%%%%%%%%%%%%%%%%%%%%%%%%%%%%%%%%%%%%%%%%%%%%%%%%%%%%%%%%%%%

Having set up the 2PI effective action in our truncation scheme, we can now derive the equations of motion for the classical fields and the two-point functions. These are correspond to stationary configurations of $\Gamma_{\rm 2PI}$:
\begin{equation}
    \frac{\delta \Gamma_{\rm 2PI}}{\delta \varphi^a}=0\,,\quad 
    \frac{\delta \Gamma_{\rm 2PI}}{\delta \chi^a} = 0 \quad {\rm and} \quad
    \frac{\delta \Gamma_{\rm 2PI}}{\delta \Delta^{ba}_{\beta\alpha}} = 0.
    \label{eq:stationarity_cond}
\end{equation}
From now on, we will always set $a=1$ in the field equations, and then omit the Keldysh index for notational simplicity. In the Hartree approximation~\cref{eq:Gamma2_HT}, there are no contributions to the field equations from $\Gamma_2^{\rm HT}$, and we find:
\begin{subequations}
\begin{align}
    &\Big(\Box_x-\sqmup + \sfrac{1}{6}\lpp\varphi^2_\scRB(x)
    +\sfrac12\lpc\chi^2_\scRB(x)\nonumber\\
    &\qquad\qquad
    +\sfrac12\lpp\Delta^{\!\scRB}_{\varphi\varphi}(x,x)
    +\sfrac12\lpc\Delta^{\!\scRB}_{\chi\chi}(x,x)\Big)\varphiB(x)
    +\lpc\,\Delta^{\!\scRB}_{\varphi\chi}(x,x)\chiB(x) = 0 \,,
    \label{eq:EOM_phi}
\\
    &\Big(\Box_x-\sqmuc
    +\sfrac{1}{6}\lcc\chi^2_\scRB(x)
    +\sfrac12\lpc\varphi^2_\scRB(x)\nonumber\\
    &\qquad\qquad
    +\sfrac12\lcc\Delta^{\!\scRB}_{\chi\chi}(x,x)
    +\sfrac12\lpc\Delta^{\!\scRB}_{\varphi\varphi}(x,x)\Big)\chiB(x)
    +\lpc\,\Delta^{\!\scRB}_{\varphi\chi}(x,x)\varphiB(x) = 0\,.
    \label{eq:EOM_chi}
\end{align}
\label{eq:EOM_bare_fields}
\end{subequations}

\vskip-0.3cm \noindent
The stationarity condition corresponding to a variation with respect to the propagator can be written for general propagators in matrix form as:
\begin{align}
    &\Big(\Box_x\,\delta_{\alpha\gamma}+\mathcal{M}^2_{\alpha\gamma,a}
    \Big)i\Delta^{\!\scRB ab}_{\gamma\beta}(x,y)
    - c\int_z \Pi^{{\rm B}ac}_{\alpha\gamma}(x,z)\Delta^{\!{\rm B} cb}_{\gamma\beta}(z,y) = \delta^{(4)}(x-y)\delta^{ab},
    \label{eq:EOM_Delta}
\end{align}
where again $\alpha, \beta, \gamma \in \{\varphi,\chi\}$ and $a,b,c$ are the Keldysh indices, with a sum over repeated indices being assumed. Moreover, we have denoted $\mathcal{M}^2_{\alpha\gamma,a}\equiv \mathcal{M}^2_{\alpha\gamma}(\varphiB^a,\chiB^a)$ and defined the self-energy matrix 
\begin{equation}
    \Pi^{{\rm B}ac}_{\alpha\gamma}(x,y) \equiv  2iac\frac{\delta\Gamma_2}{\delta\Delta^{{\rm B} ca}_{\gamma\alpha}(y,x)}.
    \label{eq:self-energy}
\end{equation}
In the full dynamical case, the equations~\cref{eq:EOM_bare_fields,eq:EOM_Delta} are strongly coupled and must be solved simultaneously, typically using numerical methods. However, we focus on the constant background solution to the field equations, whereby the propagators can be expressed as functionals of the fields, i.e., $\Delta \equiv \Delta[\varphi,\chi]$. Reinserting these into the effective action~\cref{eq:Gamma_2PI} allows one to obtain the corresponding \textit{2PI improved effective potential} for constant field configurations, which is the goal of our study. To do so, however, one must address the various divergences in the local correlation functions by a proper renormalization procedure. 

%%%%%%%%%%%%%%%%%%%%%%%%%%%%%%%%%%%%%%%%%%%%%%%%%%%%%%%%%%%%%%%%%%%%%%%%%%%%%%%%%%%%%%%%%%%%%%%%%%%%%%%%%%%%%%%%
%%%%%%%%%%%%%%%%%%%%%%%%%%%%%%%%%%%%%%%%%%%%%%%%%%%%%%%%%%%%%%%%%%%%%%%%%%%%%%%%%%%%%%%%%%%%%%%%%%%%%%%%%%%%%%%%
%
\section{Renormalization}
\label{sec:renorm}
%
%%%%%%%%%%%%%%%%%%%%%%%%%%%%%%%%%%%%%%%%%%%%%%%%%%%%%%%%%%%%%%%%%%%%%%%%%%%%%%%%%%%%%%%%%%%%%%%%%%%%%%%%%%%%%%%%
%%%%%%%%%%%%%%%%%%%%%%%%%%%%%%%%%%%%%%%%%%%%%%%%%%%%%%%%%%%%%%%%%%%%%%%%%%%%%%%%%%%%%%%%%%%%%%%%%%%%%%%%%%%%%%%%

Due to the resummed nature of the propagators in the 2PI formalism, the standard BPHZ procedure \cite{10.1007/BF02392399, Bogoliubov:1957gp, Hepp:1966eg, Zimmermann:1969jj}, used to determine the structure of the divergences in the 1PI formalism of standard quantum field theory, is not sufficient for us (cf.~\cite{Berges:2005hc} and references therein). Consistent renormalization with a finite number of counterterms can still be carried out by defining auxiliary vertex functions with their independent renormalization conditions, as discussed extensively in~\cite{Berges:2005hc}, and demonstrated in~\cite{Banik:2023nqm} for an on-shell renormalization scheme. In this work, we will use the method developed in~\cite{Fejos:2007ec,Patkos:2008sg,Patkos:2008ik}, and later applied in~\cite{Pilaftsis:2013xna,Pilaftsis:2017enx,Kainulainen:2021eki}, where the auxiliary $n$-point functions are renormalized in the $\widebar{\rm MS}$-scheme. The $\widebar{\rm MS}$-parameters are then related to physical parameters as in~\cite{Kainulainen:2021eki}.

We begin by defining the renormalized fields and propagators (without subscripts) in terms of the bare ones. In the 2PI approach, the classical fields and propagators need different wave-function renormalization factors:
\begin{equation}
    \varphiB  \equiv Z^\frac12_\lin{\varphi,\!2}\,\varphi\,,\quad 
        \chiB \equiv Z^\frac12_\lin{\rchi,\!2}\, \chi 
    \quad {\rm and} \quad
    \Delta^{\rm B}_{\alpha\beta} \equiv 
     Z^\frac12_\lin{\alpha,\!0} Z^\frac12_\lin{\beta,\!0} \Delta_{\alpha\beta},
\label{eq:bare_vs_renormalized_fields_and_propagators}
\end{equation}
where $\alpha,\beta \in \{\varphi,\chi\}$. Different truncations for $\Gamma_2$ give different approximations for the effective action $\hat \Gamma_{\rm 1PI}[\varphi]$, which re-sum infinite subsets of 1PI diagrams and the associated counterterms. A complication that arises is that these non-perturbative resummations treat the various $n$-point functions (vertices) differently depending on how many legs in the vertex correspond to external fields and how many to internal propagators. As a result, one is forced to introduce independent renormalized {\em and} non-renormalized parameters associated with the different types of $n$-point functions, which then also require independent renormalization conditions. Here we only need to consider the 2- and 4-point functions, for which we define:
\begin{align}
    \mu^2_\lin{{\rm B}\alpha,i} &\equiv \mu^2_{\alpha i} + \delta\mu^2_\lin{\alpha,i}, 
    \qquad \alpha = \varphi,\chi \quad {\rm and} \quad i=0,2
    \\ 
    \lambda^\luin{i}_\lin{{\rm B}\alpha\beta} &\equiv 
    \labidx{i} + \delta \labidx{i}, 
    \qquad \alpha,\beta = \varphi,\chi \quad {\rm and} \quad i=0,2,
\label{eq:bare_vs_renormalized_couplings_0}
\end{align}
where the index $i$ counts the number of classical fields connected to the vertex. We will adhere to this notation throughout the rest of this work. The bare parameters maintain their key property of being scale-independent. Of course, all $n$-point functions and the associated parameters with different $i$ would be identical if $\Gamma_2$ was evaluated exactly, but this is never possible in practice.

It is useful to split the renormalized Lagrangian into the renormalized parameters and effective counterterms that incorporate the contribution from the wave-function renormalization factors. We define:
\begin{subequations}
\begin{align}
    \mu^2_\lin{\alpha,\!i} \!+\! \delta^\duin{i}_\lin{\mu\alpha} \;
    &\; \equiv \;
    Z_\lin{\alpha,\! i} (\mu^2_{\alpha i} + \delta\mu^2_\lin{\alpha,i}),
    \\[0.1ex] 
    \slaa{i}
    &\;\equiv\;  Z_\lin{\alpha,\!i}Z_\lin{\alpha,\!0} (\lambda^\luin{i}_\lin{\alpha\alpha} \!+\! 
                 \delta \lambda^\luin{i}_\lin{\alpha\alpha}),
    \\[0.1ex] 
    \slaa{4}
    &\;\equiv\;  Z^2_\lin{\alpha,\!2} (\lambda^\luin{4}_\lin{\alpha\alpha} \!+\! 
                 \delta \lambda^\luin{4}_\lin{\alpha\alpha}),
    \\[0.1ex] 
    \slpc{0}
    &\;\equiv\; Z_\lin{\varphi,\!0}Z_\lin{\rchi,\!0} 
                (\lpcidx{0} \!+\! \delta\lpcidx{0}),
    \\[0.1ex] 
    \slpc{2}
    &\;\equiv\; Z_\lin{\varphi,\!0}  Z^\linb{1/2}_\lin{\varphi,\!2} Z^\linb{1/2}_\lin{\rchi,\!2} 
                (\lpcidx{2} \!+\! \delta\lpcidx{2}),
    \\[0.1ex] 
    \slpc{4}
    &\;\equiv\; Z_\lin{\varphi,\!2}Z_\lin{\rchi,\!2} 
                (\lpcidx{4} \!+\! \delta\lpcidx{4}),
   \\[0.1ex] 
    \sblpc{0}
    &\;\equiv\; Z^\linb{1/2}_\lin{\varphi,\!0} Z^\linb{1/2}_\lin{\rchi,\!0} 
                Z^\linb{1/2}_\lin{\varphi,\!0} Z^\linb{1/2}_\lin{\rchi,\!0} 
                (\blpcidx{0} \!+\! \delta\blpcidx{0}),
    \\[0.1ex] 
    \sblpc{2}
    &\;\equiv\; Z^\linb{1/2}_\lin{\varphi,\!0} Z^\linb{1/2}_\lin{\rchi,\!0}
                Z^\linb{1/2}_\lin{\varphi,\!2} Z^\linb{1/2}_\lin{\rchi,\!2} 
                (\blpcidx{2} \!+\! \delta\blpcidx{2}),
    \end{align}
    \label{eq:bare_vs_renormalized_couplings_1}
\end{subequations}

\vskip-0.5cm\noindent
where in the first two equations again $\alpha \in \{\varphi,\chi\}$ and $i=1,2$. 
A practical reason for these definitions is that renormalization conditions directly set the counterterms $\delta^\duin{i}_\lin{\mu\alpha}$, $\dlabidx{i}$ and the wave-function renormalization factors $Z_\lin{\alpha,\! i} \equiv 1 + \delta^\duin{i}_\alpha$, rather than $\delta\labidx{i}$. Second, the {\em kernels} that appear on the left-hand sides of the equations in~\cref{eq:bare_vs_renormalized_couplings_1} appear frequently in the following analysis. 

%%%%%%%%%%%%%%%%%%%%%%%%%%%%%%%%%%%%%%%%%%%%%%%%%%%%%%%%%%%%%%%%%%%%%%%%%%%%%%%%%%%%%%%%%%%%%%%%%%%%%%%%%%%%
%
\subsection{Renormalized 2PI action}
%
%%%%%%%%%%%%%%%%%%%%%%%%%%%%%%%%%%%%%%%%%%%%%%%%%%%%%%%%%%%%%%%%%%%%%%%%%%%%%%%%%%%%%%%%%%%%%%%%%%%%%%%%%%%%

Rewriting the bare fields, propagators, and parameters in terms of their renormalized counterparts, using the notations defined above, we find the following renormalized 2PI effective action in complex Keldysh time variables:
\begin{align}
    \Gamma_{\rm{2PI}}[\varphi,\chi;\Delta] = 
    & {\cal S}_{\rm cl}[\varphi,\chi] 
    - \frac{i}{2} \text{Tr} \ln\left[ Z_0\,\Delta\right] 
    + \frac{i}{2} \text{Tr} [\Delta^{-1}_{0} \Delta] 
\nonumber \\
    &+ \delta {\cal S}_{\rm cl}[\varphi,\chi]
     + \frac{i}{2}\text{Tr} [\delta \Delta^{-1}_{0}\Delta]
     + \Gamma_{2}[\varphi,\chi,\Delta\, ;\slab{i}]\,.
    \label{eq:Gamma_2PI_renorm}
\end{align}
In the real time representation ${\cal S}_{\rm cl}$ is the same as in~\cref{eq:Bare_classical_action}, with the fields replaced by the renormalized ones and with $\mu^2_{\alpha} \to \mu^2_\lin{\alpha,\!2}$ and $\lambda_{\alpha\beta} \to \lambda_\lin{\alpha\beta}^\luin{4}$. The classical counterterm action has the same structure, with the functional form
\begin{align}
    \delta {\cal S}_{\rm cl}[\varphi,\chi] \equiv 
    &\int_x
    \Big(  \sfrac{1}{2} \delta^\duin{2}_\lin{\varphi} (\partial_{\mu} \varphi)^2
         + \sfrac{1}{2} \delta^\duin{2}_\lin{\chi}    (\partial_{\mu} \chi)^2
         + \sfrac{1}{2} \delta^\duin{2}_{\mu\varphi} \varphi^2 
         + \sfrac{1}{2} \delta^\duin{2}_{\mu\chi}    \chi^2 
\nonumber \\
    &\qquad - \sfrac{1}{4!}\dlppidx{4} \varphi^4
            - \sfrac{1}{4!}\dlccidx{4} \chi^4 
            - \sfrac{1}{4} \dlpcidx{4} \varphi^2\chi^2 \Big)\,.
 \label{eq:classical_counter_term_action}
 \end{align}
Next, the free renormalized inverse propagator $i\Delta_0^{-1}$ has the same form as the free bare inverse propagator~\cref{eq:inv_prop}, only with $\mu^2_{\alpha} \to \mu^2_\lin{\alpha,\!0}$ and $\lambda_{\alpha\beta} \to \labidx{2}$. Note the difference in the indices in this replacement rule compared to the one used for the renormalized classical action. The corresponding counterterm part is given by 
\begin{align}
    i\delta(\Delta^{ab}_{0;\,\alpha\beta})^{-1}(x,y) = 
    - \Big( (Z^\frac12_\lin{\alpha,\! 0} Z^\frac12_\lin{\beta,\! 0}-1)\Box_x
    + \delta\mathcal{M}^2_{\alpha\beta}(\varphi^a,\chi^a)\Big)\,\delta^{(4)}(x-y)\delta ^{ab}
\label{eq:inv_ct_prop}
\end{align}
where the counterterm mass matrix is 
\begin{align}
    \delta\mathcal{M}^2(\varphi,\chi) = 
    \begin{pmatrix}
    -\delta^\duin{0}_\lin{\varphi} + \frac12\dlppidx{2}\varphi^2 + \frac12\dlpcidx{2}\chi^2 
    & \dblpcidx{2}\varphi\chi
    \\[2mm]
    \dblpcidx{2}\varphi\chi 
    & -\delta^\duin{0}_\lin{\chi} + \frac12\dlccidx{2}\chi^2+ \frac12\dlpcidx{2}\varphi^2
   \end{pmatrix}.	
\label{eq:ct_mass_matrix}
\end{align}
Again, one should pay attention to the difference in indices in counterterms between~\cref{eq:classical_counter_term_action} and~\cref{eq:ct_mass_matrix}. Finally, the Hartree approximation interaction term, written explicitly in terms of renormalized parameters and counterterms, is $\Gamma_2^{\rm HT} = \sum_\pm a\Gamma^{\rm ht}_{2}[\Delta^{aa}_{\alpha\beta}]$, where
\begin{align}
    \Gamma_2^{\rm ht}[\Delta_{\alpha\beta}] = 
    &-\frac{1}{8}\klpp{0} \int_x \Delta_{\varphi\varphi}^2
     -\frac{1}{8}\klcc{0} \int_x \Delta_{\chi\chi}^2
    \nonumber \\
    &-\frac{1}{4 }\klpc{0} \int_x \Delta_{\varphi\varphi}\Delta_{\chi\chi}
     -\frac{1}{4}\kblpc{0} \int_x \big(\Delta_{\varphi\chi}+\Delta_{\chi\varphi}\big)^2 \,.
    \label{eq:Gamma2_HT_renorm}
\end{align}
From~\cref{eq:self-energy} we see that this action gives rise to the following singular self-energy terms:
\begin{align}
    \Pi^{ac}_{\varphi\varphi}(x,y) 
    &= - \frac{ic}{2}\Big(\klpp{0} \Delta_{\varphi\varphi}(x,x) 
                        + \klpc{0} \Delta_{\varphi\chi}(x,x)  \Big) \delta^{ac} \delta^{(4)}(x-y),
    \nonumber \\    
    \Pi^{ac}_{\chi\chi}(x,y) 
    &= - \frac{ic}{2}\Big(\klcc{0} \Delta_{\chi\chi}(x,x) 
                        + \klpc{0} \Delta_{\chi\varphi}(x,x) \Big) \delta^{ac} \delta^{(4)}(x-y),
    \nonumber \\[0.7ex]    
    \Pi^{ac}_{\varphi\chi}(x,y) 
    &= - ic \kblpc{0} i\Delta_{\varphi\chi}(x,x) \delta^{ac} \delta^{(4)}(x-y),
    \label{eq:self-energy2}
\end{align}
where we have again left out the Keldysh indices from the local correlation functions, and used $\Delta_{\varphi\chi} = \Delta_{\chi\varphi}$ through symmetry.

%%%%%%%%%%%%%%%%%%%%%%%%%%%%%%%%%%%%%%%%%%%%%%%%%%%%%%%%%%%%%%%%%%%%%%%%%%%%%%%%%%%%%%%%%%%%%%%%%%%%%%%%%%
\paragraph{Renormalization conditions.}
%%%%%%%%%%%%%%%%%%%%%%%%%%%%%%%%%%%%%%%%%%%%%%%%%%%%%%%%%%%%%%%%%%%%%%%%%%%%%%%%%%%%%%%%%%%%%%%%%%%%%%%%%%

Let us now denote the zero-temperature minimum of the effective potential in field space by $(v,w)$. We will renormalize the propagator at this configuration, at zero momentum $p^2=0$, so that our renormalization conditions are
\begin{align}
    i(\Delta^{11}_{\alpha\beta}(p))^{-1} = p^2 - \hat m^2_{\alpha\beta} \,
    &,\quad \frac{\partial}{\partial p^2}i(\Delta^{11}_{\alpha\beta}(p))^{-1}\Big|_{p^2 = 0}= 1\,,
    \label{eq:os_renorm_prop}
    \\
    \frac{\delta \Gamma_{\rm 2PI}}{\delta \varphi}\bigg|_{p^2=0;\,(v,w)} = 0\,&, \quad \frac{\delta \Gamma_{\rm 2PI}}{\delta \chi}\bigg|_{p^2=0;\,(v,w)} = 0\,.
    \label{eq:os_renorm_field}
\end{align}
Here $\hat m^2$ is the symmetric mass matrix associated with the two-point function at the renormalization point. Its eigenvalues are the $p^2=0$ masses of the classical correlation function. They offer a convenient parametrization for the theory, but their connection to physical pole masses must be worked out separately. Moreover, conditions~\cref{eq:os_renorm_prop,eq:os_renorm_field} alone are far from sufficient to renormalize the theory. As we demonstrate, a significant part of the 2PI renormalization procedure can be carried out for arbitrary field configurations and for general out-of-equilibrium conditions, based on the idea of the cancellation of sub-divergences, with no reference to~\cref{eq:os_renorm_prop,eq:os_renorm_field}.  

%%%%%%%%%%%%%%%%%%%%%%%%%%%%%%%%%%%%%%%%%%%%%%%%%%%%%%%%%%%%%%%%%%%%%%%%%%%%%%%%%%%%%%%%%%%%%%%%%%%%%%%%%%
%
\subsection{Renormalized two-point function} 
\label{sec:renormalized_delta}
%
%%%%%%%%%%%%%%%%%%%%%%%%%%%%%%%%%%%%%%%%%%%%%%%%%%%%%%%%%%%%%%%%%%%%%%%%%%%%%%%%%%%%%%%%%%%%%%%%%%%%%%%%%%

Our procedure starts by simply writing the bare equations of motion~\cref{eq:EOM_phi,eq:EOM_chi,eq:EOM_Delta} in terms of renormalized parameters and variables. We begin from the EOM for the propagator~\cref{eq:EOM_Delta}. Given the singular Hartree self-energy functions~\cref{eq:self-energy2}, we can write it as:
\begin{equation}
(Z_\lin{\alpha,0}\Box_x \delta_{\alpha\gamma} + m^2_{\alpha\gamma}(x) )\Delta^{1a}_{\gamma\beta}(x,y) = -i\delta^{1a}\delta_{\alpha\beta}\delta^{(4)}(x-y),
\label{eq:Delta_eom_direct_space}
\end{equation}
where we chose the first Keldysh index to be 1, and the elements of the $m^2_{\alpha\beta}$ matrix, which in general is different from $\hat m^2_{\alpha\beta}$, are defined as 
\begin{subequations}
\begin{align}
    m_{\varphi\varphi}^2 &= 
    - (\ssqmupF{0}) 
     + \sfrac12 \klpp{2} \varphi^2 
     + \sfrac12 \klpc{2} \chi^2
    \nonumber \\[0.7ex]
    & + \sfrac12 \klpp{0} \Delta_{\varphi\varphi} 
      + \sfrac12 \klpc{0} \Delta_{\chi\chi}\,, 
    \label{eq:m2_phiphi}
    \\[1.4ex]
    m_{\chi\chi}^2 &= 
    -(\ssqmucF{2})
     + \sfrac12 \klcc{2} \chi^2 
     + \sfrac12 \klpc{2} \varphi^2
    \nonumber \\[0.7ex]
    & + \sfrac12 \klcc{0} \Delta_{\chi\chi} 
      + \sfrac12 \klpc{0} \Delta_{\varphi\varphi}\,,
    \label{eq:m2_chichi}
    \\[1.4ex]
    m_{\varphi\chi}^2 &= 
    \kblpc{2} \varphi\chi
    +\kblpc{0}\Delta_{\varphi\chi}\,.
\label{eq:Dm2_phichi}
\end{align}
\label{eq:mass_squared}
\end{subequations}
\vskip-0.4cm\noindent
We suppressed the space-time coordinates in classical fields $\varphi(x)$ and $\chi(x)$ and in local correlation functions $\Delta_{\alpha\beta}(x,x)$, but at this point, they can indeed be fully general functions. The key observation is that if the theory is to be renormalizable, the mass squared matrix elements~\cref{eq:mass_squared} must be finite for arbitrary classical field configurations and in an arbitrary environment, {\em e.g.}~for any, even a non-equilibrium $\Delta_{\alpha\beta}(x,x)$. 

%%%%%%%%%%%%%%%%%%%%%%%%%%%%%%%%%%%%%%%%%%%%%%%%%%%%%%%%%%%%%%%%%%%%%%%%%%%%
\paragraph{The divergence structure of the local correlation function.}
%%%%%%%%%%%%%%%%%%%%%%%%%%%%%%%%%%%%%%%%%%%%%%%%%%%%%%%%%%%%%%%%%%%%%%%%%%%%

To proceed, we have to find out the divergence structure of the local correlation functions. This step is crucial in the cancellation of subdivergences method~\cite{Fejos:2007ec}, which undercuts most of the complexity involved in the systematic approach of~\cite{Berges:2005hc}. To this end we for now assume $\varphi$ and $\chi$ and $\Delta_{\alpha\beta}$ are constants. We can then Fourier transform the equation~\cref{eq:Delta_eom_direct_space}, to obtain
\begin{equation}
(Z_\lin{\alpha,\!0}\, p^2 - m^2_{\alpha\beta}) \Delta^{1a}_{\alpha\beta} = i\delta^{1a}\delta_{\alpha\beta}.
\label{eq:Delta_eom_momentum_space}
\end{equation}
From here we already see that the second renormalization condition in~\cref{eq:os_renorm_prop} imposes $Z_\lin{\alpha,0} = 1$. This reflects the fact that the singular Hartree self-energy functions~\cref{eq:self-energy2} are $p^2$-independent in the momentum space representation.

The mass matrix $m^2_{\alpha\beta}$ is real and symmetric, and it can be diagonalized by an orthogonal transformation. In this diagonal basis, equation~\cref{eq:Delta_eom_momentum_space} allows the solution:
\begin{equation}
\Delta^\lin{\!11}_{ii} = \frac{i}{p^2-m_i^2+i\epsilon} + 2\pi f_i(p_0) \delta(p^2-m_i^2).
\label{eq:propagator_solutions}
\end{equation}
For the Wightman function $\Delta^\lin{\!12}_{ii} = \Delta^\lin{\!<}$, only the spectral solution is allowed. The functions $\Delta^\lin{\!11}$ and $\Delta^\lin{\!<}$ are exactly related: $\Delta^\lin{\!11}=\Delta^r + \Delta^\lin{\!<}$, and since the retarded function $\Delta^{r}$ does not have a local limit, one finds $\Delta^\lin{\!11}(x,x)=\Delta^\lin{\!<}(x,x)$. In thermal equilibrium the functions $f_i(p_0)$ would go to the Bose-distributions $f_i(p_0) \to f_{\rm BE,i}(|p_0|/T)$. Certainly in the thermal case, but also in any physical non-equilibrium situation, the integral over the spectral part of the solution~\cref{eq:propagator_solutions} is finite. The only divergence in $\Delta_{ii}(x,x)$ then arises from the integral over the vacuum Feynman propagator in~\cref{eq:propagator_solutions}. This contribution can be evaluated using dimensional regularization:
\begin{align}
    \Delta_{0}(m_i^2) \equiv Q^{2\epsilon}\int\frac{d^dp}{(2\pi)^d}\frac{i}{p^2-m_i^2+i\epsilon} &= -\frac{m^2_i}{16\pi^2\epsilon} + \frac{m^2_i}{16\pi^2}\left[\ln\left(\frac{m^2_i}{Q^2}\right)-1\right]
    \nonumber \\
    &\equiv m_i^2\DeltaEps + \DeltaFO(m_i^2,Q),
    \label{eq:A0_int}
\end{align}
with the usual definitions: $2\epsilon = 4-d$ and $Q \equiv 4\pi e^{-\gamma_E}\,\bar{Q}$ being an arbitrary renormalization scale. We can thus identify the divergent and finite pieces in Eq.~\cref{eq:A0_int}: 
\begin{align}
    \Delta_{ii}(x,x)\equiv m^2_i \DeltaEps + \DeltaFO(m^2_i,Q) 
    + \delta \Delta_{{\rm F}ii}(x,x)\,,
    \label{eq:correlator_decomp}
\end{align}
where $\delta \Delta_{{\rm F}ii}$ denotes the finite contribution to the local correlation function from the spectral part of the solution~\cref{eq:propagator_solutions}. In this paper, we are only interested in the case where $\delta \Delta_{{\rm F}ii}$ represents the finite temperature correction, but the result is obviously more general than that. 

The same matrix $U_{\theta}$, that diagonalizes the mass matrix $m_{\alpha\beta}^2\to {\rm diag}(m_1^2,m_2^2)$, also diagonalizes the local correlation function $\Delta_{\alpha\beta} \to \Delta_{\rm diag} \equiv {\rm diag}(\Delta_1, \Delta_2)$. Explicitly,
\begin{equation}
    \Delta_{\rm diag} = U_{\theta}\, \Delta\,U^T_{\theta}\,,\quad {\rm with} \quad U_\theta \equiv 
    \begin{pmatrix}
        \cos \theta &-\sin \theta \\
        \sin \theta &\phantom{-} \cos \theta
    \end{pmatrix}
    \label{eq:prop_diag}
\end{equation} 
and $\tan 2\theta = 2m^2_{\varphi\chi}/(m^2_{\chi\chi}-m^2_{\varphi\varphi})$. This gives the relations 
\begin{subequations}
    \begin{align}
        \Delta_{\varphi\varphi} &= \sfrac12\left[\Delta_1 + \Delta_2 
                                  +\cos 2\theta \left(\Delta_1 - \Delta_2\right)\right],
    \\[0.5ex]
        \Delta_{\chi\chi} &= \sfrac12\left[\Delta_1 + \Delta_2 
                                  -\cos 2\theta \left(\Delta_1 - \Delta_2\right) \right],
    \\[0.5ex]
        \Delta_{\varphi\chi} &= \sfrac{1}{2}\sin 2\theta\left(\Delta_2 - \Delta_1\right).
    \end{align}
    \label{eq:basis_convert_prop}%
\end{subequations}
Obviously, similar relations hold for the mass-squared matrix. It is now easy to see that the divergent part of $\Delta_{\alpha\beta}$ is just $m^2_{\alpha\beta}\Delta_\epsilon$, so that we can write the whole local correlation function as
\begin{equation}
\Delta_{\alpha\beta}(x,x) \equiv m^2_{\alpha\beta}\Delta_\epsilon + \DeltaF_{\alpha\beta},
\label{eq:Delta_divided}
\end{equation}
where $\DeltaF_{\alpha\beta}\equiv \Delta^{\alpha\beta}_{\rm FO} + \delta \DeltaF_{\alpha\beta}$. 

%%%%%%%%%%%%%%%%%%%%%%%%%%%%%%%%%%%%%%%%%%%%%%%%%%%%%%%%%%%%%%%%%%%%%%%%%%%%%%%%%%%%%%%%%%%%%%%%%%%%%%%%%%%%%%%%
\subsubsection{Cancellation of subdivergences}
%%%%%%%%%%%%%%%%%%%%%%%%%%%%%%%%%%%%%%%%%%%%%%%%%%%%%%%%%%%%%%%%%%%%%%%%%%%%%%%%%%%%%%%%%%%%%%%%%%%%%%%%%%%%%%%%

The cancellation of subdivergences method imposes the idea explained above with an important twist: after inserting the split~\cref{eq:Delta_divided} into equations~\cref{eq:mass_squared}, one assumes that {\rm only the manifestly finite terms make up the finite mass-squared matrix} $m_{\alpha\beta}^2$. That is:
\begin{subequations}
\begin{align}
    \sqmpp &\equiv -\sqmupF{0}
    +\sfrac12\lppidx{2} \varphi^2 
    +\sfrac12\lpcidx{2} \chi^2
    +\sfrac12\lppidx{0}\DeltaF_\lin{\varphi\varphi}
    +\sfrac12\lpcidx{0}\DeltaF_\lin{\chi\chi}\,, 
    \label{eq:sqmphiphi}
    \\[1.2ex]
    \sqmcc &\equiv -\sqmucF{0}
    +\sfrac12\lccidx{2} \chi^2 
    +\sfrac12\lpcidx{2} \varphi^2
    +\sfrac12\lccidx{0}\DeltaF_\lin{\chi\chi} 
    +\sfrac12\lpcidx{0}\DeltaF_\lin{\varphi\varphi}\,,
    \label{eq:sqmchichi}
    \\[1.2ex]
    \sqmpc &\equiv \blpcidx{2}\varphi\chi
    +\blpcidx{0}\DeltaF_\lin{\varphi\chi}. 
    \label{eq:sqmphichi}
\end{align}
\label{eq:sqm_mixedbasis}
\end{subequations}

\vskip-0.3cm
\noindent 
For this to work, all counterterm contributions must cancel along with the divergent parts of the local correlation functions. This imposes the following cancellation conditions:
\begin{subequations}
\begin{align}
- &\delta^\duin{0}_{\mu\varphi}
     + \sfrac12\dlppidx{2} \varphi^2 
     + \sfrac12\dlpcidx{2} \chi^2
     + \sfrac12\dlppidx{0} \DeltaF_\lin{\varphi\varphi}
     + \sfrac12\dlpcidx{0} \DeltaF_\lin{\chi\chi} 
    \nonumber \\[1ex]
    &
    \qquad \qquad \qquad \qquad 
    +\sfrac12\klpp{0} \DeltaEps \sqmpp
    +\sfrac12\klpc{0} \DeltaEps \sqmcc = 0\,,
    \label{eq:cancel_Deltapp}
\\[1.5ex]
- &\delta^\duin{0}_{\mu\chi}
    + \sfrac12\dlccidx{2} \varphi^2 
    + \sfrac12\dlpcidx{2} \chi^2
    + \sfrac12\dlccidx{0} \DeltaF_\lin{\chi\chi}
    + \sfrac12\dlpcidx{0} \DeltaF_\lin{\varphi\varphi}
    \nonumber \\[1ex]
    &
    \qquad \qquad \qquad \qquad
    +\sfrac12\klcc{0} \DeltaEps\sqmcc
    +\sfrac12\klpc{0} \DeltaEps\sqmpp = 0\,,
    \label{eq:cancel_Deltacc}
\\[1.5ex]
 &\dblpcidx{2} \varphi \chi
    +\dblpcidx{0}\DeltaF_\lin{\varphi\chi} 
    + \kblpc{0}\DeltaEps\sqmpc= 0\,.
    \label{eq:cancel_Deltapc}
    \end{align}
    \label{eq:cancel_Delta_subdiv}
\end{subequations}

\vskip-0.4cm\noindent
Inserting the expressions~\cref{eq:sqm_mixedbasis} to the $\DeltaEps m^2_{\alpha\beta}$-terms, and demanding that potential divergences related to different functional structures cancel independently, one obtains systems of equations to determine the counterterms. For example, the equation~\cref{eq:cancel_Deltapp} can be written as $A + B\varphi^2 + C\chi^2 + D\DeltaF_\lin{\varphi\varphi} + E\DeltaF_\lin{\chi\chi} = 0$, where all coefficients are constant combinations of mass parameters, couplings, counterterms and $\DeltaEps$. Clearly, the only way to make this expression vanish for arbitrary field configurations is that each coefficient vanishes separately. This gives us the following five independent renormalization conditions:
\begin{align}
 \delta^\duin{0}_{\mu\varphi} 
    + \sfrac12 \sqmupF{0} \klpp{0} \DeltaEps 
    + \sfrac12 \sqmucF{0} \klpc{0} \DeltaEps &= 0\,,
\nonumber \\[1ex]
 \dlppidx{i}
    + \sfrac12 \lppidx{i} \klpp{0} \DeltaEps 
    + \sfrac12 \lpcidx{i} \klpc{0} \DeltaEps &= 0 \,,
\nonumber \\[1ex]
 \dlpcidx{i}
    + \sfrac12 \lpcidx{i} \klpp{0} \DeltaEps 
    + \sfrac12 \lccidx{i} \klpc{0} \DeltaEps &= 0,
\label{eq:cancel_Delta}
\end{align}
with $i=0,2$. Equation~\cref{eq:cancel_Deltacc} can be treated similarly and it gives conditions that can be obtained from~\cref{eq:cancel_Delta} by switching $\varphi \leftrightarrow \chi$ everywhere. Finally equation~\cref{eq:cancel_Deltapc} gives two more conditions:
\begin{align}
 \dblpcidx{i}
    + \sfrac12 \blpcidx{i} \kblpc{0} \DeltaEps &= 0 \,,
\label{eq:cancel_Deltab}
\end{align}
again with $i=0,2$.

%%%%%%%%%%%%%%%%%%%%%%%%%%%%%%%%%%%%%%%%%%%%%%%%%%%%%%%%%%%%%%%%%%%%%%%%%%%%%%%%%%%%%%%%%%%%%%%%%%%%%%%%%%%%%%%%%%%
\paragraph{Relations among auxiliary couplings.}
%%%%%%%%%%%%%%%%%%%%%%%%%%%%%%%%%%%%%%%%%%%%%%%%%%%%%%%%%%%%%%%%%%%%%%%%%%%%%%%%%%%%%%%%%%%%%%%%%%%%%%%%%%%%%%%%%%%

In addition to giving a finite correlation function, equations~\cref{eq:cancel_Delta,eq:cancel_Deltab} have other important consequences. For example, from \cref{eq:cancel_Deltab}, we can see that if couplings $\blpcidx{0}$ and $\blpcidx{2}$ are set to be the same at some scale $Q_0$, they remain equal for all $Q$. One can prove this by solving the renormalization group equations explicitly, but the result can be seen directly as follows: given the equality of couplings at $Q_0$, equations~\cref{eq:cancel_Deltab} imply that also the counterterms $\dblpcidx{i}$, and hence the kernels $\blpcidx{i}+\dblpcidx{i}$ are the same for $i=0,2$ at $Q_0$. As is evident from~\cref{eq:bare_vs_renormalized_couplings_1}, these kernels are related to the bare couplings $\blpcidx{i}+\delta\blpcidx{i}$ by multiplicative combinations of $Z_\lin{\alpha,\!i}$-factors, and since the wave-function renormalization factors are finite and scale-independent in the Hartree approximation\footnote{We already showed that $Z_\lin{\alpha,\!0}=1$ and in Appendix~\cref{app:Z2_wave_function_ren_factors} we show also that the $Z_\lin{\alpha,\!2}$ are finite and scale independent.}, these kernels determine the renormalization group running to this order. Given the equality of the couplings at $Q_0$ and the same running equations, they must be equal for all $Q$. One can indeed choose these couplings to be equal at the $\widebar{\rm MS}$-renormalization scale, and so we can set:
\begin{equation}
\blpcidx{2} = \blpcidx{0} \quad \Leftrightarrow \quad 
\blpcidx{2}\!+\!\dblpcidx{2} = \blpcidx{0}\!+\!\dblpcidx{0}.
\label{eq:blambda2_blambda0_relation}
\end{equation}
Comparing pairwise the second and fourth and the third and fifth equations in~\cref{eq:cancel_Delta}, as well as their counterparts in an equivalent set of equations where $\varphi$ and $\chi$ are exchanged, one can similarly show that it is also possible to set
\begin{equation}
   \labidx{2} = \labidx{0}   
     \quad \Leftrightarrow \quad 
   \slab{2}   = \slab{0}.
\label{eq:lambda2_lambda0_relation}
\end{equation}
These equations significantly reduce the number of independent auxiliary parameters and the amount of necessary bookkeeping in what follows.

%%%%%%%%%%%%%%%%%%%%%%%%%%%%%%%%%%%%%%%%%%%%%%%%%%%%%%%%%%%%%%%%%%%%%%%%%%%%%%%%%%%%%%%%%%%%%%%%%%%%%%%%%%%%%%%%%%%
\paragraph{Renormalized equations of motion for Wightman functions.}
%%%%%%%%%%%%%%%%%%%%%%%%%%%%%%%%%%%%%%%%%%%%%%%%%%%%%%%%%%%%%%%%%%%%%%%%%%%%%%%%%%%%%%%%%%%%%%%%%%%%%%%%%%%%%%%%%%%

Before continuing the renormalization process further, we point out that the above results generalize to spatially varying classical fields and correlation functions, which are each solved dynamically from their evolution equations. Indeed, the division of the mass-squared matrix into~\cref{eq:sqm_mixedbasis} and the proof of vanishing of~\cref{eq:cancel_Delta_subdiv} by the conditions~\cref{eq:cancel_Delta} go through as such for spacetime-dependent configurations. The key observation is that the local correlation function can still be written as~\cref{eq:Delta_divided}, with the only divergence given by $m^2_{\alpha\beta}(x)\DeltaEps$. The finite part of the correlation functions is solved directly from the renormalized equation of motion for the Wightman function, which is now simply given by:
\begin{equation}
(\Box_x \delta_{\alpha\gamma} + m^2_{\alpha\gamma}(x) )\Delta^<_{\gamma\beta}(x,y) = 0.
\label{eq:Delta_EOM_xspace}
\end{equation}
This proves the finiteness of the two-point function and its equation of motion in the Hartree approximation. However, we still need to connect the various auxiliary $\widebar{\rm MS}$-parameters appearing in the mass-squared matrix elements to some physical parameters.

%%%%%%%%%%%%%%%%%%%%%%%%%%%%%%%%%%%%%%%%%%%%%%%%%%%%%%%%%%%%%%%%%%%%%%%%%%%%%%%%%%%%%%%%%%%%%%%%%%%%%%%%%%%%%%%%%
\paragraph{Renormalized mass parameters.} 
%%%%%%%%%%%%%%%%%%%%%%%%%%%%%%%%%%%%%%%%%%%%%%%%%%%%%%%%%%%%%%%%%%%%%%%%%%%%%%%%%%%%%%%%%%%%%%%%%%%%%%%%%%%%%%%%%

The elements of the mass matrix in~\cref{eq:Delta_EOM_xspace} are finite and solvable from the coupled set of gap equations~\cref{eq:sqm_mixedbasis}, when the finite part of the local correlation function in~\cref{eq:Delta_divided} is given. This mass matrix coincides with the renormalized mass matrix $\hat m^2_{\alpha\beta}$, when $(\varphi,\chi) = (v,w)$, corresponding to the minimum of the effective action and $\delta \DeltaF_{\alpha\beta}=0$ (an obvious, although not a mandatory choice). That is,
\begin{equation}
\hat m^2_{\alpha\beta} = m^2_{\alpha\beta}(v,w;\DeltaFO).
\label{eq:mhatmatrix}
\end{equation}
The diagonal elements of this equation can be used to write the Lagrangian parameters $\sqmupF{0}$ and $\sqmucF{0}$ in terms of $\hat m^2_{\alpha\alpha}$ and the (yet undetermined) vacuum expectation values $(v,w)$. From the off-diagonal equation~\cref{eq:sqmphichi} at the renormalization point, one can derive a connection between the couplings $\blpcidx{0}$ and $\lpcidx{4}$:
\begin{equation}
\blpcidx{0}= \frac{\lpcidx{4}}{1+\lpcidx{4}\frac{\delta\hat\Delta}{\delta \hat m^2}},
\label{eq:lpc0_vslpc4}
\end{equation}
where $\delta\hat\Delta \equiv \hDeltaFO(\hat m_2^2)-\hDeltaFO(\hat m_1^2)$ and $\delta \hat m^2 \equiv \hat m_2^2-\hat m_1^2$. This exhausts the cancellation of subdivergences and the renormalization conditions~\cref{eq:os_renorm_prop} for the propagator. To derive further connections between couplings and between the mass parameters and the vacuum expectation values (VEVs), we need to renormalize the field equations.

%%%%%%%%%%%%%%%%%%%%%%%%%%%%%%%%%%%%%%%%%%%%%%%%%%%%%%%%%%%%%%%%%%%%%%%%%%%%%%%%%%%%%%%%%%%%%%%%%%%%%%%%%%%%%%%%
%
\subsection{Renormalized field equations}
%
%%%%%%%%%%%%%%%%%%%%%%%%%%%%%%%%%%%%%%%%%%%%%%%%%%%%%%%%%%%%%%%%%%%%%%%%%%%%%%%%%%%%%%%%%%%%%%%%%%%%%%%%%%%%%%%%

We now complete the renormalization procedure by determining the counterterms $\dlabidx{4}$ and $\delta^\duin{2}_\lin{\mu\alpha}$. We do this by imposing the cancellation of subdivergences principle on the renormalized versions of the field equations~\cref{eq:EOM_phi,eq:EOM_chi}:
\begin{subequations}
\begin{align}
    & \Big( Z_\lin{\varphi,\!2}\Box_x - (\ssqmupF{2})
           + \sfrac{1}{6} \klpp{4} \varphi^2 
           + \sfrac{1}{2} \klpc{4} \chi^2 
    \nonumber \\
    & + \sfrac{1}{2} \klpp{2} \Delta_{\varphi\varphi}  
      + \sfrac{1}{2} \klpc{2} \Delta_{\chi\chi} \Big) \varphi
                   + \kblpc{2} \Delta_{\varphi\chi} \chi = 0,
    \label{eq:min_cond_phi}
    \\[1ex]
    & \Big( Z_\lin{\chi,\!2}\Box_x - (\ssqmucF{2})
      + \sfrac{1}{6}\klcc{4} \chi^2 
      + \sfrac{1}{2}\klpc{4} \phi^2 
    \nonumber \\
    & + \sfrac{1}{2}\klcc{2} \Delta_{\chi\chi} 
      + \sfrac{1}{2}\klpc{2} \Delta_{\varphi\varphi} \Big) \chi
                 + \kblpc{2} \Delta_{\varphi\chi}\varphi = 0 \,.
    \label{eq:min_cond_chi}
\end{align}
\end{subequations}
One works with these equations similarly to the propagator equation above, inducing the split~\cref{eq:Delta_divided} and using expressions~\cref{eq:sqm_mixedbasis} in the emerging $m^2_{\alpha\beta}\DeltaEps$-terms\footnote{The finite mass matrix $\bar m^2_{\alpha\beta}$ arising here in fact initially differs from $m^2_{\alpha\beta}$, such that now all auxiliary couplings on the {\em r.h.s.}~of~\cref{eq:sqm_mixedbasis} have index 2. However, using~\cref{eq:blambda2_blambda0_relation,eq:lambda2_lambda0_relation} the mass matrices become equal after all with $\bar m^2_{\alpha\beta} \to m^2_{\alpha\beta}$, and we thus do not make this distinction here.}. This results in a large number of equations analogous to~\cref{eq:cancel_Delta,eq:cancel_Deltab}, that give more relations among the auxiliary couplings. Most of them are degenerate with~\cref{eq:blambda2_blambda0_relation,eq:lambda2_lambda0_relation} and we do not display them here, but we do give the list of the kernels in \eqref{eq:limiting_kernels} of Appendix~\cref{app:Z2_wave_function_ren_factors}, using the conditions~\cref{eq:blambda2_blambda0_relation,eq:lambda2_lambda0_relation}.
The only new relation comes from setting the coefficients of the linear $\varphi$-term in~\cref{eq:min_cond_phi} and the linear $\chi$-term in~\cref{eq:min_cond_chi} to zero after the split. They give rise to constraints:
\begin{align}
 \delta^\duin{2}_{\mu\varphi} 
    + \sfrac12 \sqmupF{2} \klpp{0} \DeltaEps 
    + \sfrac12 \sqmucF{2} \klpc{0} \DeltaEps &= 0,
\nonumber \\[1ex]
 \delta^\duin{2}_{\mu\chi} 
    + \sfrac12 \sqmucF{2} \klcc{0} \DeltaEps 
    + \sfrac12 \sqmupF{2} \klpc{0} \DeltaEps &= 0.
\end{align}
Combined with the first equation in~\cref{eq:cancel_Delta}, and its analog with $\varphi$ and $\chi$ interchanged, these allow us to set 
\begin{equation}
    \mu^{2}_\lin{\alpha,\!2} \!+\! \delta^\duin{2}_{\mu\alpha} \;=\; 
    \mu^{2}_\lin{\alpha,\!0} \!+\! \delta^\duin{0}_{\mu\alpha}, 
    \quad {\rm with} \quad \alpha = \varphi, \chi.
\label{eq:mu2mu0_relation}
\end{equation}
Together with~\cref{eq:lambda2_lambda0_relation,eq:blambda2_blambda0_relation} this equation completes the degeneracy of parameters corresponding to no external classical fields and two external classical fields in the Hartree approximation.

Given~\cref{eq:blambda2_blambda0_relation,eq:lambda2_lambda0_relation,eq:mu2mu0_relation}, we can make a shortcut in the renormalization process if we, in addition, define:
\begin{subequations}
\begin{align}
    \sfrac16\klaa{4} & \;=\; - \sfrac13\laaidx{4} + \sfrac12\klaa{2},
    \\[0.2ex]
    \sfrac12\klpc{4} & \;=\;\;\; - \lpcidx{4} + \sfrac12\klpc{0} + \sblpc{0}.
\end{align}
\label{eq:ct4-0_relations}
\end{subequations}

\vskip-0.4cm\noindent
Using these relations and~\cref{eq:blambda2_blambda0_relation,eq:lambda2_lambda0_relation,eq:mu2mu0_relation}, it is straightforward to show that the potentially divergent terms remaining after the split cancel and that~\cref{eq:min_cond_phi,eq:min_cond_chi} reduce to:
\begin{align}
     Z_\lin{\varphi,\!2}\Box_x \varphi 
    + m^2_{\varphi\varphi}\varphi 
    + m^2_{\varphi\chi}\chi 
    & \;=\; \sfrac{1}{3}\lppidx{4}\varphi^3
                      + \lpcidx{4}\chi^2\varphi,
    \nonumber \\
     Z_\lin{\chi,\!2}\Box_x \chi 
    + m^2_{\chi\chi}\chi 
    + m^2_{\varphi\chi}\varphi 
    & \;=\; \sfrac{1}{3}\lccidx{4}\chi^3
                      + \lpcidx{4}\varphi^2\chi.
\label{eq:eom_fields_final}
\end{align}
The wave-function renormalization constants $Z_\lin{\alpha,\!2}$ are still undefined. Their evaluation is a somewhat lengthy process, and since we do not need them in deriving the Hartree effective potential, we relegate their derivation to Appendix~\cref{app:Z2_wave_function_ren_factors}.

The reader might be wondering how we arrived at definitions~\cref{eq:ct4-0_relations}. At this point, one can think of them as a guess\footnote{\label{fn:guess} In fact, the couplings $\labidx{4}$ can be understood as mere placeholders for solutions to Eqs.~\cref{eq:ct4-0_relations}, which will be properly identified only later, when we connect the auxiliary parameters to physical ones. However, this eventually leads to a correct identification of the physical parameters and makes the notation cleaner in the following discussion.}, motivated by a will to get the field equations renormalized by the same counterterms that renormalized the two-point equation of motion, and to get them into the specific form~\cref{eq:eom_fields_final}. Indeed, at the tree-level~\cref{eq:eom_fields_final} would be equivalent with
\begin{equation}
Z_\lin{\alpha,\!2}\Box_x\eta  - \frac{{\rm d}V}{{\rm d}\eta} = 0,
\label{eq:eom_fields_withV}
\end{equation}
for $\eta = \varphi,\chi$. We shall later see that this is also true at the Hartree level: equations~\cref{eq:eom_fields_final} and~\cref{eq:eom_fields_withV} are equivalent when $V$ is identified with the Hartree effective potential.

%%%%%%%%%%%%%%%%%%%%%%%%%%%%%%%%%%%%%%%%%%%%%%%%%%%%%%%%%%%%%%%%%%%%%%%%%%%%%%%%%%%%%%%%%%%%%%%%%%%%%%%%%%%
\paragraph{Connecting vacuum expectation values to mass matrix.}
%%%%%%%%%%%%%%%%%%%%%%%%%%%%%%%%%%%%%%%%%%%%%%%%%%%%%%%%%%%%%%%%%%%%%%%%%%%%%%%%%%%%%%%%%%%%%%%%%%%%%%%%%%%

Equations~\cref{eq:eom_fields_final} restricted to solutions that are constants in space and time correspond to our renormalization conditions~\cref{eq:os_renorm_field}. These solutions allow us to find relations between the mass matrix $\hat m^2_{\alpha\beta}$~\cref{eq:mass_matrix} and the vacuum expectation values of the fields at the global minimum. Indeed, employing~\cref{eq:eom_fields_final} at $(\varphi,\chi)=(v,w)$ we find
\begin{align}
      \Big[\hsqmpp-\frac{\lpp^{(4)}}{3}v^2\Big]v
    + \Big[\hsqmpc-\lpcidx{4}\,vw\Big]w = 0\,,
\nonumber\\
       \Big[\hsqmcc-\frac{\lccidx{4}}{3}v^2\Big]w
     + \Big[\hsqmpc-\lpcidx{4}\,vw \Big]v 
    = 0\,. 
\label{eq:field_eqn_rel}
\end{align}
In the square parentheses, one recovers the usual tree-level relations among elements of the mass matrix, the quartic couplings, and the values of the background field:
\begin{equation}
    \hat{m}^2_{\varphi\varphi} = \frac13\lppidx{4}v^2\,,\quad \hat{m}^2_{\chi\chi} = \frac13\lccidx{4}w^2\,,\quad {\rm and} \quad
    \hat{m}^2_{\varphi\chi} = \lpcidx{4}vw\,.
    \label{eq:mass_quartic_vev}
\end{equation}
Using~\cref{eq:basis_convert_prop} and \cref{eq:sqmphichi} one can show in particular that the equation for the off-diagonal mass-matrix element is consistent with~\cref{eq:lpc0_vslpc4}. 

At this point, we have succeeded in renormalizing our fields and correlation functions and their equations of motion, parametrizing them in terms of $p^2=0$ masses of the propagator and two sets of couplings: $\labidx{0}$ and  $\labidx{4}$. In order to further narrow down the number of parameters, we must compute the effective action and relate our parameters to the various $n$-point functions derived from it. However, it is useful to first study the running of the various auxiliary $\widebar {\rm MS}$-couplings and mass parameters.

%%%%%%%%%%%%%%%%%%%%%%%%%%%%%%%%%%%%%%%%%%%%%%%%%%%%%%%%%%%%%%%%%%%%%%%%%%%%%%%%%%%%%%%%%%%%%%%%%%%%%%%%%%%%%%%%%
%
\subsection{Running $\widebar{\rm MS}$-parameters}
\label{sec:running_of_couplings}
%
%%%%%%%%%%%%%%%%%%%%%%%%%%%%%%%%%%%%%%%%%%%%%%%%%%%%%%%%%%%%%%%%%%%%%%%%%%%%%%%%%%%%%%%%%%%%%%%%%%%%%%%%%%%%%%%%%

In this section, we study the evolution of the auxiliary couplings and mass parameters with the renormalization scale. As usual, the running equations are defined by demanding that the bare parameters are scale invariant,
\begin{equation}
    Q\frac{\partial}{\partial Q}\Big[Q^{2\epsilon}(\labidx{i}\!+\!\delta\labidx{i})\Big]_{\epsilon \to 0}  = 0\,, \quad 
    Q\frac{\partial}{\partial Q}\Big[Q^{2\epsilon}
    (\mu^{2}_\lin{\alpha,\!i}+\delta\mu^{2}_\lin{\alpha,\!i})\Big]_{\epsilon \to 0}  = 0\,.
\label{eq:rge_running_equations}
\end{equation}
We make two observations: First, because all wave-function renormalization factors are finite and scale-independent in the Hartree approximation, we can replace $\delta\lambda^\luin{i}_\lin{\alpha\beta} \to \delta^\duin{i}_\lin{\lambda\alpha\beta}$ and $\delta\mu^{2}_\lin{\alpha,\!i} \to \delta^\duin{i}_{\mu\alpha}$ in~\cref{eq:rge_running_equations} and work with kernels defined in~\cref{eq:bare_vs_renormalized_couplings_1}. Second, as was explained above, the renormalized parameters with indices (0) and (2) can be chosen equal in the Hartree approximation. This was because the equality of {\em e.g.}~couplings $\lambda^\luin{i}_\lin{\alpha\beta}$ at one scale imposes the equality of counterterms $\delta^\duin{i}_\lin{\lambda\alpha\beta}$ and hence equal running, as a result of the first observation we made. As a result, we will give explicit solutions in most cases only for the index (0) parameters.

The counterterms and the relevant kernels can be easily solved from~\cref{eq:cancel_Delta,eq:cancel_Deltab}. First, from~\cref{eq:cancel_Deltab} we find, keeping the dependence on the number of classical fields $i$,
\begin{equation}
 \sblpc{i} = \frac{\blpcidx{i}}{1 + \blpcidx{0}\DeltaEps},
\label{eq:cancel_Deltapp_3}
\end{equation}
where $i=1,2$. From this, it is straightforward to show that
\begin{equation}
    Q\frac{\partial \blpcidx{i}}{\partial Q} = \frac{\blpcidx{0}\blpcidx{i}}{8\pi^2}.
\label{eq:running_barcoupling0}
\end{equation}
Equations~\cref{eq:running_barcoupling0} are easily integrated, first for $i=0$ and then, using the solution for $\blpcidx{0}(Q)$, for $\blpcidx{2}(Q)$. One finds:
\begin{equation}
    \blpcidx{i}(Q) = \frac{\blpcidx{i}(Q_0)}{1 - \blpcidx{0}(Q_0)L_{QQ_0}} \,,
    \qquad L_{QQ_0} \equiv   \frac{1}{8\pi^2} \ln\Big(\dfrac{Q}{Q_0}\Big).
    \label{eq:running_barcoupling0_soln}
\end{equation}
This proves explicitly our earlier argument: choosing $\blpcidx{2}(Q_0)=\blpcidx{0}(Q_0)$ implies that the couplings are the same for all $Q$. The proof works similarly for other couplings and mass parameters, and so we concentrate only on index (0) couplings from now on.  

\begin{figure}[t!]
    \centering
    \includegraphics[width=0.95\columnwidth]{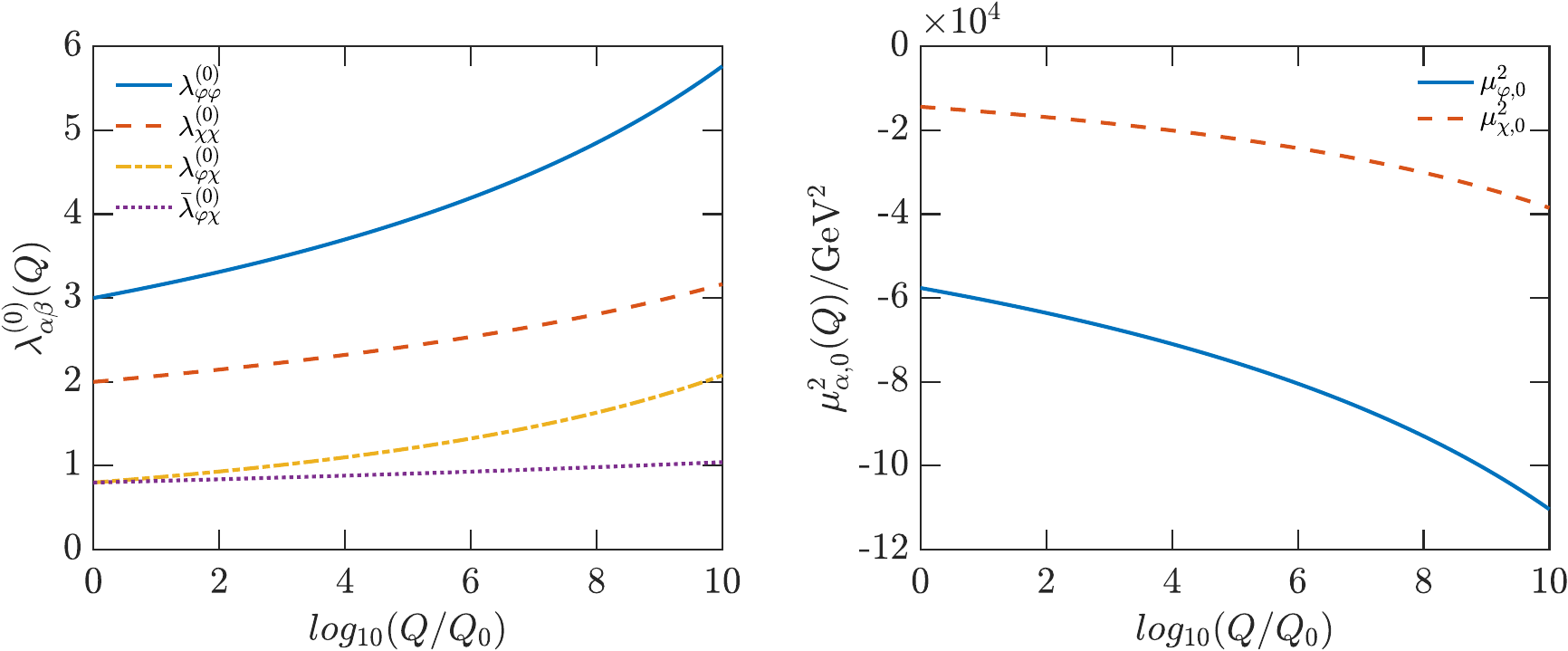}
    \caption{The renormalization scale running of the auxiliary couplings 
    $\lambda_{\alpha\beta}^{(0)}(Q)$ \textit{(left)} and mass parameters $\mu^2_{\alpha,0}(Q)$ \textit{(right)} in the Hartree approximation. We have chosen the parameters at the initial scale $Q_0 = 100$~GeV as 
    $\{\lambda^{(0)}_{\varphi\varphi},\lambda^{(0)}_{\chi\chi},\lambda^{(0)}_{\varphi\chi},\oline{\lambda}^{(0)}_{\varphi\chi}\} = \{3,2,0.8,0.8\}$ and $\{\mu^2_{\varphi,0},\mu^2_{\chi,0}\} = \{-240^2,-120^2\}\,{\rm GeV}^2$.}
    \label{fig:running_couplings}
\end{figure}

Next, from~\cref{eq:cancel_Delta} and from its equivalent with $\varphi$ and $\chi$ exchanged, we find: 
\begin{align}
     \slaa{0} &= \frac{1}{D}\big(\laaidx{0} + A\DeltaEps\big),
      \nonumber \\
     \slpc{0} &= \frac{\lpcidx{0}}{D},
\label{eq:coupled_eqs_1}
\end{align}
where $\alpha \in \{\varphi,\chi\}$ and we defined
\begin{align}
A &\;\equiv\; \sfrac12\big(\lppidx{0}\lccidx{0} - (\lpcidx{0})^2\big)
\nonumber \\[1ex]
D &\;\equiv\; 1 + \sfrac12(\lppidx{0}+\lccidx{0})\DeltaEps + \sfrac12 A \DeltaEps^2.
\label{eq:AD}
\end{align}
These are strongly coupled equations. They can be solved, however, by first observing that they imply the following simple equations for the ratios of couplings and the $A$-factor: 
\begin{equation}
        Q \frac{\partial}{\partial Q} \Big(\frac{\laaidx{0}}{A}\Big) = -\frac{1}{8\pi^2}\,,
        \quad  
        Q \frac{\partial}{\partial Q} \Big(\frac{\lpcidx{0}}{A}\Big) = 0,
\label{eq:running_couplings0}
\end{equation}
which immediately give
\begin{align}
\frac{\laaidx{0}}{A}(Q) &= \frac{\laaidx{0}}{A}(Q_0) - L_{QQ_0},
\nonumber\\
\frac{\lpcidx{0}}{A}(Q) &= \frac{\lpcidx{0}}{A}(Q_0).
\label{eq:running_couplings1}
\end{align}
Here, $L_{QQ_0}$ is the logarithm function defined in~\cref{eq:running_barcoupling0_soln}. Equation~\cref{eq:running_couplings1} is now just a set of three algebraic equations for three unknown couplings. They can be easily solved to yield
\begin{subequations}
\begin{align}
  \laaidx{0}(Q) &= \frac{\laaidx{0}(Q_0)}{D_{QQ_0}}\Big(1-\frac{A(Q_0)}{\laaidx{0}(Q_0)}L_{QQ_0}\Big)\,,
\\
  \lpcidx{0}(Q) &= \frac{\lpcidx{0}(Q_0)}{D_{QQ_0}}\,,
\end{align}
\label{eq:running_couplings0_soln}
\end{subequations}
where
\begin{equation}
    D_{Q Q_0} \equiv 1 - \sfrac12(\lppidx{0}(Q_0)+\lccidx{0}(Q_0))L_{QQ_0} + \sfrac12A(Q_0)L_{QQ_0}^2.
\end{equation}
The running of the couplings $\labidx{0}$ with the renormalization scale is shown in the left panel of Fig.~\cref{fig:running_couplings}. 

%%%%%%%%%%%%%%%%%%%%%%%%%%%%%%%%%%%%%%%%%%%%%%%%%%%%%%%%%%%%%%%%%%%%%%%%%%%%%%%%%%%%%%%%%%%%%%%%%%%%%%%%%%%%%%%%
\paragraph{Constant couplings with index (4).}
%%%%%%%%%%%%%%%%%%%%%%%%%%%%%%%%%%%%%%%%%%%%%%%%%%%%%%%%%%%%%%%%%%%%%%%%%%%%%%%%%%%%%%%%%%%%%%%%%%%%%%%%%%%%%%%%

Lastly, consider the quartic couplings associated with four classical fields. Using relations~\cref{eq:ct4-0_relations}, it is easy to show that they obey trivial renormalization running equations:
\begin{align}
    Q\frac{\partial\laaidx{4}}{\partial Q} = 0\,,
    \label{eq:running_couplings4}
\end{align}
for all $\alpha,\beta$. These couplings thus do not run. This behavior was also observed in the single-field case~\cite{Kainulainen:2021eki}. This makes the $\labidx{4}$ a good choice to parametrize the theory: because they are constants, they are related to physical parameters by scale-independent constant relations. This is indeed how we shall proceed in what follows: we will first relate $\labidx{0}$'s to $\labidx{4}$'s and then compute $\labidx{4}$'s in terms of the physical couplings.

%===========================================================================================================
%===========================================================================================================
%===========================================================================================================
%===========================================================================================================

%%%%%%%%%%%%%%%%%%%%%%%%%%%%%%%%%%%%%%%%%%%%%%%%%%%%%%%%%%%%%%%%%%%%%%%%%%%%%%%%%%%%%%%%%%%%%%%%%%%%%%%%%%%%%%%%
\paragraph{Running mass parameters.}
%%%%%%%%%%%%%%%%%%%%%%%%%%%%%%%%%%%%%%%%%%%%%%%%%%%%%%%%%%%%%%%%%%%%%%%%%%%%%%%%%%%%%%%%%%%%%%%%%%%%%%%%%%%%%%%%

In a similar manner, one finds the renormalization group equations for the mass parameters. First, from~\cref{eq:cancel_Delta} one finds the equations for the kernels:
\begin{align}
\mu^{2}_\lin{\alpha,\!0} + \delta^\duin{0}_{\mu\alpha} 
= \frac{1}{D}\big( \mu^{2}_\lin{\alpha,\!0} + \sfrac12 B_\alpha\DeltaEps \big), 
\label{eq:cancel_mu2}
\end{align}
where $D$ was given in~\cref{eq:AD} and
\begin{align}
    B_\varphi = \lccidx{0}\sqmupF{0} - \lpcidx{0}\sqmucF{0}\,, \quad 
    B_\chi = \lppidx{0}\sqmucF{0} - \lpcidx{0}\sqmupF{0}\,.
\end{align}
From these equations, it is easy to find the running equations:
\begin{align}
    Q\frac{\partial \sqmupF{0}}{\partial Q} = \frac{\lppidx{0}\,\sqmupF{0}+\lpcidx{0}\,\sqmucF{0}}{16\pi^2}\,,\quad 
    Q\frac{\partial \sqmucF{0}}{\partial Q} = \frac{\lccidx{0}\,\sqmucF{0}+\lpcidx{0}\,\sqmupF{0}}{16\pi^2}\,.
    \label{eq:running_sqmu}%
\end{align}
These can also be solved exactly, with the following results:
\begin{align}
    \sqmupF{0}(Q) = \frac{\lppidx{0} B_\varphi(Q_0)+\lpcidx{0} B_\chi(Q_0)}{2A\, D_{QQ_0}}\,, \quad 
    \sqmucF{0}(Q) = \frac{\lccidx{0} B_\chi(Q_0)+\lpcidx{0} B_\varphi(Q_0)}{2A\, D_{QQ_0}}\,,
    \label{eq:running_sqmu_soln}
\end{align}
where $A$ and the couplings on the {\em r.h.s.}~of the equation are assumed to be functions of $Q$. The evolution of the mass parameters as a function of the renormalization scale is shown in the right panel of Fig.~\cref{fig:running_couplings}.
 
%%%%%%%%%%%%%%%%%%%%%%%%%%%%%%%%%%%%%%%%%%%%%%%%%%%%%%%%%%%%%%%%%%%%%%%%%%%%%%%%%%%%%%%%%%%%%%%%%%%%%%%%%%%%%%%%%%%%
\section{Hartree Effective Potential}
\label{sec:eff_pot}
%%%%%%%%%%%%%%%%%%%%%%%%%%%%%%%%%%%%%%%%%%%%%%%%%%%%%%%%%%%%%%%%%%%%%%%%%%%%%%%%%%%%%%%%%%%%%%%%%%%%%%%%%%%%%%%%%%%%

We now have all the tools necessary to construct the renormalized 2PI effective potential in the Hartree approximation. As mentioned earlier, we focus on the adiabatic limit where the two-point functions can be expressed as functions of the background fields. To this end, we consider the resummed masses defined in~\cref{eq:sqm_mixedbasis}, as functions of constant, but otherwise arbitrary background fields $(\varphi,\chi)$. We can then write down the 2PI effective potential in the Hartree approximation as follows:
\begin{align}
    V_{\rm HT}(\varphi,\chi)
    &= V_0(\varphi,\chi) - \frac{i}{2\mathcal{V}}{\rm Tr}\ln\Delta^{-1} - \frac{i}{2\mathcal{V}}{\rm Tr}[(\Delta_{0}^{-1}+\delta\Delta_{0}^{-1})\,\Delta] -\frac{\Gamma^{\rm ht}_2}{\mathcal{V}}\,,
    \label{eq:Vht}
\end{align}
where $\mathcal{V}$ is the space-time volume. We can then assemble the various pieces, starting with $V_0(\varphi,\chi)$, which denotes the sum of the classical action and counterterm action divided by the space-time volume:
\begin{align}
    V_0(\varphi,\chi) =
    & -\frac{1}{2}(\sqmupF{2}\!+\!\delta^\duin{2}_{\mu\varphi})\varphi^2 
       -\frac{1}{2}(\sqmucF{2}\!+\!\delta^\duin{2}_{\mu\chi})\chi^2 
    \nonumber \\
    &  +\frac{1}{4!}\klpp{4}\varphi^4
       + \frac{1}{4!}\klcc{4}\chi^4  + \frac{1}{4}\klpc{4} \varphi^2\chi^2 
    \nonumber \\
    = & -\frac{1}{12}\lpp^{(4)}\varphi^4 -\frac{1}{12}\lccidx{4}\chi^4 -\frac{1}{2}\lpcidx{4}\varphi^2\chi^2\,,
\end{align}
where, in the last step, we used~\cref{eq:ct4-0_relations} and~\cref{eq:limiting_kernels} and eventually took the limit $\epsilon \to 0$.  The second term in~\cref{eq:Vht} corresponds to the usual fluctuation determinant contribution and gives the familiar result
\begin{align}
    -\frac{i}{2\mathcal{V}}{\rm Tr}\ln\Delta^{-1} 
    &= -\frac{i}{2}\displaystyle\sum_{i =1}^2\int_q \ln \Delta_i^{-1} 
     = \displaystyle\sum_{i =1}^2 \left\{ \frac{m^4_i}{4}\DeltaEps 
       + \frac{m^4_i}{64\pi^2}\left[\ln\Big(\frac{m^2_i}{Q^2}\Big) - \frac{3}{2} \right] \right\},
\label{eq:second_term}
\end{align}
where $m_i^2$ are the eigenvalues of the mass matrix that are solutions to the gap-equation~\cref{eq:sqm_mixedbasis} with arbitrary constant fields. Finally, the last two terms of~\cref{eq:Vht} can be combined together, and give
\begin{align}
    - \frac{i}{2\mathcal{V}}{\rm Tr}[(\Delta_{0}^{-1}\!+\!\delta\Delta_{0}^{-1})\,\Delta] 
      - \frac{\Gamma^{\rm ht}_2}{\mathcal{V}} 
      = &-\frac{1}{8}\klpp{0}\Delta_{\varphi\varphi}^2
         -\frac{1}{8}\klcc{0}\Delta_{\chi\chi}^2 
\nonumber \\ 
       &-\frac{1}{4}\klpc{0}\Delta_{\varphi\varphi}\Delta_{\chi\chi}
        -\frac{1}{2}(\blpcidx{0}+\delta\blpcidx{0})\Delta_{\varphi\chi}^2\,.
\label{eq:last_terms}
\end{align}
Again, using the result~\cref{eq:limiting_kernels} and rotating to the diagonal basis, one can show that the divergent parts cancel out between~\cref{eq:second_term} and~\cref{eq:last_terms}. Combining the finite terms that survive in the limit $\epsilon\to 0$, we eventually find:
\begin{align}
    V_{\rm HT}(\varphi, \chi) &= -\frac{1}{12}\lpp^{(4)}\varphi^4 -\frac{1}{12}\lccidx{4}\chi^4 -\frac{1}{2}\lpcidx{4}\varphi^2\chi^2  -
    \displaystyle \sum_{i=1}^{2}\frac{m^4_i}{64\pi^2}\left[\ln\Big(\frac{m^2_i}{Q^2}\Big)-\frac12\right] \nonumber 
    \\
    &\quad +\frac{1}{4A} \big( \lccidx{0}\,m^4_{\varphi\varphi}+\lppidx{0}\,m^4_{\chi\chi}-2\lpcidx{0}\sqmpp\sqmcc \big)
    + \frac{m^4_{\varphi\chi}}{2\blpcidx{0}}\,.
    \label{eq:Vht_final}
\end{align}
This is the vacuum effective potential in the Hartree approximation with two scalar fields. Despite the apparent dependence on the renormalization scale, the effective potential~\cref{eq:Vht_final} is in fact independent of $Q$. This can be proved by first showing that the masses are scale-invariant, which we demonstrate explicitly in Appendix~\cref{app:scale_invariance}. Given $\partial_Q m^2 = 0$ and the running equations~\cref{eq:running_barcoupling0},~\cref{eq:running_couplings0,eq:running_couplings4}, it is straightforward to  prove that $\partial_Q V_{\rm HT} = 0$. 

%%%%%%%%%%%%%%%%%%%%%%%%%%%%%%%%%%%%%%%%%%%%%%%%%%%%%%%%%%%%%%%%%%%%%%%%%%%%%
\subsection{Connection to physical parameters}
\label{subsec:phys_params}
%%%%%%%%%%%%%%%%%%%%%%%%%%%%%%%%%%%%%%%%%%%%%%%%%%%%%%%%%%%%%%%%%%%%%%%%%%%%%

Similar to the gap-equation~\cref{eq:sqm_mixedbasis} and the equations of motion~\cref{eq:eom_fields_final}, the Hartree potential~\cref{eq:Vht_final} is also parametrized by two different sets of parameters: $\labidx{0}$ and $\labidx{4}$ (remember that $\blpcidx{0}$ was already related to $\lpcidx{4}$ by equation~\cref{eq:lpc0_vslpc4}). We will now establish the connection between these parameters and relate them to the physical observables. To begin, we use equations~\cref{eq:sqm_mixedbasis} repeatedly to manipulate the terms in the last line of~\cref{eq:Vht_final}, such that they can be combined with the last term in the first line. In this way we can eventually rewrite~\cref{eq:Vht_final} as follows:
\begin{align}
    V_{\rm HT}(\varphi,\chi) = 
    &-\frac12\sqmupF{0}\varphi^2 -\frac12\sqmucF{0}\chi^2
    +\frac{1}{4!}(3\lppidx{0}-2\lppidx{4})\varphi^4 
    +\frac{1}{4!}(3\lccidx{0}-2\lccidx{4})\chi^4 
    \nonumber \\
    & + \frac14(\lpcidx{0}+ 2\blpcidx{0}-2\lpcidx{4})\varphi^2\chi^2 + \frac{1}{4}{\rm Tr}\left(m_0^2 \Delta\right)-\frac{1}{128\pi^2}{\rm Tr}\left(m^4\right),
    \label{eq:VHT_traceform}
\end{align} 
where $m^2$ is the mass matrix defined from~\cref{eq:sqm_mixedbasis} and the trace invokes summation over the field indices. Finally, $m_0^2$ refers to the symmetric tree-level mass matrix with components
\begin{subequations}
\begin{align}
    m^2_{0\varphi\varphi} 
    &= -\sqmupF{0}+\sfrac12\lppidx{0}\varphi^2+\sfrac12\lpcidx{0}\chi^2\,,
    \\
    m^2_{0\chi\chi} 
    &= -\sqmucF{0}+\sfrac12\lccidx{0}\chi^2   +\sfrac12\lpcidx{0}\varphi^2\,,
    \\
    m^2_{0\varphi\chi} 
    &= \blpcidx{0}\varphi\chi. 
\end{align}
\label{eq:mass0matrix}%
\end{subequations}
Note that in~\eqref{eq:VHT_traceform}, the terms that are polynomial in the fields collapse to the usual tree-level potential for $\labidx{0}=\labidx{4} = \lambda_\lin{\alpha\beta}$ and $\blpcidx{0}=\lambda_{\varphi\chi}$. 

Equation~\cref{eq:VHT_traceform} provides a practical form for taking field derivatives of the potential. For example, for the $\varphi$-derivative one immediately finds:
\begin{equation}
\frac{\partial V_{\rm HT}}{\partial \varphi} 
= m^2_{0\varphi\varphi}\varphi + m^2_{0\varphi\chi}\chi 
- \frac13\lppidx{4}\varphi^3 -\lpcidx{4}\varphi\chi^2 
+\frac14{\rm Tr}\Big[m^2_0 \frac{\partial \Delta}{\partial\varphi} + \Delta\frac{\partial m^2_0}{\partial\varphi}-\frac{m^2}{16\pi^2}\frac{\partial m^2}{\partial \varphi}\Big].
\end{equation}
Then, using the identity
\begin{equation}
m^2_{0\varphi\varphi}\varphi + m^2_{0\varphi\chi}\chi 
= m^2_{\varphi\varphi}\varphi + m^2_{\varphi\chi}\chi -\sfrac12{\rm Tr}\Big[\Delta\frac{\partial m_0^2}{\partial\varphi}\Big],
\end{equation}
one can rewrite $\partial V_{\rm HT}/\partial \varphi$ replacing tree-level mass with the full mass $m^2_{0\alpha\beta}\to m^2_{\alpha\beta}$ and the trace with a new trace-term that can be proved to vanish:
\begin{equation}
{\rm Tr}\big[...\big] \to {\rm Tr}\Big[
   m^2_0 \frac{\partial \Delta}{\partial\varphi} 
   - \Delta\frac{\partial m^2_0}{\partial\varphi}
   - \frac{m^2}{16\pi^2}\frac{\partial m^2}{\partial \varphi} 
   \Big] = 0.
\label{eq:vanishing_trace}
\end{equation}
To show this, it is useful to first observe that~\cref{eq:vanishing_trace} vanishes if one replaces $m^2_0 \to m^2$. Then, subtracting this vanishing expression,~\cref{eq:vanishing_trace} can be expressed solely in terms of $\Delta$ and the difference $m^2-m_0^2$. The vanishing of the trace becomes evident after one rewrites also $m^2-m_0^2$ in terms of $\Delta$ using~\cref{eq:sqm_mixedbasis}. 

A similar derivation can also be done for the $\chi$-derivative, and so we eventually find:
\begin{subequations}
\begin{align}
        \frac{\partial V_{\rm HT}}{\partial \varphi} 
        &= \Big(\sqmpp -\frac13\lppidx{4}\varphi^2   \Big)\varphi
         + \Big(\sqmpc        -\lpcidx{4}\varphi\chi \Big)\chi ,
\label{eq:dVHTdphi}
\\
        \frac{\partial V_{\rm HT}}{\partial \chi} 
        &= \Big(\sqmcc - \frac13\lccidx{4}\chi^2      \Big) \chi
         + \Big(\sqmpc -        \lpcidx{4}\varphi\chi \Big) \varphi \,.
\label{eq:dVHTdchi}
\end{align}
\label{eq:dVHT}
\end{subequations}

\vskip-0.4cm\noindent
These equations hold for any field values. They prove our earlier assertion that the field equations~\cref{eq:eom_fields_final} can be written in terms of the Hartree potential as~\cref{eq:eom_fields_withV}, and finally that the vacuum-expectation values in~\cref{eq:field_eqn_rel} indeed correspond to the minimum of the potential.

%%%%%%%%%%%%%%%%%%%%%%%%%%%%%%%%%%%%%%%%%%%%%%%%%%%%%%%%%%%%%%%%%%%%%%%%%%%%%%%%%%%%%%%%%%%%%%%%%%%%%%
\paragraph{Connection between the auxiliary couplings.}
%%%%%%%%%%%%%%%%%%%%%%%%%%%%%%%%%%%%%%%%%%%%%%%%%%%%%%%%%%%%%%%%%%%%%%%%%%%%%%%%%%%%%%%%%%%%%%%%%%%%%%

Equation~\cref{eq:mass_quartic_vev} only provides a link between the $p^2\!=\!0$ masses, $\labidx{4}$-couplings and the vacuum expectation values $v,w$ at the minimum of the potential. We get more constraints between couplings when we identify the mass matrix $\hat m^2_{\alpha\beta}$ with the matrix of second derivatives of the potential at the minimum. Differentiating equations~\cref{eq:dVHT} a second time with respect to the fields, one finds:
\begin{subequations}
\begin{align}
         \frac{\partial^2 V_{\rm HT}}{\partial \varphi^2} 
       & = \sqmpp 
         + \Big[\Big(\frac{\partial \sqmpp}{\partial \varphi}-\lppidx{4}\varphi\Big)\varphi
         +      \Big(\frac{\partial \sqmpc}{\partial \varphi}-\lpcidx{4}\chi\Big)\chi  \Big]\,,
    \\
         \frac{\partial^2 V_{\rm HT}}{\partial \chi^2} 
       & = \sqmcc 
         + \Big[\Big(\frac{\partial \sqmpc}{\partial \chi}-\lpcidx{4}\varphi\Big)\varphi
         +      \Big(\frac{\partial \sqmcc}{\partial \chi}-\lccidx{4}\chi\Big)\chi  \Big]\,,
    \\
         \frac{\partial^2 V_{\rm HT}}{\partial \varphi\partial \chi} 
       & = \sqmpc 
         + \Big[\Big(\frac{\partial \sqmpp}{\partial \chi}-\lpcidx{4}\chi\Big)\varphi
         +      \Big(\frac{\partial \sqmpc}{\partial \chi}-\lpcidx{4}\varphi\Big)\chi  \Big] .
\label{eq:d1VHTc}
\end{align}
\label{eq:d2VHT}
\end{subequations}

\vskip-0.3cm\noindent
These expressions are again valid for arbitrary fields. We now define:
\begin{equation}
    \frac{\partial^2 V_{\rm HT}}{\partial \varphi^2}\bigg|_{(v,w)} = \hat{m}^2_{\varphi\varphi}\,,\quad \frac{\partial^2 V_{\rm HT}}{\partial \chi^2}\bigg|_{(v,w)} = \hat{m}^2_{\chi\chi}\,, \quad  \frac{\partial^2 V_{\rm HT}}{\partial \varphi\,\partial \chi}\bigg|_{(v,w)} = \hat{m}^2_{\varphi\chi}\,,
    \label{eq:d2VHT_mass}
\end{equation}
which implies that the terms in square brackets of~\cref{eq:d2VHT} equate to 0 at the minimum. This gives three equations connecting the $\labidx{0}$ and the $\labidx{4}$ couplings\footnote{The derivative $\partial^2 V_{\rm HT}/\partial \chi\partial \varphi$ apparently gives another condition, but this is in the end is degenerate with~\cref{eq:d1VHTc}.}: 
\vskip-0.3cm \noindent
\begin{subequations}
\begin{align}
  \Big(\frac{\partial \hsqmpp}{\partial \varphi}-\lppidx{4}v\Big)v
+ \Big(\frac{\partial \hsqmpc}{\partial \varphi}-\lpcidx{4}w\Big)w &= 0 \,,
\\
  \Big(\frac{\partial \hsqmpc}{\partial \chi}-\lpcidx{4}v\Big)v 
+ \Big(\frac{\partial \hsqmcc}{\partial \chi}-\lccidx{4}w\Big)w &= 0\,,
\\
  \Big(\frac{\partial \hsqmpp}{\partial \chi}-\lpcidx{4}w\Big)v
+ \Big(\frac{\partial \hsqmpc}{\partial \chi}-\lpcidx{4}v\Big)w &= 0\,.
\end{align}
\label{eq:lambda_connection}
\end{subequations}

\vskip-0.3cm \noindent
Equations~\cref{eq:lambda_connection} are linear in $\labidx{4}$, so it is easy to solve them to obtain $\labidx{4}$ in terms of $\labidx{0}$. In contrast, due to the iterative nature of the gap-equation~\cref{eq:sqm_mixedbasis}, they can not be solved analytically for $\labidx{0}[\labidx{4}]$, but they are easily solved numerically.

Let us point out that equations~\cref{eq:lambda_connection} guarantee, through equations~\cref{eq:full_propagator_final,eq:Valphazero} derived in Appendix~\cref{app:Z2_wave_function_ren_factors}, that $\hat m^2_{\alpha\beta}$ is also the $p^2\!=\!0$ mass matrix of the full propagator of the theory, as defined in~\cref{eq:FullPropagator_1}. Another point to make is that relations~\cref{eq:lambda_connection} are scale-invariant because $\smash{\partial_Q m^2_{\alpha\beta}=0}$. This means that~\cref{eq:lambda_connection} are consistent with the renormalization group running of $\labidx{0}$ and $\blpcidx{0}$. That is, whatever scale $Q_0$ one uses for $\Delta^{\rm FO}_{\alpha\beta}$ in the gap equation defines the scale for the $\widebar{\rm MS}$ parameters through~\cref{eq:lambda_connection}, and changing that scale accounts for their running according to equations found in Sec.~\cref{sec:running_of_couplings}.

%===========================================================================================================
%===========================================================================================================
\begin{figure}[t!]
    \centering
    \includegraphics[width=0.8\columnwidth]{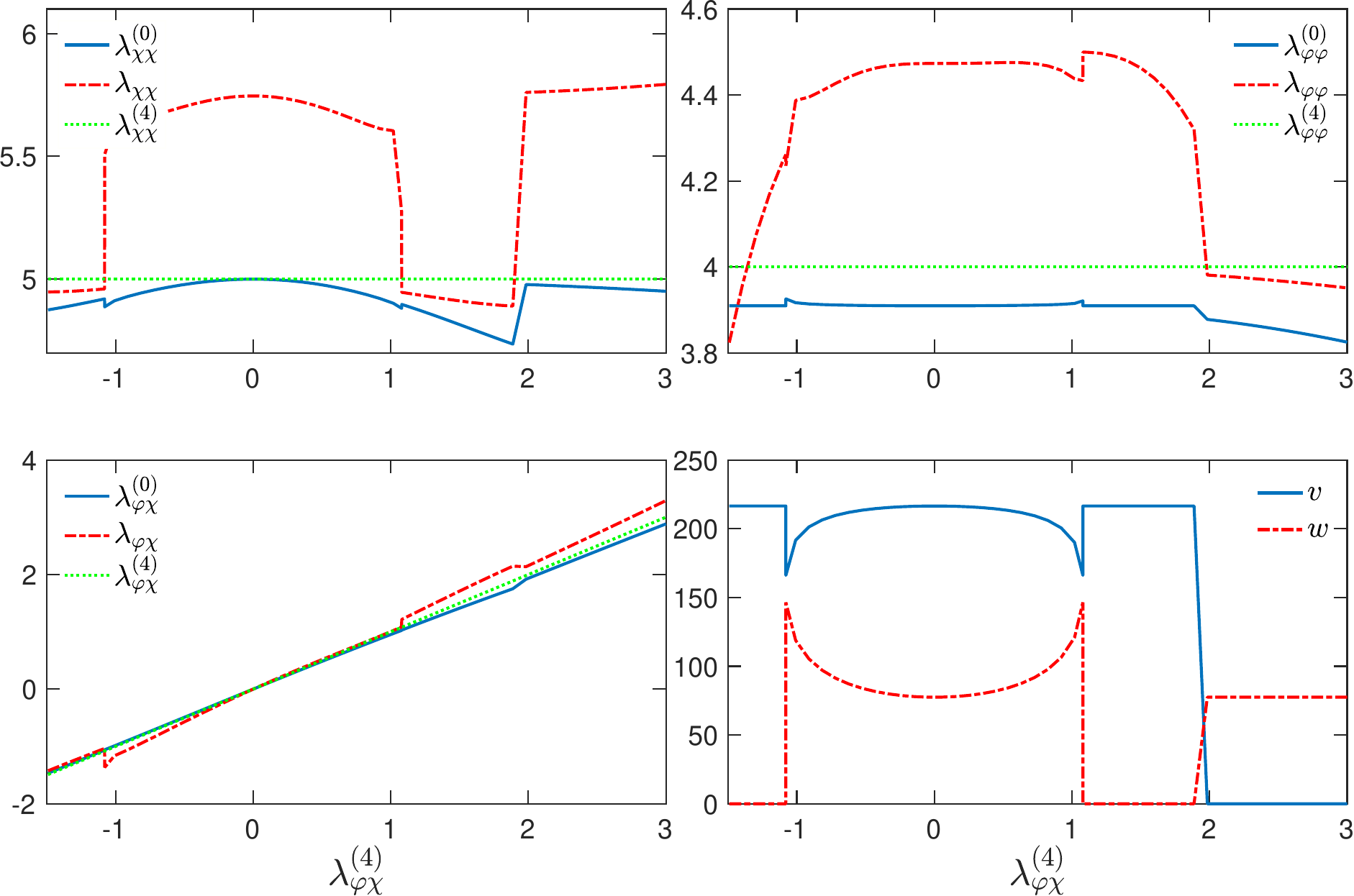}
    \caption{\textit{Upper panels and the lower left panel:} The values of the various auxiliary couplings $\lambda_{\alpha\beta}^{(0)}$ and the physical couplings $\hat\lambda_{\alpha\beta}$, as a function of the auxiliary coupling $\lambda_{\varphi\chi}^{(4)}$ and for $\lambda_{\varphi\varphi}^{(4)}=4$ and $\lambda_{\chi\chi}^{(4)} = 5$. \textit{Lower right panel:} The position of the global minimum $(v,w)$ for these parameters. We have assumed $Q_0 = 100$ GeV and the mass eigenvalues $\{\hat{m}_1,\hat{m}_2\} = \{250,100\}$~GeV.}
    \label{fig:couplings_as_function_of_lambda4pc}
\end{figure} 
%===========================================================================================================
%===========================================================================================================

%%%%%%%%%%%%%%%%%%%%%%%%%%%%%%%%%%%%%%%%%%%%%%%%%%%%%%%%%%%%%%%%%%%%%%%%%%%%%%%%%%%%%%%%%%%%%%%%%%%%%%
\paragraph{Connection to physical couplings.}
%%%%%%%%%%%%%%%%%%%%%%%%%%%%%%%%%%%%%%%%%%%%%%%%%%%%%%%%%%%%%%%%%%%%%%%%%%%%%%%%%%%%%%%%%%%%%%%%%%%%%%

With Equation~\cref{eq:lambda_connection} giving $\labidx{0}=\labidx{0}[\labidx{4}]$, we have our theory entirely parametrized by the $p^2\!=\!0$ masses of the (full) propagator and the scale-independent auxiliary couplings $\labidx{4}$. These couplings play the role of the tree-level couplings for the 2PI effective potential in the sense that conditions~\cref{eq:mass_quartic_vev} and~\cref{eq:d2VHT_mass} are equivalent to the usual perturbative renormalization conditions defined to keep couplings fixed to their tree-level values. To connect $\labidx{4}$ to physical observables, we consider the fourth-order derivatives of the potential,
\begin{equation}
    \frac{\partial^4 V_{\rm HT}}{\partial \varphi^4}\bigg|_{(v,w)} = \hat{\lambda}_{\varphi\varphi}\,,\quad \frac{\partial^4 V_{\rm HT}}{\partial \chi^4}\bigg|_{(v,w)} = \hat{\lambda}_{\chi\chi}\,, \quad  \frac{\partial^4 V_{\rm HT}}{\partial \varphi^2\,\partial \chi^2}\bigg|_{(v,w)} = \hat{\lambda}_{\varphi\chi}\,.
\label{eq:d4VHT}
\end{equation}
The couplings $\hat\lambda_\lin{\alpha\beta}$ are the full four-point functions of the theory at zero external momentum, and can be connected to physical 2-to-2 scattering rates as in usual perturbation theory. Equations~\cref{eq:d4VHT} can be solved numerically as follows: given a set of couplings $\smash{\hat{\lambda}_\lin{\alpha\beta}}$ one makes a guess $\labidx{4g}$, solves the corresponding $\labidx{0g}$ from~\cref{eq:lambda_connection}, and then evaluates~\cref{eq:d4VHT} to get $\smash{\hat{\lambda}^g_\lin{\alpha\beta}}$. After this one adjusts the guess $\labidx{4g}$ until $\smash{\hat{\lambda}^g_\lin{\alpha\beta}}$ matches $\smash{\hat{\lambda}_\lin{\alpha\beta}}$ with the requested accuracy. We will use this procedure to compute the auxiliary couplings from $\smash{\hat{\lambda}_\lin{\alpha\beta}}$ in the following sections.

We illustrate the difference between various couplings in Fig.~\cref{fig:couplings_as_function_of_lambda4pc}, where we plot $\smash{\hat{\lambda}_{\alpha\beta}}$ and $\labidx{0}$ as a function of $\lpcidx{4}$, for fixed $\{\lppidx{4},\lccidx{4}\}= \{4,5\}$, and $\{m_1,m_2\}= \{250,100\}$~GeV. The jumps, most prominently seen in the $\chi\chi$- and $\varphi\varphi$-couplings, are caused by jumps in the position of the global minimum, as indicated in the lower right panel. Indeed, for fixed masses, the VEVs are derived quantities. For large negative $\lpcidx{4}$ the minimum is on $\varphi$-axis. For roughly $-1<\lpcidx{4}<1$ it moves off-axis, after which it goes back to $\varphi$-axis before finally jumping to $\chi$-axis for $\lpcidx{4}>2$. While the differences are not large overall, as expected for the perturbative coupling range, the complicated structure underlines the need for a careful identification of physical parameters in any realistic model.

%%%%%%%%%%%%%%%%%%%%%%%%%%%%%%%%%%%%%%%%%%%%%%%%%%%%%%%%%%%%%%%%%%%%%%%%%%%%%%%%%%%%%%%%%%%%%%%%%%%%%%
\paragraph{Final remarks.}
%%%%%%%%%%%%%%%%%%%%%%%%%%%%%%%%%%%%%%%%%%%%%%%%%%%%%%%%%%%%%%%%%%%%%%%%%%%%%%%%%%%%%%%%%%%%%%%%%%%%%%

We finish this section with a few comments, starting from the numerical implementation. 
In principle, one can compute~\cref{eq:d4VHT} numerically starting from~\cref{eq:dVHT}, but this brings accuracy issues due to the finite precision of $m^2_{\alpha\beta}$, solved numerically from the gap equation. To ameliorate the problem, we solved the field derivatives of $m^2_{\alpha\beta}$ analytically and computed~\cref{eq:d4VHT} from~\cref{eq:d2VHT_mass} using more robust second-order numerical derivatives. We provide the analytical expressions for the derivatives in Appendix~\cref{app:mass-derivatives}. The second numerical issue concerns solving~\cref{eq:d4VHT} on the axis, when one of the VEVs is zero. This can be done through a limiting procedure using a very small, but non-zero VEV very close to the axis. For more precision, one has to expand the conditions~\cref{eq:lambda_connection} to higher order before taking the limit VEV$\to 0$. We have implemented both approaches, but we do not display the somewhat complicated formulae for the expanded conditions~\cref{eq:lambda_connection} here. 

Finally, we comment on our use of relations~\cref{eq:ct4-0_relations}. As we explained in Footnote~\cref{fn:guess}, the couplings $\labidx{4}$ can be thought of as mere placeholders for the solutions to~\cref{eq:ct4-0_relations} in terms of the kernel functions, which we could now insert in~\cref{eq:dVHT,eq:d2VHT_mass}. Since we already know that $\slab{0}\to 0$ as $\epsilon \to 0$, this corresponds to setting $\labidx{4} \to -\sfrac12\klab{4}$. Making this relabeling would have had no physical effect, however, as $\labidx{4}$'s were but intermediaries in the process of parameterizing the theory in terms of $\hat\lambda_\lin{\alpha\beta}$. So we see that the definition~\cref{eq:ct4-0_relations} is just a convenient choice motivated by the fact that the resulting couplings $\labidx{4}$ preserve the usual tree-level relations.

%%%%%%%%%%%%%%%%%%%%%%%%%%%%%%%%%%%%%%%%%%%%%%%%%%%%%%%%%%%%%%%%%%%%%%%%%%%%%%%%%%%%%%%%%%%%%%%%%%%%%%
%%%%%%%%%%%%%%%%%%%%%%%%%%%%%%%%%%%%%%%%%%%%%%%%%%%%%%%%%%%%%%%%%%%%%%%%%%%%%%%%%%%%%%%%%%%%%%%%%%%%%%
%
\section{Finite-Temperature Phase Transitions}
\label{sec:PT}
%
%%%%%%%%%%%%%%%%%%%%%%%%%%%%%%%%%%%%%%%%%%%%%%%%%%%%%%%%%%%%%%%%%%%%%%%%%%%%%%%%%%%%%%%%%%%%%%%%%%%%%%
%%%%%%%%%%%%%%%%%%%%%%%%%%%%%%%%%%%%%%%%%%%%%%%%%%%%%%%%%%%%%%%%%%%%%%%%%%%%%%%%%%%%%%%%%%%%%%%%%%%%%%

With the renormalization procedure complete and the connection to physical parameters established, we now study thermal corrections to the Hartree effective potential in this section. In the 2PI approach, these are incorporated primarily through solutions to the gap equation~\eqref{eq:sqm_mixedbasis}, which yield \textit{consistently resummed} thermal masses. These avoid the pitfall of relying on the high-temperature limit, while simultaneously improving the convergence of the perturbative expansion at finite temperature, as discussed in the Introduction. We then present the standard one-loop effective potentials employed in the literature, and close this section with a brief review of phase transition dynamics. 

%%%%%%%%%%%%%%%%%%%%%%%%%%%%%%%%%%%%%%%%%%%%%%%%%%%%%%%%%%%%%%%%%%%%%%%%%%%%%%%%%%%%%%%%%%%%%%%%%%%%%%
%
\subsection{Thermally-corrected Hartree potential}
%
%%%%%%%%%%%%%%%%%%%%%%%%%%%%%%%%%%%%%%%%%%%%%%%%%%%%%%%%%%%%%%%%%%%%%%%%%%%%%%%%%%%%%%%%%%%%%%%%%%%%%%

Extending~\cref{eq:d2VHT} to finite temperature is essentially trivial due to the generality of the renormalization procedure explained in Sec.~\cref{sec:renormalized_delta}.  We first present the standard definitions of the bosonic, one-loop thermal integrals:
\begin{equation}
    \mathcal{J}(x)\equiv \frac{1}{2\pi^2}\,{\rm Re}\int_0^{\infty} dy \,y^2\,\ln\left(1 - e^{\sqrt{y^2+x-i\varepsilon}}\right)\,,\quad {\rm and} \quad \mathcal{I}(x) \equiv 2 \frac{\partial \mathcal{J}(x)}{\partial x}\,.
    \label{eq:thermal_int}
\end{equation}
Here, the infinitesimal imaginary part $i\varepsilon$ ensures the correct branch of the logarithm is chosen for $x <0$. Then, based on the decomposition in \eqref{eq:correlator_decomp}, we can identify the finite-temperature correction as
\begin{equation}
    \delta \Delta_{{\rm F}ii} (T) = T^2\mathcal{I}\Big(\frac{m^2_{Ti}}{T^2}\Big)\,.
    \label{eq:A0_int_thermal}
\end{equation}
The mass matrix elements in~\cref{eq:sqm_mixedbasis} automatically pick up thermal corrections through the finite pieces: $\DeltaF_{\alpha\beta}\equiv \Delta^{\alpha\beta}_{\rm FO} + \delta \DeltaF_{\alpha\beta}$. We emphasize this by denoting the finite-temperature mass eigenvalues by $m^2_i(T) \equiv m^2_{Ti}$  in~\cref{eq:A0_int_thermal}. With this consideration, one can repeat the calculation in Sec.~\cref{sec:eff_pot} to arrive at the following result for the \textit{thermally-corrected, scale-independent} effective potential in the Hartree approximation:
\begin{equation}
     V_{\rm HT}(\varphi, \chi;T) = V_{\rm HT}(\varphi,\chi)\Big|_{m^2 \to m^2_T} + \displaystyle \sum_{i=1}^{2} \Big\{T^4\mathcal{J}\Big(\frac{m^2_{Ti}}{T^2}\Big)
    -\frac{m^2_{Ti}}{2}\Big[T^2\,\mathcal{I}\Big(\frac{m^2_{Ti}}{T^2}\Big) \Big]\Big\},
    \label{eq:Vht_thermal}
\end{equation}
where $V_{\rm HT}(\varphi,\chi)$ is the vacuum effective potential derived in~\cref{eq:Vht_final}. Note that the vacuum part also receives thermal corrections, being computed now with the thermal masses. Thus,~\cref{eq:Vht_thermal} is systematically renormalized and can be viewed as a consistently resummed potential to super-daisy level. In this approach, thermal corrections affecting all modes are treated on an equal footing, allowing for a smooth continuation between the non-relativistic and relativistic regimes. In this respect, the Hartree potential is superior to the usual ring-resummed potentials, where this feature needs to be introduced in a more ad hoc fashion.

%%%%%%%%%%%%%%%%%%%%%%%%%%%%%%%%%%%%%%%%%%%%%%%%%%%%%%%%%%%%%%%%%%%%%%%%%%%%%%%%%%%%%%%%%%%%%%%%%%%%
\subsection{Standard one-loop potentials}
%%%%%%%%%%%%%%%%%%%%%%%%%%%%%%%%%%%%%%%%%%%%%%%%%%%%%%%%%%%%%%%%%%%%%%%%%%%%%%%%%%%%%%%%%%%%%%%%%%%%

We now introduce our potentials for comparison. The standard one-loop thermal effective potential, without thermal resummation, is given by:
\begin{align}
	V_{\rm 1L}(\varphi,\chi;T) = &-\frac12\mu^2_{\varphi}\varphi^2 +\frac{1}{4!}\lambda_{\varphi\varphi}\varphi^4
	-\frac12\mu^2_{\chi}\chi^2 +\frac{1}{4!}\lambda_{\chi\chi}\chi^4+\frac{1}{4}\lambda_{\varphi\chi}\varphi^2\chi^2  \nonumber 	\\
	&+V_{\rm CW}(\varphi,\chi) + V_{\rm ct}(\varphi,\chi)  + \displaystyle \sum_{i=1}^{2} T^4\mathcal{J}\Big(\frac{m^2_{0,i}}{T^2}\Big)\,.
    \label{eq:V1L}
\end{align}
Here, the mass parameters and couplings appearing in the first line of~\cref{eq:V1L} are the tree-level ones. The Coleman-Weinberg (CW) potential and the associated counterterm (ct) potential are given by
\begin{align}
	&V_{\rm CW}(\varphi,\chi) =  \frac{1}{64\pi^2}\displaystyle \sum_{i=1}^{2} m^4_{0,i}\Big[\ln\Big(\frac{m^2_{0,i}}{Q^2}\Big)-\frac32\Big] ,
    \label{eq:VCW}
    \\
	&V_{\rm ct}(\varphi,\chi) = 
    - \frac12\delta_{\mu\varphi}\varphi^2 
    -\frac12\delta_{\mu\chi}\chi^2 
    +\frac{1}{4!}\delta_{\lambda\varphi\varphi}\varphi^4
	+\frac{1}{4!}\delta_{\lambda\chi\chi}\chi^4
    +\frac{1}{4}\delta_{\lambda\varphi\chi}\varphi^2\chi^2\,.
    \label{eq:Vct}
\end{align}
The masses $m^2_{0,i}(\varphi,\chi)$ are the eigenvalues of the tree-level mass matrix that can be obtained from~\cref{eq:mass0matrix}, setting $\labidx{0}\to \lambda_\lin{\alpha\beta}$ and $\mu^2_\lin{\alpha\!,0}\to \mu^2_\alpha$, and the counterterm potential is defined by the renormalization conditions:
\begin{equation}
\frac{\partial (V_{\rm CW}\! +\! V_{\rm ct})}{\partial \alpha} = 0
\quad {\rm and} \quad
\frac{\partial^2 (V_{\rm CW}\! +\! V_{\rm ct})}{\partial \alpha\,\partial\beta} = 0,
\label{eq:one-loop-renormalization-conditions}
\end{equation}
where $\alpha,\beta \in \{\varphi,\chi\}$. The counterterms are most easily calculated numerically from~\cref{eq:one-loop-renormalization-conditions}. Note that the renormalized one-loop potential can not be written in the form presented in Refs.~\cite{Curtin:2014jma,Carena:2019une,Lewicki:2021pgr,Ellis:2022lft} for mixing scalars, because the counterterm Lagrangian is strictly polynomial in the fields.

There are two common approaches to extend~\cref{eq:V1L} to include thermal resummations. In the Parwani approximation~\cite{Parwani:1991gq}, one replaces the mass matrix elements $m^2_{0,\alpha\beta}(\varphi,\chi)$ with their thermally corrected counterparts at the lowest order in couplings and in the high-$T$ expansion:
\begin{equation}
        m^2_{0,\varphi\varphi} \to m^2_{0,\varphi\varphi} + \frac{1}{24}(\lpp\!+\!\lpc)T^2\,,\quad 
        m^2_{0,\chi\chi} \to  m^2_{0,\chi\chi} + \frac{1}{24}(\lcc\!+\!\lpc)T^2\,.
\label{eq:thermalhighTmasses}
\end{equation}
Diagonalizing the resulting mass matrix, one finds thermal mass eigenvalues $m^2_{0,i}(\varphi,\chi;T)$, which are used to replace the zero-temperature eigenvalues in $V_{\rm CW}$ and in the ${\cal J}$-functions appearing in~\cref{eq:V1L}. This feature is similar to the Hartree potential, where the vacuum part is also calculated with thermally corrected masses. The difference is that in the Parwani approximation, the masses are not consistently resummed, and they are usually computed only to leading order in the high-temperature expansion of the $\cal{I}$-function. 

In the ring approximation~\cite{Carrington:1991hz, Arnold:1992rz}, thermal resummation is restricted to zero Matsubara modes, where it represents the screening of the long-distance zero modes by the short-distance physics. This is physically the more sound assumption, motivated by a clear hierarchy of scales. It yields the following result:
\begin{align}
    V_{\rm ring}(\varphi,\chi;T)=V_{\rm 1L}(\varphi,\chi;T)+\frac{T}{12\pi}\sum_{i = 1}^2{\rm Re}\left\{m^3_{0,i}(\varphi,\chi)-m^3_{0,i}(\varphi,\chi;T)\right\}\,.
\label{eq:ring_correction}
\end{align}
In this case, there is no correction to the vacuum potential. Only the zero-mode, which vanishes in the $T=0$ limit, is replaced by the resummed term. While the ring-potential is well motivated in the high-$T$ limit, it can not be used for low temperatures, where the ring-correction~\cref{eq:ring_correction} does not display the correct Boltzmann suppression. In contrast to the one-loop resummations, the Hartree potential~\cref{eq:Vht_thermal} is a self-consistent treatment for all temperatures, from the high-temperature expansion to a correctly Boltzmann-suppressed expression at low temperatures.

%%%%%%%%%%%%%%%%%%%%%%%%%%%%%%%%%%%%%%%%%%%%%%%%%%%%%%%%%%%%%%%%%%%%%%%%%%%%%%%%%%%%%%%%%%%%%%%%%%%%%%
%
\subsection{Phase transitions}
%
%%%%%%%%%%%%%%%%%%%%%%%%%%%%%%%%%%%%%%%%%%%%%%%%%%%%%%%%%%%%%%%%%%%%%%%%%%%%%%%%%%%%%%%%%%%%%%%%%%%%%%

In this work, we are interested in \textit{first-order} phase transitions, which have abundant 
phenomenological applications. A first-order PT is usually associated with the presence of degenerate minima separated by a substantial thermally induced barrier at the critical temperature $T_c$, and the transition progresses through nucleation of bubbles of the true phase. The computation of the tunneling rate is a complex problem, for which several different methods exist, including lattice simulations~\cite{Moore:2001vf,Gould:2022ran,Gould:2024chm,Gould:2025wec,Hirvonen:2025hqn}, advanced Langer methods~\cite{Langer:1969bc,Moore:2000jw,Croon:2020cgk,Ekstedt:2023sqc,Gould:2021ccf,Gould:2023ovu} and 2PI methods~\cite{Batini:2023zpi,Carosi:2024lop}. In this paper, we are mainly interested in comparing results obtained via different resummations for the effective potential, and adopt a simple estimate where the probability of bubble formation is quantified by the following nucleation rate per Hubble volume~\cite{Coleman:1977py,Callan:1977pt, Linde:1980tt, Linde:1981zj}:
\begin{equation}
    \Gamma_{\rm N}(T) = A(T) \exp \left(-\frac{S_3(T)}{T}\right)\,,
    \label{eq:nucl_rate}
\end{equation}
where the pre-factor $A(T)$ can be approximated, primarily on dimensional grounds, as~\cite{Linde:1980tt,Linde:1981zj}
\begin{equation}
    A(T) \approx T^4\left(\frac{S_3}{2\pi T}\right)^{\frac32}\,.
\end{equation}
The quantity $S_3$, appearing in the exponent of~\cref{eq:nucl_rate} and in the above approximation for the pre-factor, is the $O(3)$-symmetric Euclidean action of the scalar fields, 
\begin{align}
    S_3(T) = 4\pi\int_0^\infty dr \, r^2 
    \left[ \frac{Z_\lin{\varphi,\!2}}{2}\Big(\frac{d\varphi}{dr}\Big)^2
          +\frac{Z_\lin{\rchi,\!2}}{2}  \Big(\frac{d\chi}{dr}   \Big)^2 + V(\varphi,\chi;T)\right],
\label{eq:S3}
\end{align}
evaluated at the configuration that minimizes the action. Here $V(\varphi,\chi;T)$ is the effective potential at finite temperature, provided in either ring-resummed, Parwani, or Hartree approximation, as under consideration. We have also included the constant wave-function renormalization factors, which are evaluated in Appendix~\cref{app:Z2_wave_function_ren_factors} for the Hartree approximation. In the other approximations we compare to, $Z_\lin{\alpha,\!2}=1$. The bounce configuration that minimizes~\cref{eq:S3} is a solution to the classical equations of motion:
\begin{align}
    \frac{d^2 \varphi}{dr^2} + \frac{2}{r}\frac{d\varphi}{dr} 
    = \frac{1}{Z_\lin{\varphi,\!2}}\frac{\partial V}{\partial \varphi} \,,\qquad 
    \frac{d^2 \chi}{dr^2} + \frac{2}{r}\frac{d\chi}{dr} 
    = \frac{1}{Z_\lin{\rchi,\!2}}\frac{\partial V}{\partial \chi}\,,
\label{eq:bounceEOM}
\end{align}
subject to the boundary conditions 
\begin{equation}
    \varphi (r \to \infty) = 0\,,\quad   \partial_r\varphi\big|_{r = 0} = 0\,,\quad \chi (r \to \infty) = 0\,,\quad   \partial_r\chi\big|_{r = 0} = 0\,.
    \label{eq:bounceEOM_bc}
\end{equation}
To obtain $S_3$ for the bounce configuration, we solve the equations of motion~\cref{eq:bounceEOM} using our own code based on the relaxation method, corroborating our results with the \verb|Mathematica| package \verb|FindBounce|~\cite{Guada:2020xnz}. This allows us to obtain the nucleation rate, from which we can determine the nucleation temperature $T_n$, which signals the onset of nucleation of bubbles of the true vacuum, through $(4\pi/3)H^{-3}(T_n) \Gamma_{\rm N}(T_n) = H(T_n)$. This corresponds to the zero of the function:
\begin{equation}
N_3(T) \equiv S_3(T)/T - \log(4\pi/3) + 4\log(H/T) - \frac32\log(S_3(T)/(2\pi T)).
\label{eq:nucl_cond}
\end{equation}
Here, assuming a radiation-dominated epoch, the Hubble rate $H$ is given by:
\begin{equation}
     H^2(T) = \frac{\rho_{\rm R}(T)}{3M^2_{\rm Pl}}\,, \quad {\rm with}\quad   \rho_{\rm R}(T) = \frac{\pi^2}{30}g_*(T) \, T^4\,,
\end{equation}
where $M_{\rm Pl} = 2.435\times 10^{18}$~GeV is the reduced Planck mass, $\rho_{\rm R}$ is the radiation energy density and $g_*$ is the relativistic degrees of freedom. As we are interested in transitions occurring at $\mathcal{O}(100~{\rm GeV})$, we take for definiteness $g_* = 106.75$ corresponding to SM degrees of freedom above the PT.

%%%%%%%%%%%%%%%%%%%%%%%%%%%%%%%%%%%%%%%%%%%%%%%%%%%%%%%%%%%%%%%%%%%%%%%%%%%%%%%%%%%%%%%%%%%%%%%%%%%%%
%%%%%%%%%%%%%%%%%%%%%%%%%%%%%%%%%%%%%%%%%%%%%%%%%%%%%%%%%%%%%%%%%%%%%%%%%%%%%%%%%%%%%%%%%%%%%%%%%%%%%
%
\section{Numerical results}
\label{sec:numerical_results}
%
%%%%%%%%%%%%%%%%%%%%%%%%%%%%%%%%%%%%%%%%%%%%%%%%%%%%%%%%%%%%%%%%%%%%%%%%%%%%%%%%%%%%%%%%%%%%%%%%%%%%%
%%%%%%%%%%%%%%%%%%%%%%%%%%%%%%%%%%%%%%%%%%%%%%%%%%%%%%%%%%%%%%%%%%%%%%%%%%%%%%%%%%%%%%%%%%%%%%%%%%%%%

In this section, we provide numerical results and compare the PT features for the effective potentials presented in the previous section. We focus on cases where the global minimum of the potentials at $T=0$ lies at $(v,0)$, and accordingly present scenarios corresponding to one-step transitions in Sec.~\cref{subsec:1stepPT}, and a two-step transition in Sec.~\cref{subsec:2stepPT}. In Sec.~\cref {subsec:GW}, we study the implications on the gravitational wave (GW) spectrum sourced from the PTs in these scenarios.

To ensure a fair comparison between potentials, we impose the same renormalization conditions for each potential. In practice, we ensure this as follows. We start from a set of couplings $\lambda_{\alpha\beta}$ parameterizing the standard one-loop potentials with renormalization conditions~\cref{eq:one-loop-renormalization-conditions}. Note that just as $\labidx{4}$ in the Hartree case, these parameters are independent of the renormalization scale, as the $Q$-dependence cancels between $V_{\rm CW}$ and $V_{\rm ct}$, given~\cref{eq:one-loop-renormalization-conditions}. We then define the physical couplings $\hat\lambda_{\alpha\beta}$ as
\begin{equation}
    \hat{\lambda}_{\varphi\varphi} 
    \equiv \frac{\partial^4 V_{\rm 1L}}{\partial \varphi^4}\bigg|_{(v,w,0)} \,,\quad \hat{\lambda}_{\chi\chi} 
    \equiv \frac{\partial^4 V_{\rm 1L}}{\partial \chi^4}\bigg|_{(v,w,0)}    \,,\quad  \hat{\lambda}_{\varphi\chi} 
    \equiv \frac{\partial^4 V_{\rm 1L}}{\partial \varphi^2\,\partial \chi^2}\bigg|_{(v,w,0)} \,.
\label{eq:d4VCW}
\end{equation}
We then compute the auxiliary couplings in the Hartree case as explained in Section~\cref{subsec:phys_params}, requiring that~\cref{eq:d4VHT} holds for $\hat\lambda_{\alpha\beta}$'s defined by~\cref{eq:d4VCW}. That is, we require that different effective actions give the same physical 4-point functions at zero external momentum. Note that the renormalization conditions~\cref{eq:field_eqn_rel,eq:d2VHT_mass} and~\cref{eq:one-loop-renormalization-conditions} already ensure that the two-point functions at $p^2=0$ are the same. However, the couplings $\lambda_{\alpha\beta}$ and $\labidx{4}$ are not equivalent, and hence neither are the VEVs corresponding to the minimum of the potential. 

%========================================================================
%========================================================================
\begin{table}[b!]
\centering
\renewcommand{\arraystretch}{1.5}
\begin{tabular}{|c|c|c|c|c|c|}
\hline
& $\lambda$ & ${\hat \lambda}$ & $\lambda^{(4)}$ & $\lambda^{(0)}(\hat m_1)$ & $\oline{\lambda}^{(0)}(\hat m_1)$\\
\hline
\hline
$\varphi \varphi $  & 6.0 & 7.1173 & 5.9170 & 5.9436 & --- \\
\hline 
$\chi \chi$  & 4.0 & 4.0000 & 4.0849 & 3.9348  & --- \\
\hline 
$\varphi \chi$  & 2.6 & 2.9645 & 2.5911 & 2.4822 & 2.6141   \\
\hline 
\end{tabular}
\caption{The various quartic couplings for our one-step transition benchmark scenario. The values for the $p^2 =0$ masses are $(\hat{m}_1,\hat{m}_2) = (250,125)$~GeV. The couplings $\lambda^{(0)}$ and $\bar\lambda^{(0)}$ were evaluated at the scale $Q=\hat m_1 = 250$ GeV.}
\label{tab:lambdas_1stepa}
\end{table}
%========================================================================
%========================================================================

%%%%%%%%%%%%%%%%%%%%%%%%%%%%%%%%%%%%%%%%%%%%%%%%%%%%%%%%%%%%%%%%%%%%%%%%%%%%%%%%%%%%%%%%%%%%%%%%%%%%%
%
\subsection{One-step transition}
\label{subsec:1stepPT}
%
%%%%%%%%%%%%%%%%%%%%%%%%%%%%%%%%%%%%%%%%%%%%%%%%%%%%%%%%%%%%%%%%%%%%%%%%%%%%%%%%%%%%%%%%%%%%%%%%%%%%%
    
We first study one-step cases, in which the transition proceeds directly from the field origin to the true vacuum $(v,0)$, where it subsequently stays until zero temperature. The conditions we use to realize a first-order transition and obtain the critical temperature $T_c$ and the critical minimum $v_c$ are then:
\begin{align}
    V(0,0,T_c) = V(v_c,0,T_c)\,,
    \quad {\rm and}\quad 
    \partial_{\varphi}V(v_c,0,T_c) = 0\,.
\end{align}
We will consider two different benchmark points corresponding to a one-step transition of this type. In the first case, we find the strongest transition with the Hartree approximation, while in the second case, the Hartree approximation gives the weakest transition. The specific parameters used for these benchmark points are provided in Tables~\cref{tab:lambdas_1stepa} and~\cref{tab:lambdas_1stepb}. In this subsection, we set $Z_\lin{\alpha,\!2}\equiv 1$ also in the Hartree case, as we find the corrections $(Z_\lin{\alpha,\!2}\!-\!1)$ to be very small, and we are interested in seeing how the results vary when using different approximations only for the potentials. We will study the effect of the wave-function renormalization factors in the next subsection, when we discuss a two-step transition.

%========================================================================
%========================================================================
\begin{figure}[t!]
    \centering
    \includegraphics[width=0.41\columnwidth]{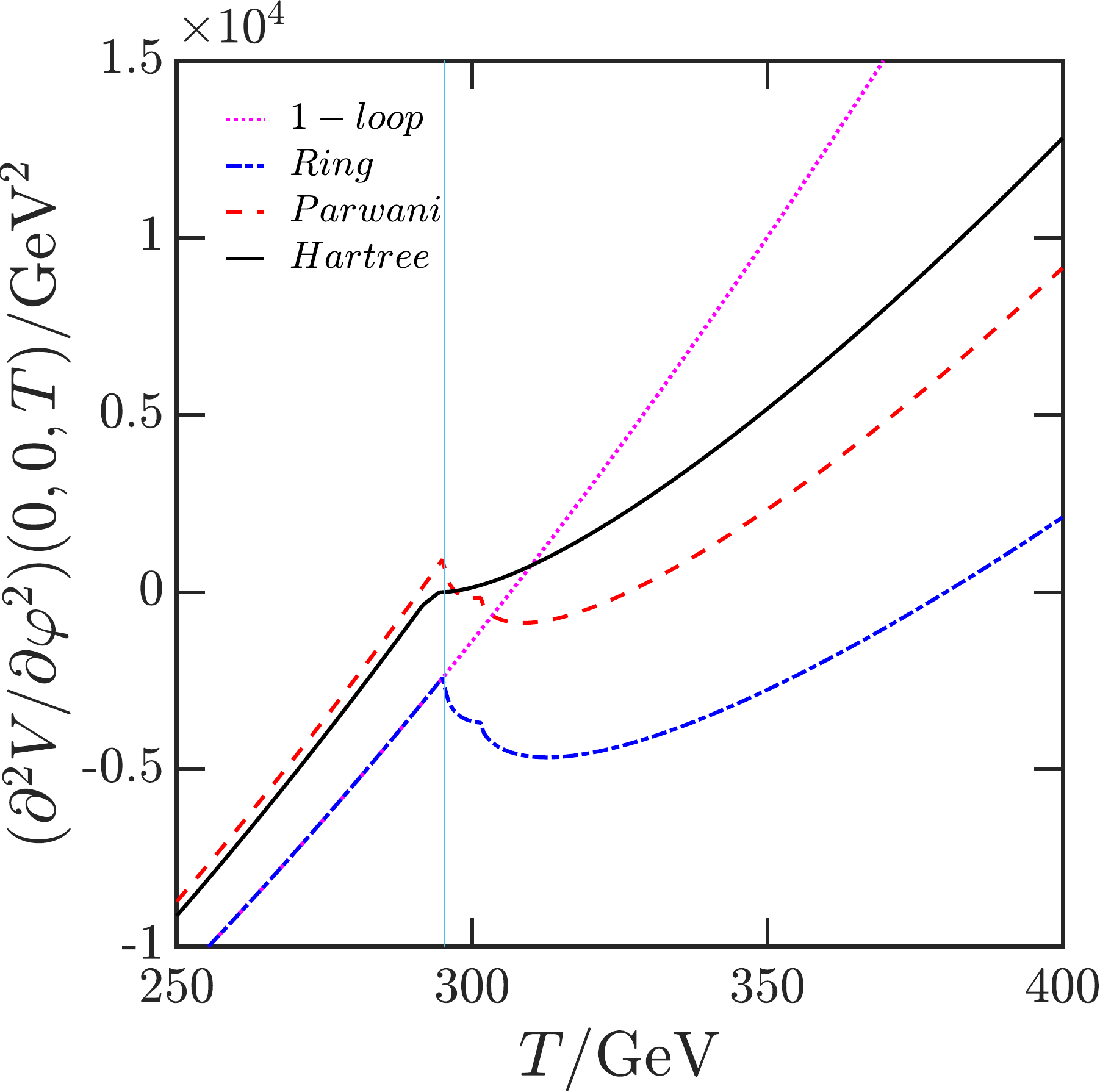}
    \qquad
    \includegraphics[width=0.41\columnwidth]{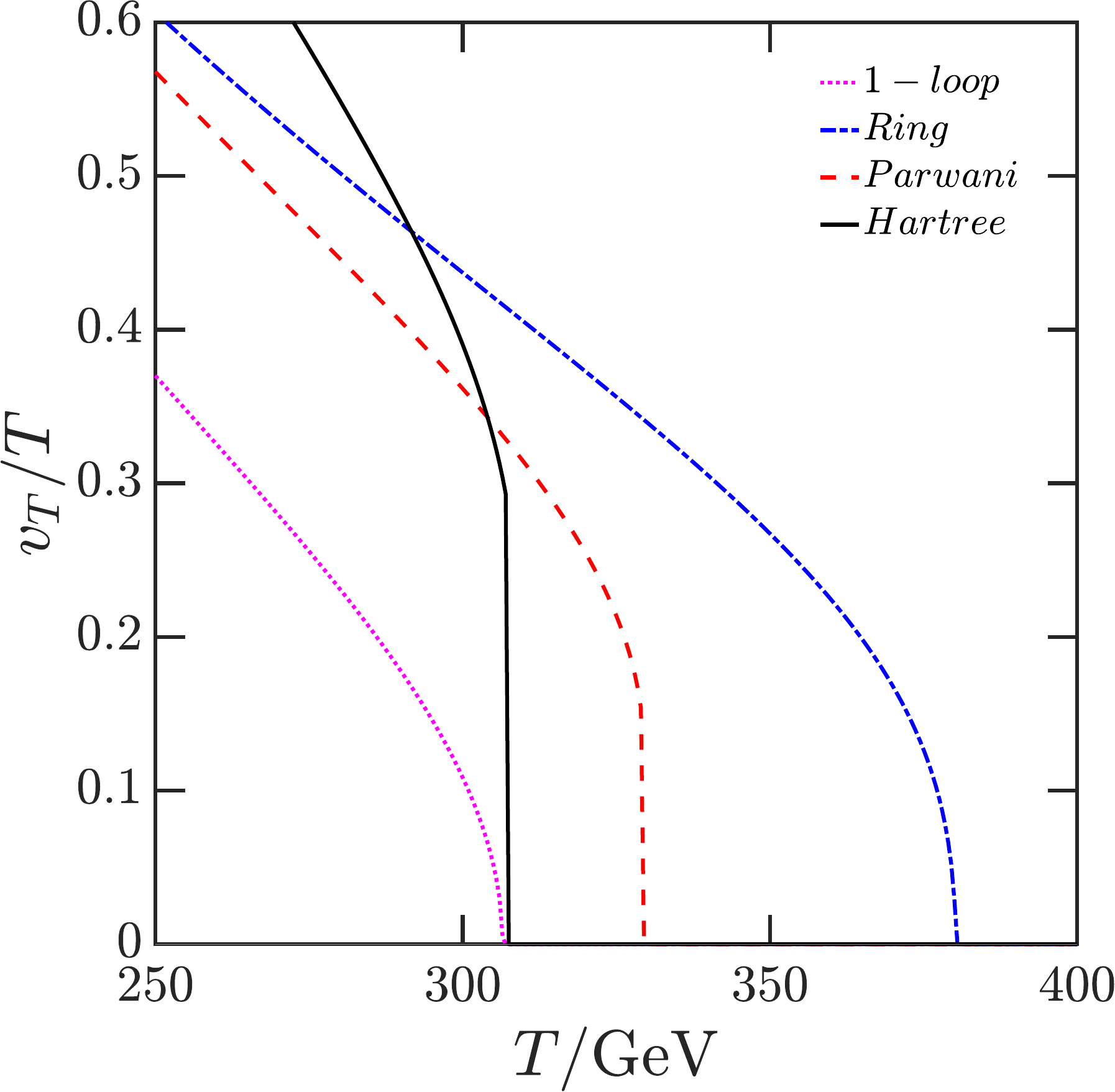}\;\;\;
    \caption{
    \textit{Left:} The thermal evolution of the second derivatives of the potential in $\varphi$-direction at the origin. The vertical cyan line indicates $T_0 = \sqrt{12/(\lambda_{\varphi\varphi}\!+\!\lambda_{\varphi\chi})}\,\hat{m}_1$, where the high-temperature limit approximated thermal mass vanishes at the origin.
    \textit{Right:}  
    The ratio $v_T/T$ as a function of the temperature, where $v_T$ is the position of the second minimum. The magenta dotted lines in both figures correspond to the non-resummed one-loop potential, shown for comparison.}
    \label{fig:d2V_voT}
\end{figure} 
%========================================================================
%========================================================================

%%%%%%%%%%%%%%%%%%%%%%%%%%%%%%%%%%%%%%%%%%%%%%%%%%%%%%%%%%%%%%%%%%%%%%%%%
\paragraph{Benchmark 1.}
%%%%%%%%%%%%%%%%%%%%%%%%%%%%%%%%%%%%%%%%%%%%%%%%%%%%%%%%%%%%%%%%%%%%%%%%%

In the left plot of Fig.~\cref{fig:d2V_voT}, we show the double derivatives of the potentials at the origin as a function of the temperature. We note sharp kinks for the Parwani and ring-resummed potentials close to the temperature $T_0$, where the thermal mass $m^2_{0,\varphi\varphi}(0,0;T_0) = 0$. This non-analytic behavior is not exhibited by the Hartree potential, as the double derivative of $V_{\rm HT}$ corresponds to a physical, consistently resummed mass. The right plot of the same figure shows the evolution of $v_T/T$, where $v_T$ is the position of the second minimum at a given temperature. The size of the jump in $v_T/T$ at the nucleation temperature is an indicator of the strength of the transition. A second minimum develops in Parwani and Hartree approximations, with the latter giving the larger jump. In contrast, the one-loop and ring approximations predict that the transition is of second order.

\begin{figure}[t!]
    \centering
    \includegraphics[width=0.41 \columnwidth]{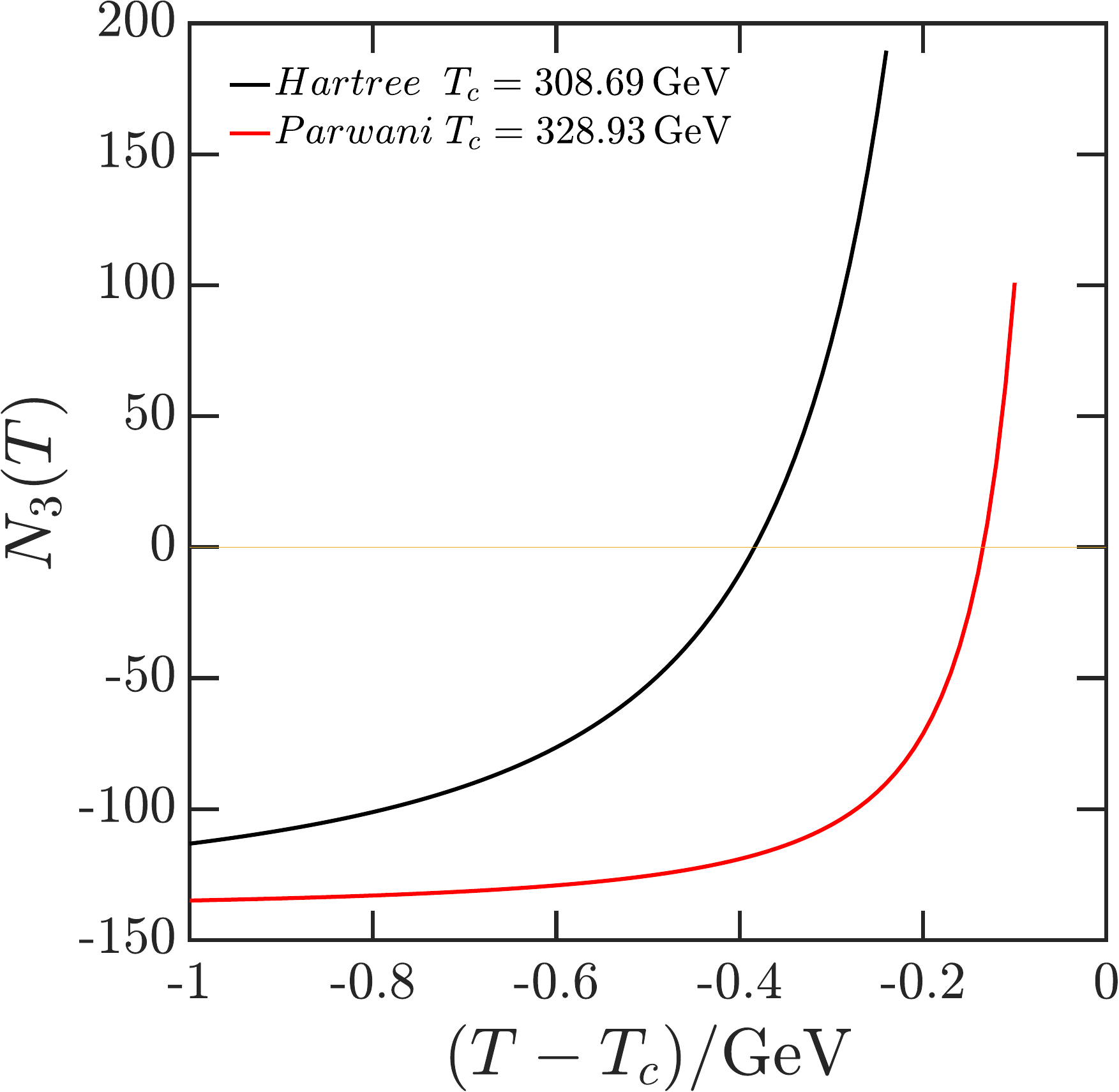}
    \qquad
    \includegraphics[width=0.405\columnwidth]{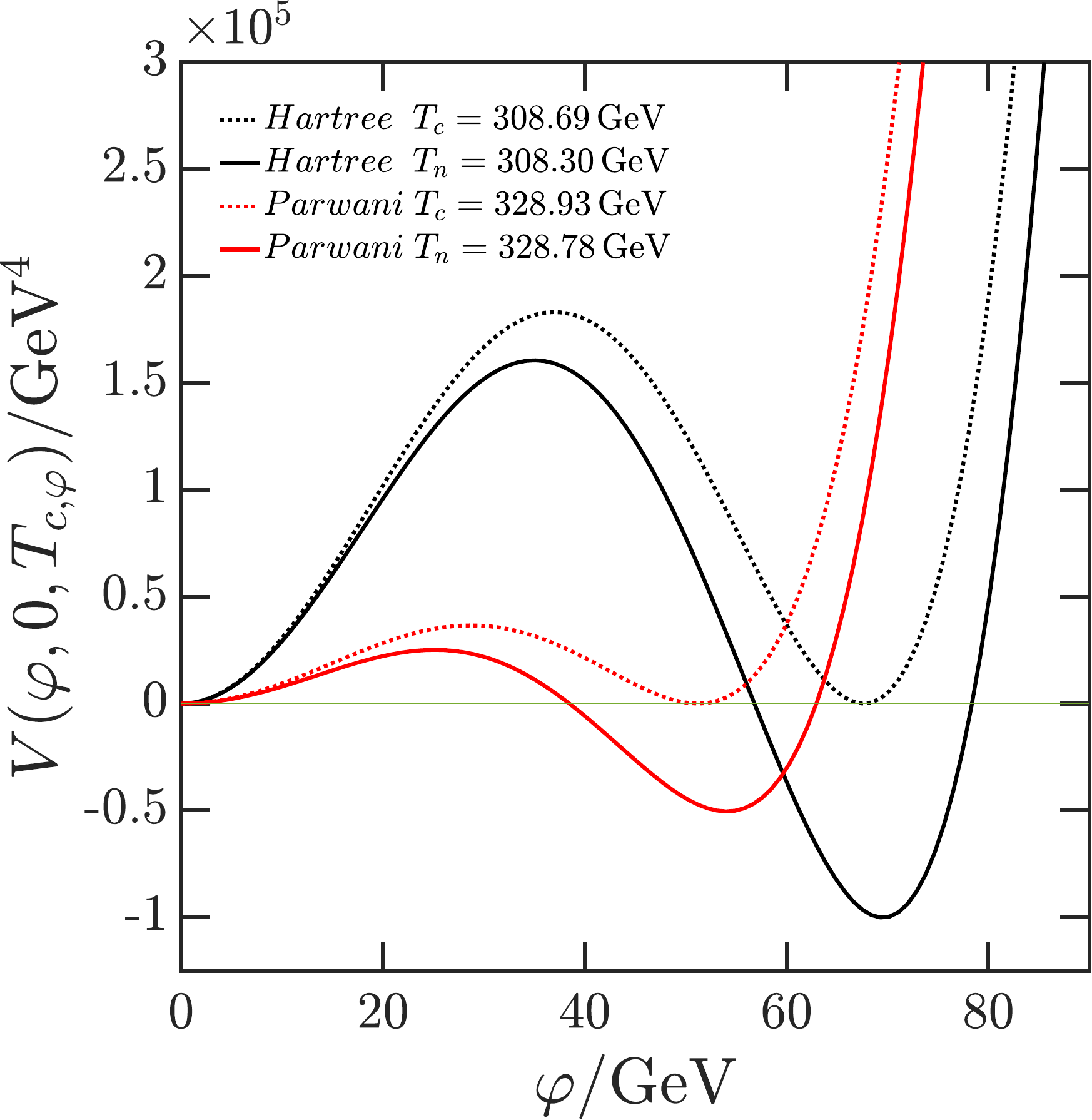}\;\;\;\;\;
    \caption{
    \textit{Left:} 
    The nucleation condition functions $N_3(T)$, defined in~\cref{eq:nucl_cond} for Hartree and Parwani potentials as a function $T-T_c$, for a one-step transition from the origin to $(v,0)$. The condition $N_3(T_n)=0$ defines the nuclation temperature $T_n$. 
    \textit{Right:} 
    The corresponding Hartree and Parwani effective potentials at their respective critical ($T_c$) and nucleation ($T_n$) temperatures.}
    \label{fig:1step_PTa}
\end{figure}

In the left plot of Fig.~\cref{fig:1step_PTa}, we show the nucleation condition function~\cref{eq:nucl_cond} obtained from the resummed potentials as functions of the distance from the critical temperature $T_c$. The condition $N_3(T_n)=0$ defines the respective nucleation temperatures $T_n$. In the right panel, we show the potentials at the critical temperatures $T_c$, where the two minima are degenerate (dotted lines), and at the nucleation temperatures (solid lines). The Hartree approximation predicts a stronger transition corresponding to a larger degree of supercooling, as seen in the left panel. It also predicts a larger $v_T/T$ and a larger latent heat, given by the potential difference between the true and false vacua at $T_n$. Both these features are clearly visible in the right panel in Fig.~\cref{fig:1step_PTa}. The main reason for the stronger transition is, of course, the much larger potential barrier between the minima in the Hartree case. 

%========================================================================
%========================================================================
%========================================================================
%========================================================================

%------------------------------------------------------------------------
%------------------------------------------------------------------------
\begin{table}[b!]
\centering
\renewcommand{\arraystretch}{1.5}
\begin{tabular}{|c|c|c|c|c|c|}
\hline
& $\lambda$ & ${\hat \lambda}$ & $\lambda^{(4)}$ & $\lambda^{(0)}(\hat{m}_2)$ & $\oline{\lambda}^{(0)}(\hat{m}_2)$\\
\hline
\hline
$\varphi \varphi $  & 2.0 & 3.0265 & 2.1636 & 2.1861 & ---    \\
\hline 
$\chi \chi$         & 4.0 & 4.0000 & 3.6847 & 3.5179 & ---    \\
\hline 
$\varphi \chi$      & 6.0 & 7.1474 & 5.9609 & 5.5596 & 6.0922 \\
\hline 
\end{tabular}
\caption{The various quartic couplings for our second one-step transition benchmark scenario. The values for the $p^2 =0$ masses in this case are $(\hat{m}_1,\hat{m}_2) = (200,425)$~GeV, and couplings $\lambda^{(0)}$ and $\bar\lambda^{(0)}$ were evaluated at the scale $Q=\hat m_2 = 425$ GeV.}
\label{tab:lambdas_1stepb}
\end{table}
%------------------------------------------------------------------------
%------------------------------------------------------------------------

%%%%%%%%%%%%%%%%%%%%%%%%%%%%%%%%%%%%%%%%%%%%%%%%%%%%%%%%%%%%%%%%%%%%
\paragraph{Benchmark 2.}
%%%%%%%%%%%%%%%%%%%%%%%%%%%%%%%%%%%%%%%%%%%%%%%%%%%%%%%%%%%%%%%%%%%%

\begin{figure}[t!]
    \centering
    \includegraphics[width=0.41 \columnwidth]{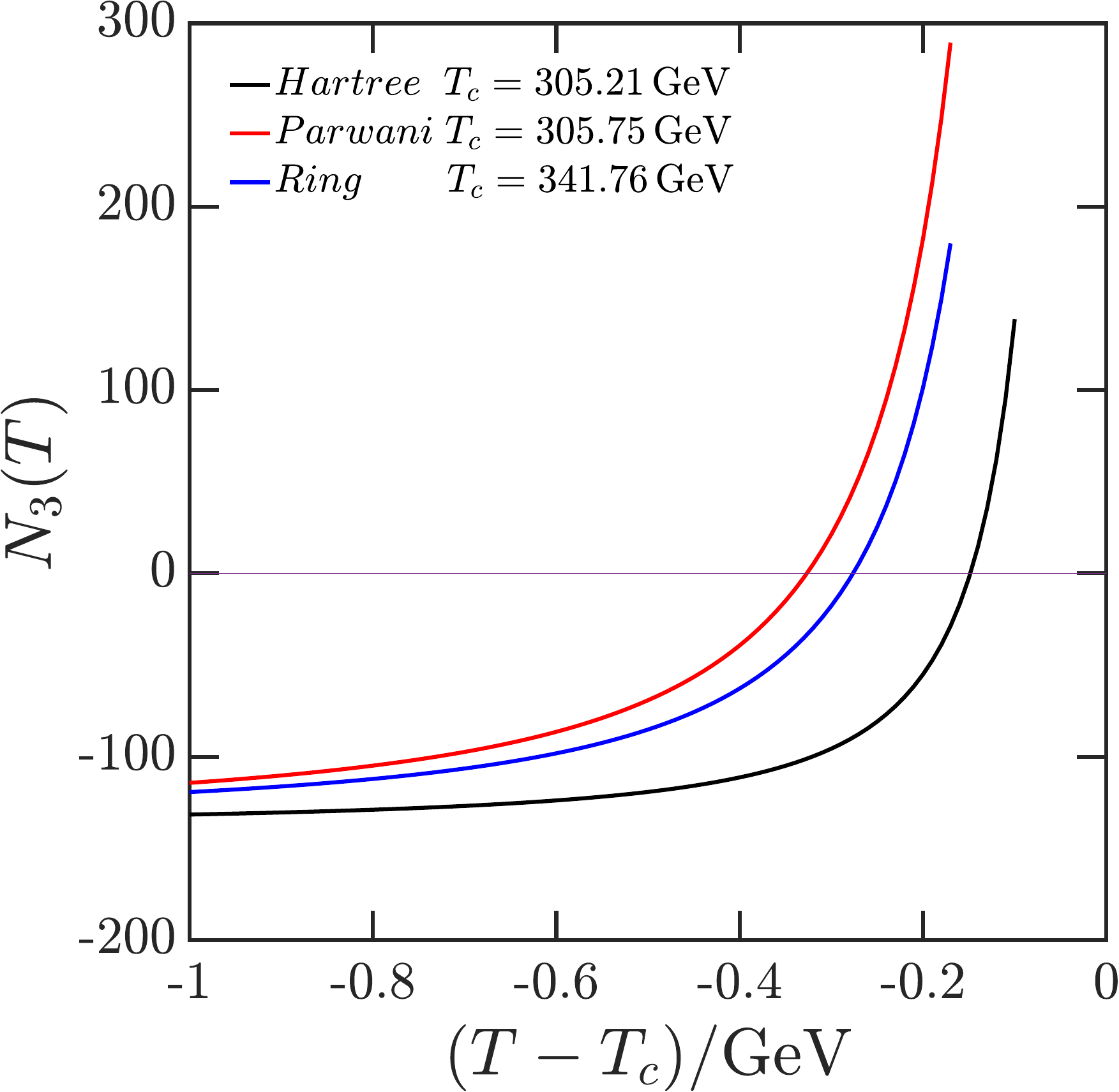}
    \qquad
    \includegraphics[width=0.405\columnwidth]{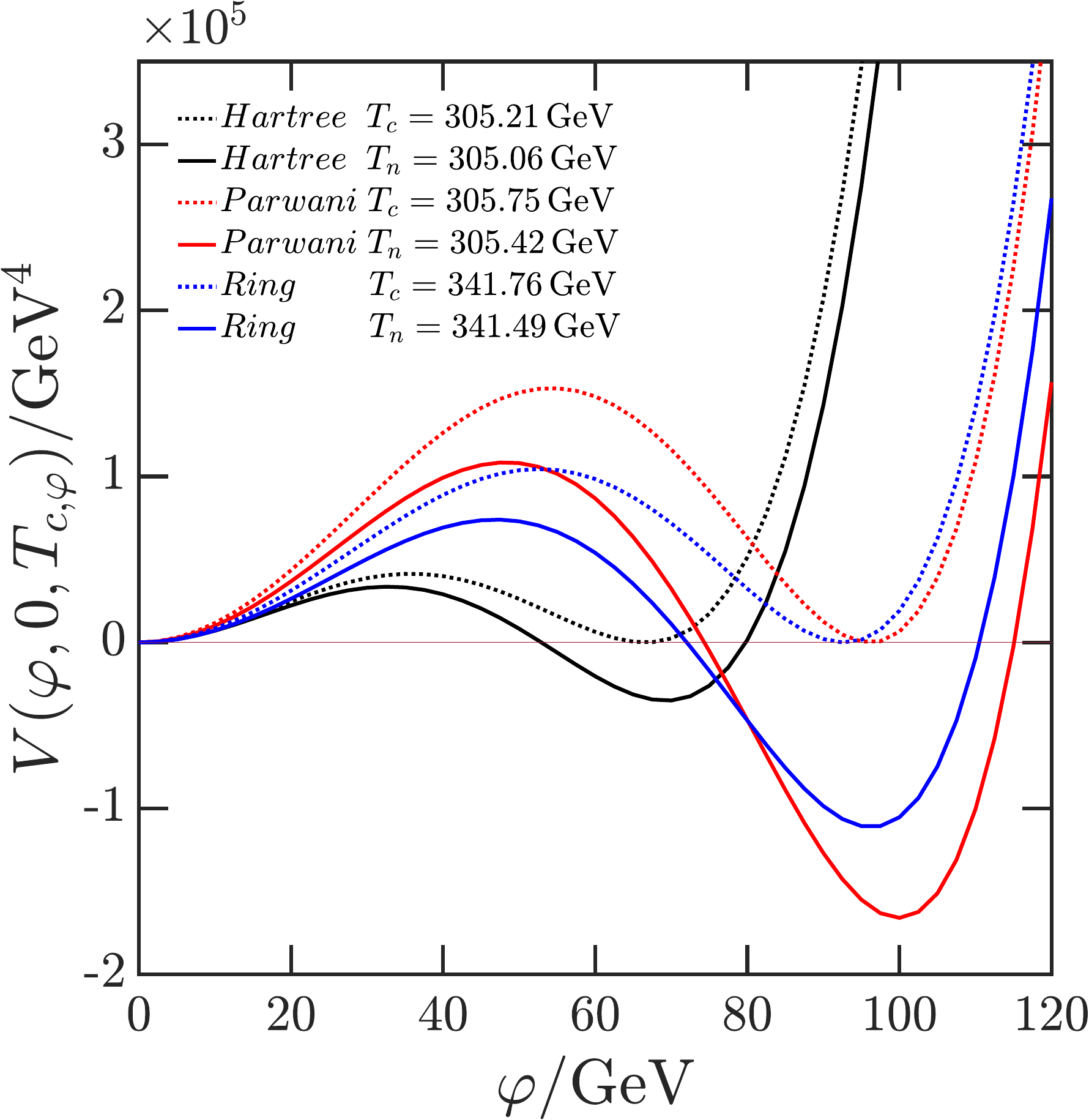}\;\;\;\;\;\;
    \caption{
    Shown are the same quantities as in Figure~\cref{fig:1step_PTa} for the Benchmark 2 case:
    \textit{Left:} 
    The nucleation condition functions $N_3(T)$ as a function $T-T_c$. 
    \textit{Right:} 
    The corresponding ring-approximated, Parwani and Hartree effective potentials at their respective critical ($T_c$) and nucleation ($T_n$) temperatures.}
    \label{fig:1step_PTb}
\end{figure}

At this stage, the reader may wonder whether the Hartree effective potential would always predict stronger transitions than other resummed potentials. This indeed seems to be the case in most randomly chosen parameter points, but it is not true in general. In Table~\cref{tab:lambdas_1stepb}, we provide the parameters for an example where the Hartree approximation gives a weaker transition. Notably, this scenario features a larger $\lpc$ and smaller $\lpp$, as well as $\hat{m}_2 > \hat{m}_1$, in contrast to the previous scenario. The nucleation conditions for this case are shown in the left plot of Fig.~\cref{fig:1step_PTb}, this time including also the ring approximation, which now also gives a first-order transition. In the right panel, we show the potentials at both critical (dotted lines) and nucleation temperatures (solid lines). We observe that the Parwani and ring-approximated effective potentials predict a much stronger transition than the Hartree case, essentially reversing the roles of approximations in comparison to the first benchmark described above. 

All transitions studied in this subsection are very weak, and the nucleation calculation is extremely well approximated by the thin-wall limit~\cite{Coleman:1977py,Anderson:1991zb}. It is difficult to make the transition stronger in the one-step transition case without violating the perturbativity of couplings. We will therefore next study the alternative two-step transition scenario.

%========================================================================
%========================================================================
%========================================================================
%========================================================================

%%%%%%%%%%%%%%%%%%%%%%%%%%%%%%%%%%%%%%%%%%%%%%%%%%%%%%%%%%%%%%%%%%%%%%%%%%
%
\subsection{Two-step transition}
\label{subsec:2stepPT}
%
%%%%%%%%%%%%%%%%%%%%%%%%%%%%%%%%%%%%%%%%%%%%%%%%%%%%%%%%%%%%%%%%%%%%%%%%%%

The stable minimum is again at the origin at very high temperatures. As the universe cools, a metastable minimum first develops in the $\chi$ direction at critical temperature $T_{c_1}$, and a transition to this minimum takes place at temperature $T_{n_1}$, given by the criterion $N_3(T_{n_1}) = 0$. As the temperature falls further, a new minimum develops on the $\varphi$-axis. Eventually, at a second critical temperature $T_c$, the minima at $(v_c,0)$ and $(0,w_c)$ become degenerate, when the potential satisfies the conditions:
\begin{align}
    V(v_c,0,T_c) = V(0,w_c,T_c)\,,
    \quad
    \partial_{\varphi}V(v_c,0,T_c) = 0
    \quad {\rm and}\quad 
    \partial_{\chi}V(0,w_c,T_c) = 0\,.
\end{align}
The transition to global minimum $(0,w_{n}) \to (v_{n},0)$ eventually happens at the nucleation temperature $T_{n}<T_{c}$, when $N_3(T_{n}) = 0$ is satisfied with the action $S_3(T_{n})$, evaluated for the nucleation path between the minima. Therefore, the transition to the true vacuum occurs in two steps: $(0,0) \to (0,w) \to (v,0)$, where the second step is of interest. 

A common feature for two-step transitions is that the parametric window for moderately strong transitions where the nucleation to the true vacuum can take place is very narrow, as an attempt to increase transition strength easily leads to a metastable state with an essentially infinite lifetime. As a result, the parameter regions for suitable two-step transitions may not overlap for different effective potentials. This appears to be the case here; we were not able to construct a single set of parameters for which a two-step transition took place with both Parwani and Hartree potentials. 

%==========================================================================================
%==========================================================================================
\begin{table}[t!]
\centering
\renewcommand{\arraystretch}{1.5}
\begin{tabular}{|c|c|c|c|c|c|c|}
\hline
& Scheme & $\lambda$ & ${\hat \lambda}$ &
$\lambda^{(4)}$ & $\lambda^{(0)}(\hat{m}_1)$ &
$\oline{\lambda}^{(0)}(\hat{m}_1)$ \\
\hline\hline
\multirow{2}{*}{$\varphi\varphi$}
& PW
& 6.0000 & 7.1055 & -- & -- & -- \\
\cline{2-7}
& HT
& 6.0000 & 7.1057 & 5.9459 & 5.9626 & -- \\
\hline
$\chi\chi$
& PW \& HT
& 2.0000 & 2.0000 & 2.0277 & 1.9748 & -- \\
\hline
\multirow{2}{*}{$\varphi\chi$}
& PW
& 1.9500 & 2.1621 & -- & -- & -- \\
\cline{2-7}
& HT
& 1.9200 & 2.1288 & 1.9160 & 1.8557 & 1.9296 \\
\hline
\end{tabular}
\caption{The various quartic couplings used for the Parwani (PW) and Hartree (HT) effective potentials for our benchmark two-step transition scenario. The values for the $p^2 =0$ masses are $(\hat{m}_1,\hat{m}_2) = (250,115)$~GeV. For the renormalization scale, we used $Q=\hat m_1$.}
\label{tab:lambdas_2step}
\end{table}
%==========================================================================================
%==========================================================================================

The parameters used in our specific example can be found in Table~\cref{tab:lambdas_2step}. The key parameter determining the strength of the second-step transition is the mixed quartic coupling $\lpc$. For the Parwani case, a suitable choice was $\lpc=1.95$. In the Hartree case, the second transition never completes for this coupling, and we used the value $\lpc = 1.92$ instead. This approximately represents the strongest transition allowed in the Hartree case, while in the Parwani case, one finds a one-step transition  $(0,0)\to (v,0)$ for $\lpc = 1.92$. We were thus not able to create a true one-to-one comparison of potentials, but we do observe that predictions differ significantly, and that the Hartree potential tends to predict a much stronger transition for a given set of parameters.

%==========================================================================================
%==========================================================================================
\begin{figure}[t!]
    \centering
    \includegraphics[width=0.43\columnwidth]{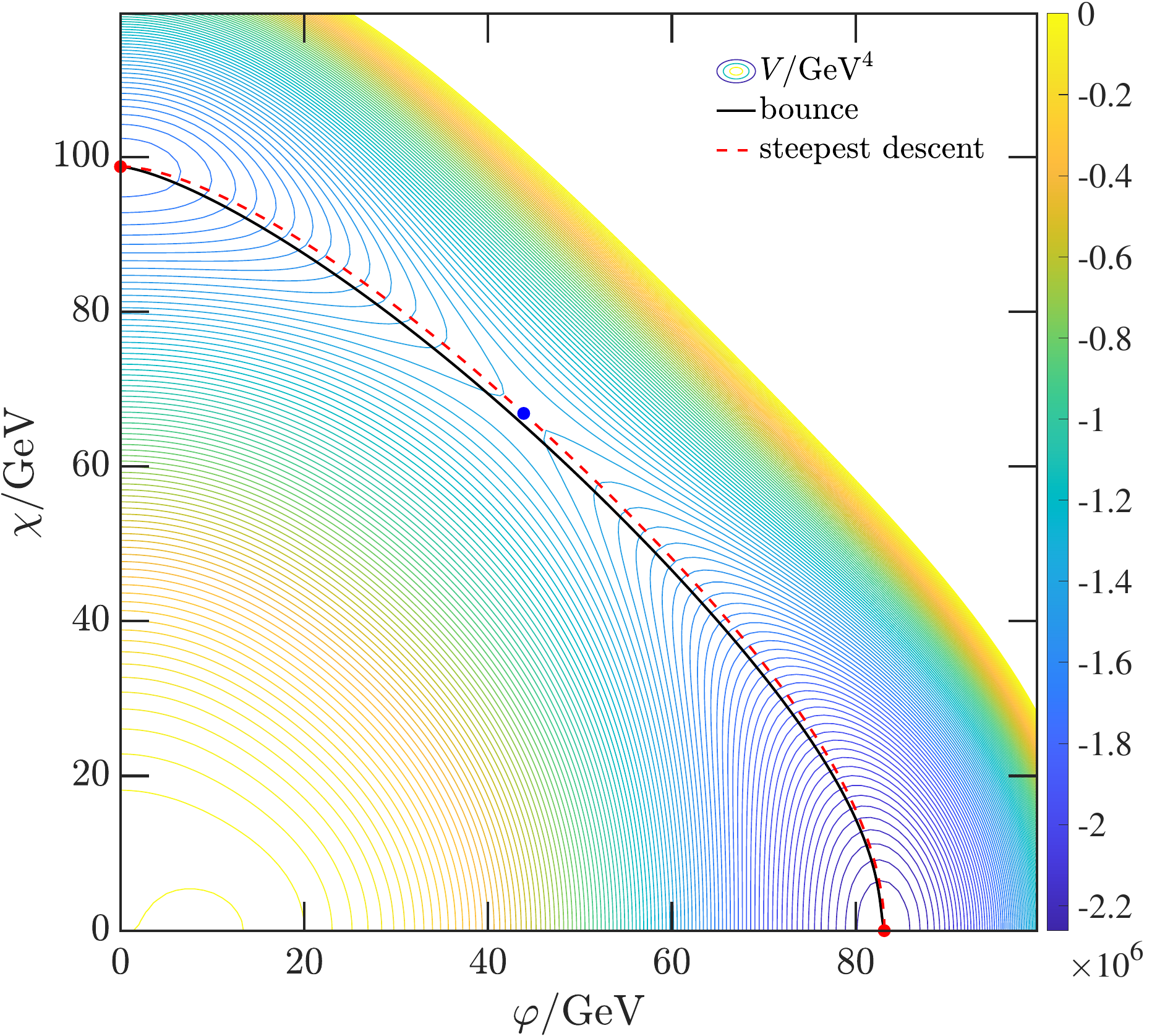}\,\,
    \includegraphics[width=0.43\columnwidth]{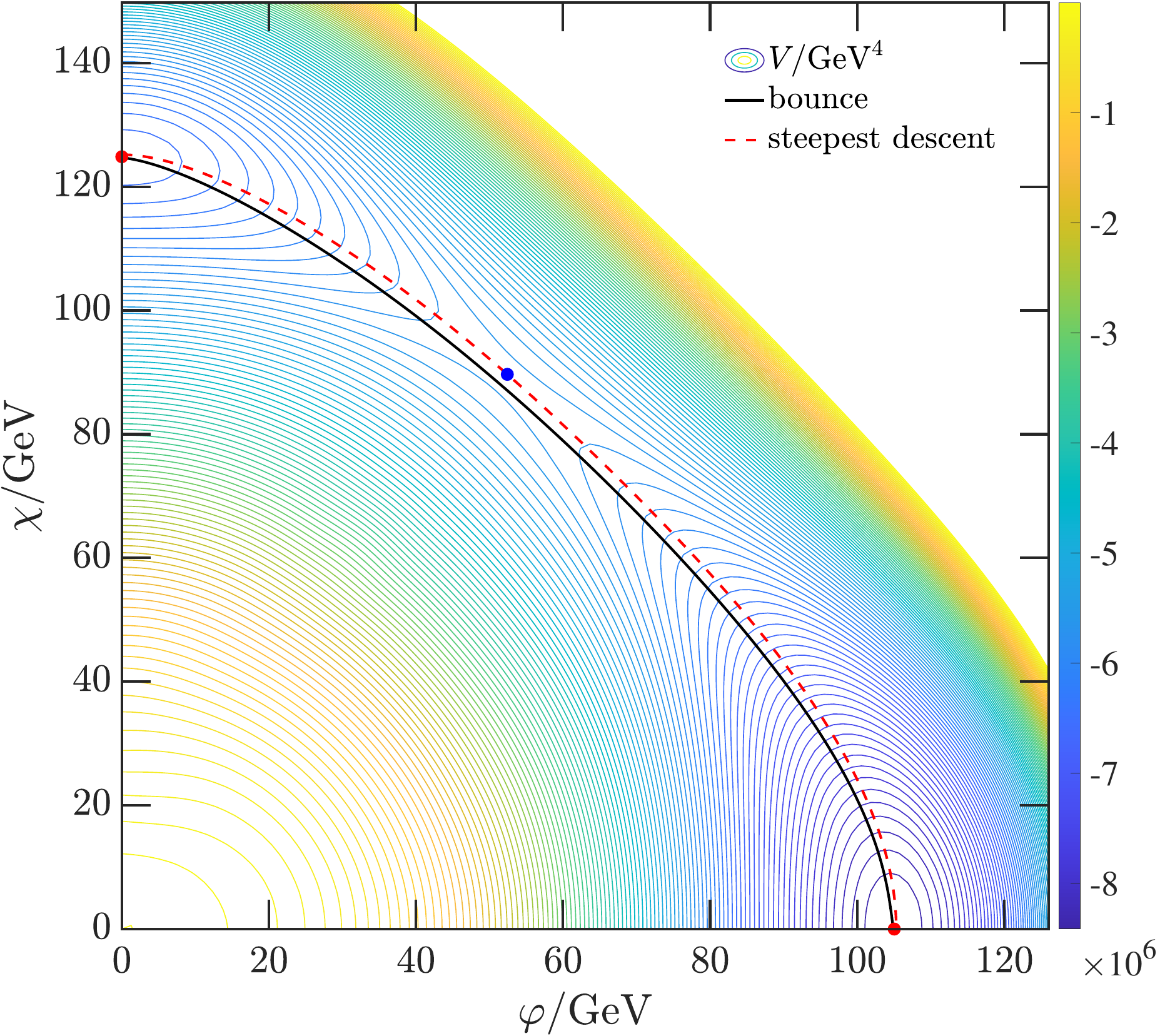}\,\,
    \caption{Contour plots of $V_{\rm PW}$ \textit{(left)} and $V_{\rm HT}$ \textit{(right)} at their respective nucleation temperatures. The trajectory of the tunneling path is shown by solid black lines and the paths of steepest descent for the potentials by red dashed lines. The saddle points of the potential are shown by blue dots and the minima on the axis by red dots.}
    \label{fig:2step_PT_nucl}
\end{figure} 
%==========================================================================================
%==========================================================================================

The first transition along the $\chi$ direction at $T_{c_1}$ is very weakly first-order for both Parwani and the Hartree cases. Below $T_c$ we solve for the tunneling path and nucleation rate~\cref{eq:nucl_rate} between the local minima on axis, and solve the nucleation temperatures $T_{n}$ from~\cref{eq:nucl_cond}. The second transition is much stronger in both cases. In Fig.~\cref{fig:2step_PT_nucl}, we show contours of the effective potentials in the Parwani (PW, left panel) and Hartree (right panel) cases, at their respective nucleation temperatures. We also show the nucleation paths (solid lines) connecting the false and true minimum. In our example, they closely trace the paths of steepest descent for the effective potential shown by dotted lines (blue dots indicate the saddle point positions).

%==========================================================================================
%==========================================================================================
\begin{figure}[t!]
    \centering
    \includegraphics[width=0.41\columnwidth]{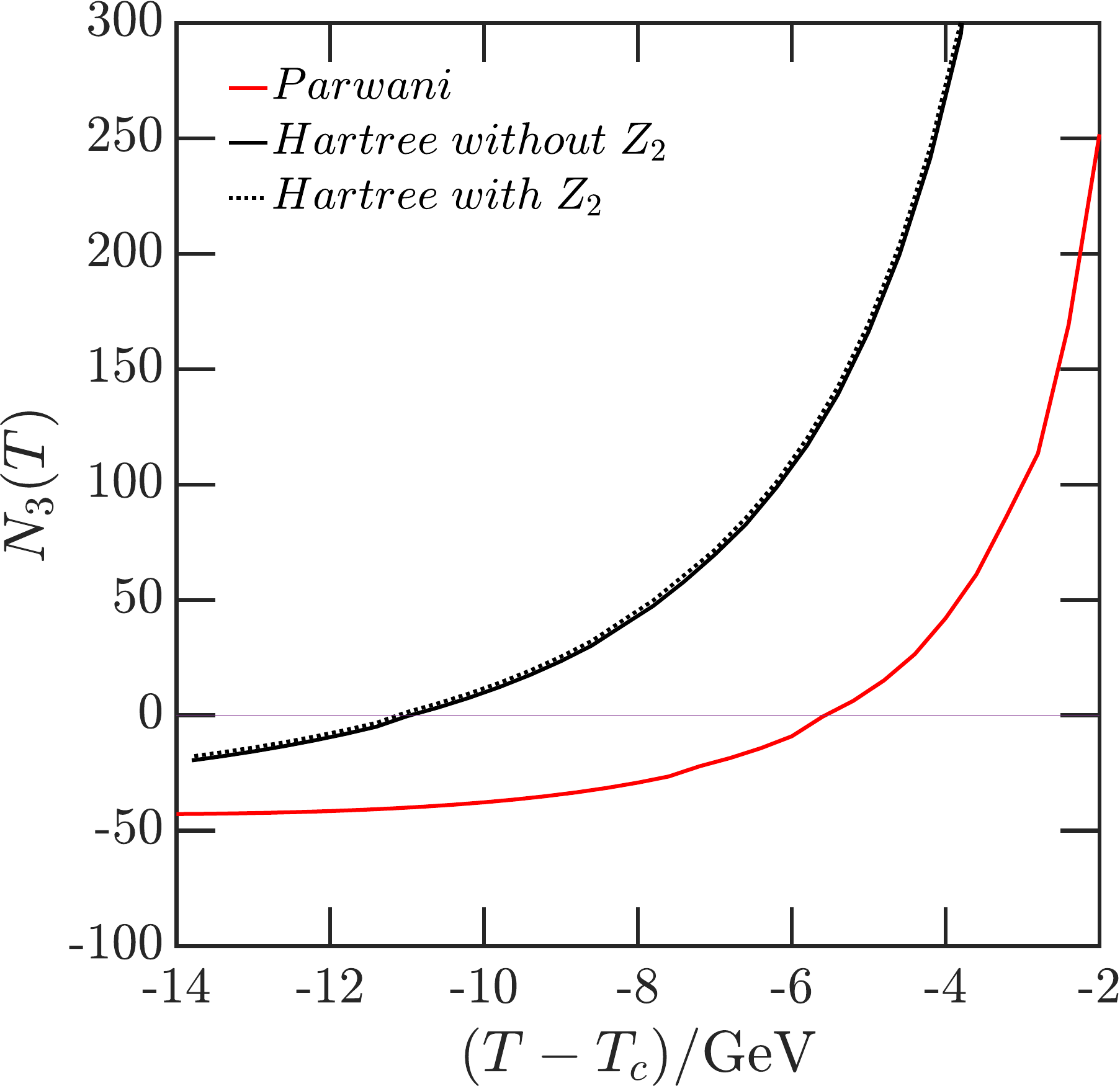}\,\,\,
    \includegraphics[width=0.405\columnwidth]{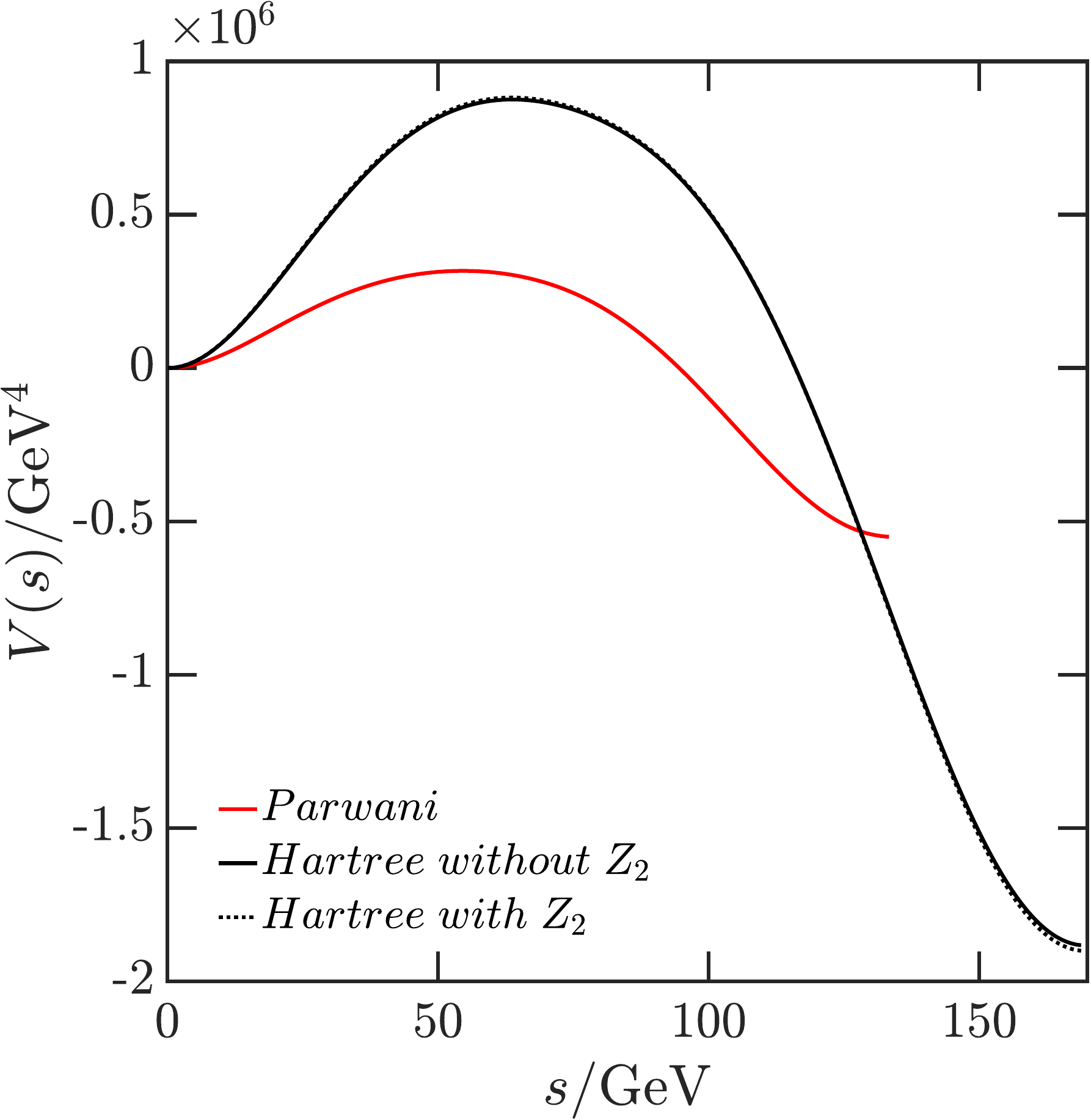} \;\;\;\;\;
    \caption{
    \textit{Left:} The nucleation condition functions $N_3(T)$. The nucleation temperatures are $T_n \approx 326.3$~GeV in the Parwani case, and $T_n \approx 299.9$~GeV in the Hartree approximation ($T_n \approx 299.8$ GeV when the finite $Z_2$-factors are included). 
    \textit{Right:} The effective potentials, at their respective $T_n$, along the nucleation path as a function of path length $s(\varphi) \equiv \int_0^{\varphi} {\rm d}{\bar \varphi} \sqrt{1+({\rm d}\chi/{\rm d}{\bar \varphi})^2}$.
    }
    \label{fig:2step_PT}
\end{figure} 
%==========================================================================================
%==========================================================================================

In the left panel of Fig.~\cref{fig:2step_PT}, we show the nucleation condition functions~\cref{eq:nucl_cond} for the Parwani and Hartree cases. The red line corresponds to the Parwani potential and the black line to the Hartree case with $Z_\lin{ \alpha,\!2}\equiv 1$. The black dotted line is the Hartree case with the constant $Z_\lin{\alpha,\!2}$ evaluated in Appendix~\cref{app:Z2_wave_function_ren_factors}, which we find to be in this case $Z_\lin{\varphi,\!2} \approx 1.0137$ and $Z_\lin{\rchi,\!2} \approx 1.0036$. The correction induced by the wave-function factors is very small and completely subdominant in comparison to the difference between different potential approximations. In the right panel, we show the effective potentials along the nucleation path as a function of the path length at their respective nucleation temperatures. The Hartree effective potential again exhibits the strongest degree of supercooling, as inferred from the potential differences between the true and false vacua in the plot in the right panel of Fig.~\cref{fig:2step_PT}, as well as from the much larger difference $T_c - T_n$ inferred from the $N_3(T)$-plots in the left panel.

%%%%%%%%%%%%%%%%%%%%%%%%%%%%%%%%%%%%%%%%%%%%%%%%%%%%%%%%%%%%%%%%%%%%%%%%%%%%%%%%%%%%%%%%%%%%%%%%%%
%
\subsection{Gravitational waves}
\label{subsec:GW}
%
%%%%%%%%%%%%%%%%%%%%%%%%%%%%%%%%%%%%%%%%%%%%%%%%%%%%%%%%%%%%%%%%%%%%%%%%%%%%%%%%%%%%%%%%%%%%%%%%%%

To demonstrate the phenomenological implications of the choice of resummation scheme, we consider stochastic gravitational wave backgrounds sourced from first-order PTs~\cite{Caprini:2019egz, Athron:2023xlk, Croon:2024mde}. First, we present the relevant parameters for computing the GW spectrum. The strength of the transition is typically quantified by the latent heat released, normalized to the radiation energy, and is given by
\begin{align}
    &\alpha(T) \equiv \frac{1}{\rho_{\rm R}(T)}\left(\Delta V(T)- \frac{T}{4}\,\frac{\partial \Delta V(T)}{\partial T}\right)\,,
    \label{eq:alpha}
\end{align}
where $\Delta V(T)$ is the potential difference between the false and true vacua and $\rho_{\rm R}(T)$ is the total energy in the radiation at temperature $T$. Next, the inverse time duration of the transition is obtained from 
\begin{align}
    \frac{\beta}{H} \equiv T\,\left[\frac{d}{dT}\left(\frac{S_3}{T}\right)\right]\,.
    \label{eq:betaoH}
\end{align}

%%%%%%%%%%%%%%%%%%%%%%%%%%%%%%%%%%%%%%%%%%
\begin{table}[t!]
\centering
\renewcommand{\arraystretch}{1.5}
\begin{tabular}{|c|c|c|c|c|}
\hline
& \multicolumn{2}{c|}{ONE-STEP} & \multicolumn{2}{c|}{TWO-STEP} \\
\hline\hline
& Parwani & Hartree & Parwani & Hartree \\
\hline
$T_*$~[GeV]  &328.8  &308.7  &326.3  &299.9  \\
\hline
$\alpha_*$   &$2.87 \times 10^{-5}$   &$6.56 \times 10^{-5}$  &$2.47 \times 10^{-5}$   &$5.83 \times 10^{-5}$   \\
\hline
$\beta/H_*$  &$7.0\times10^5$  &$2.3\times10^5$  &$5.7 \times 10^3$  &$2.8\times 10^3$  \\
\hline
\end{tabular}
\caption{Phase transition parameters for one-step (Benchmark 1) and two-step transitions obtained from the Parwani and Hartree approximations, which have been used to compute the spectra in Fig.~\cref{fig:GWs}.}
\label{tab:PT_params}
\end{table}
%%%%%%%%%%%%%%%%%%%%%%%%%%%%%%%%%%%%%%%%%%

\vskip-0.2cm
\noindent
Given that the transitions we consider in this toy model are rather weak, we focus on GWs sourced from the bulk motion in the plasma~\cite{Hindmarsh:2013xza, Giblin:2014qia, Hindmarsh:2015qta, Hindmarsh:2016lnk, Hindmarsh:2017gnf, Hindmarsh:2019phv}, neglecting the contribution from bubble collisions~\cite{Kosowsky:1992rz, Kosowsky:1992vn, Caprini:2007xq, Huber:2008hg, Weir:2016tov}, which require much stronger transitions, as well as those sourced from magneto-hydrodynamic turbulence~\cite{Kosowsky:2001xp, Dolgov:2002ra, Gogoberidze:2007an, Caprini:2009yp, Niksa:2018ofa, RoperPol:2019wvy, Kahniashvili:2020jgm, RoperPol:2021xnd,  Auclair:2022jod}, which is subject to more uncertainty. Therefore, we consider the present-day abundance of GWs arising from sound waves, which can be approximated as~\cite{Hindmarsh:2013xza, Weir:2017wfa, Caprini:2018mtu, Caprini:2019egz}
\begin{equation}
    h^2\Omega_{\rm sw}(f) = 1.2 \times 10^{-6}\, v_w \,\left(\frac{100}{g_*(T_*)}\right)^{\frac13}\,\Upsilon(y)\,\left(\frac{\beta}{H_*}\right)^{-1}
    \left(\frac{\kappa_{\rm sw}\, \alpha_*}{1+\alpha_*} \right)^2\,{\cal S}_{\rm sw}\left(f/f_{\rm sw}\right)\,,
    \label{eq:Omega_sw}
\end{equation}
where the subscript ``$*$'' indicates the PT parameters are evaluated at $T_* = T_n$\footnote{One may also consider $T_* = T_p$, where $T_p$ is the percolation temperature characterizing the completion of the PT~\cite{Ellis:2018mja, Ellis:2020nnr}. However, in the absence of significant supercooling, indicated by the values of $\alpha_* \ll 1$ we encounter, one has $T_p \approx T_n$.}. We take the following template for the spectral shape in \eqref{eq:Omega_sw}~\cite{Hindmarsh:2015qta,Hindmarsh:2016lnk,Hindmarsh:2017gnf}:
\begin{align} 
    \mathcal{S}_{\rm sw}(x) &= x^3 \left(\frac{7}{4 + 3\, x^2}\right)^{7/2},
    \label{eq:sw_spec_sim}
\end{align}
which peaks at the frequency, appropriately red-shifted to today: 
\begin{equation}
    f_{\rm sw} = \frac{8.9 \times 10^{-6}\, {\rm Hz}}{v_w}\,\left(\frac{g_*(T_*)}{100}\right)^{\frac16}\,\left(\frac{\beta}{H_*}\right)\left(\frac{T_*}{100\,{\rm GeV}}\right)\,.
    \label{eq:f_sw}
\end{equation} 
We take the bubble wall velocity appearing in~\cref{eq:Omega_sw} and~\cref{eq:f_sw} to be relativistic $v_w \sim 1$. We comment that, in practice, given the effective potential, one can determine $v_w$\footnote{For detailed studies on the calculation of $v_w$, we refer the reader to Refs.~\cite{Laurent:2022jrs,Ekstedt:2024fyq,vandeVis:2025plm}.} for each choice of resummation scheme. We defer such an investigation to future work, keeping this parameter fixed for our comparison. With this approximation, we may take the efficiency factor $\kappa_{\rm sw}$ for conversion of the energy released in the PT into the bulk fluid motion as~\cite{Steinhardt:1981ct, Espinosa:2010hh}:
\begin{equation}
    \kappa_{\rm sw} \approx \frac{\alpha_*}{0.73 + 0.083\sqrt{\alpha_*}+\alpha_*}\,.
\end{equation}
Finally, we include a suppression factor $\Upsilon(y)$ in~\cref{eq:Omega_sw} to account for the finite lifetime of the sound waves~\cite{Ellis:2020awk,Guo:2020grp}, $\tau_{\rm sw}$:
\begin{equation}
    \Upsilon(y) = 1 - \frac1y\,,\quad {\rm where} \quad y = \sqrt{1 + 2 \tau_{\rm sw}H_*}\,.
\end{equation}
We follow Ref.~\cite{Ellis:2020awk} in estimating the lifetime of the source: $\tau_{\rm sw} \sim R_*/\bar{U}_f$, where $R_* = (8\pi)^\frac13\,v_w/\beta$ is the characteristic fluid length, and $\bar{U}_f = \sqrt{3 \kappa_{\rm sw}\alpha_*/4}$ is the root-mean-square fluid-velocity~\cite{Hindmarsh:2017gnf}.

%%%%%%%%%%%%%%%%%%%%%%%%%%%%%%%%%%%%%%%%%%%%%%
\begin{figure}[t!]
    \centering
    \includegraphics[width=0.43\columnwidth]{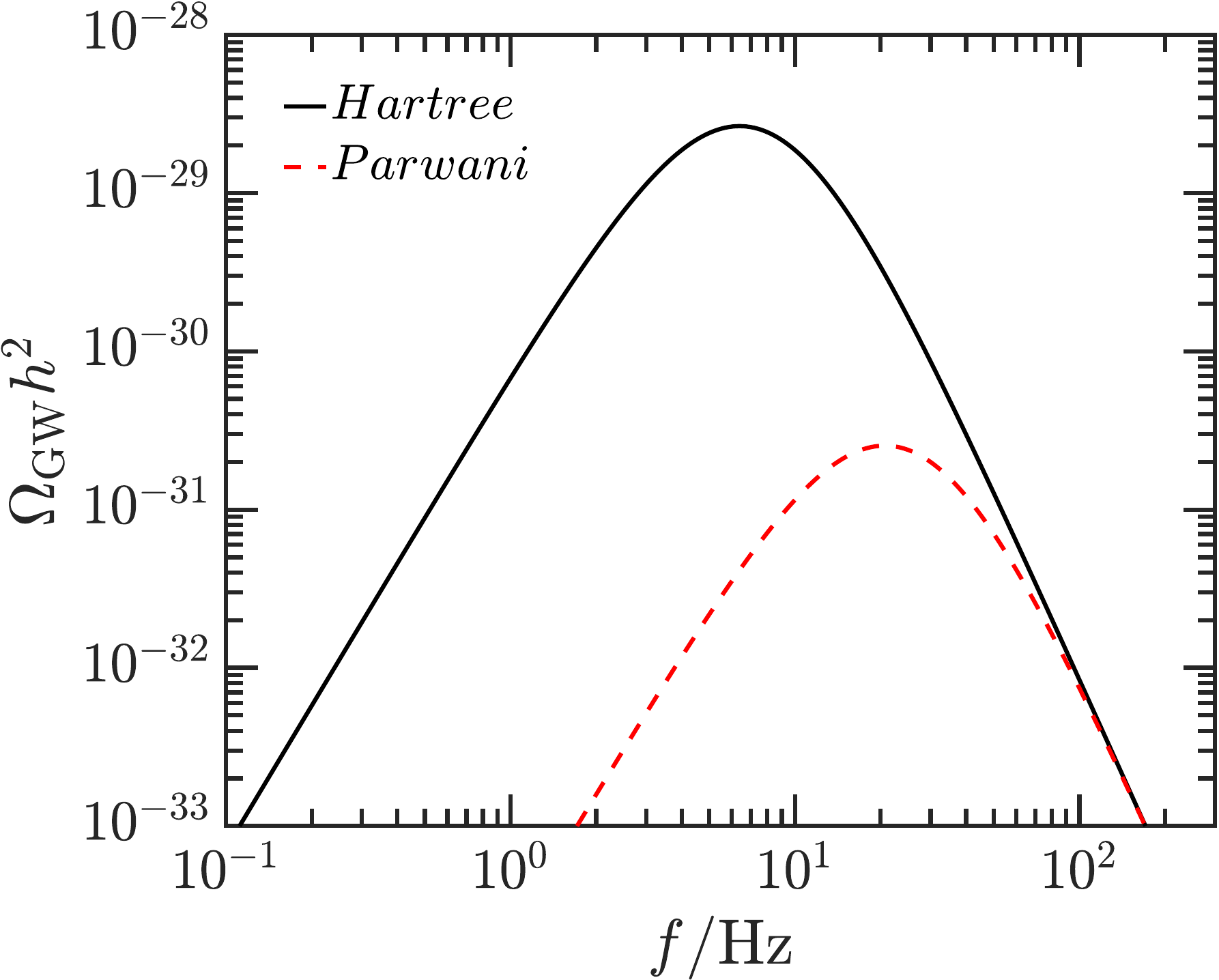}\,\,
    \includegraphics[width=0.43\columnwidth]{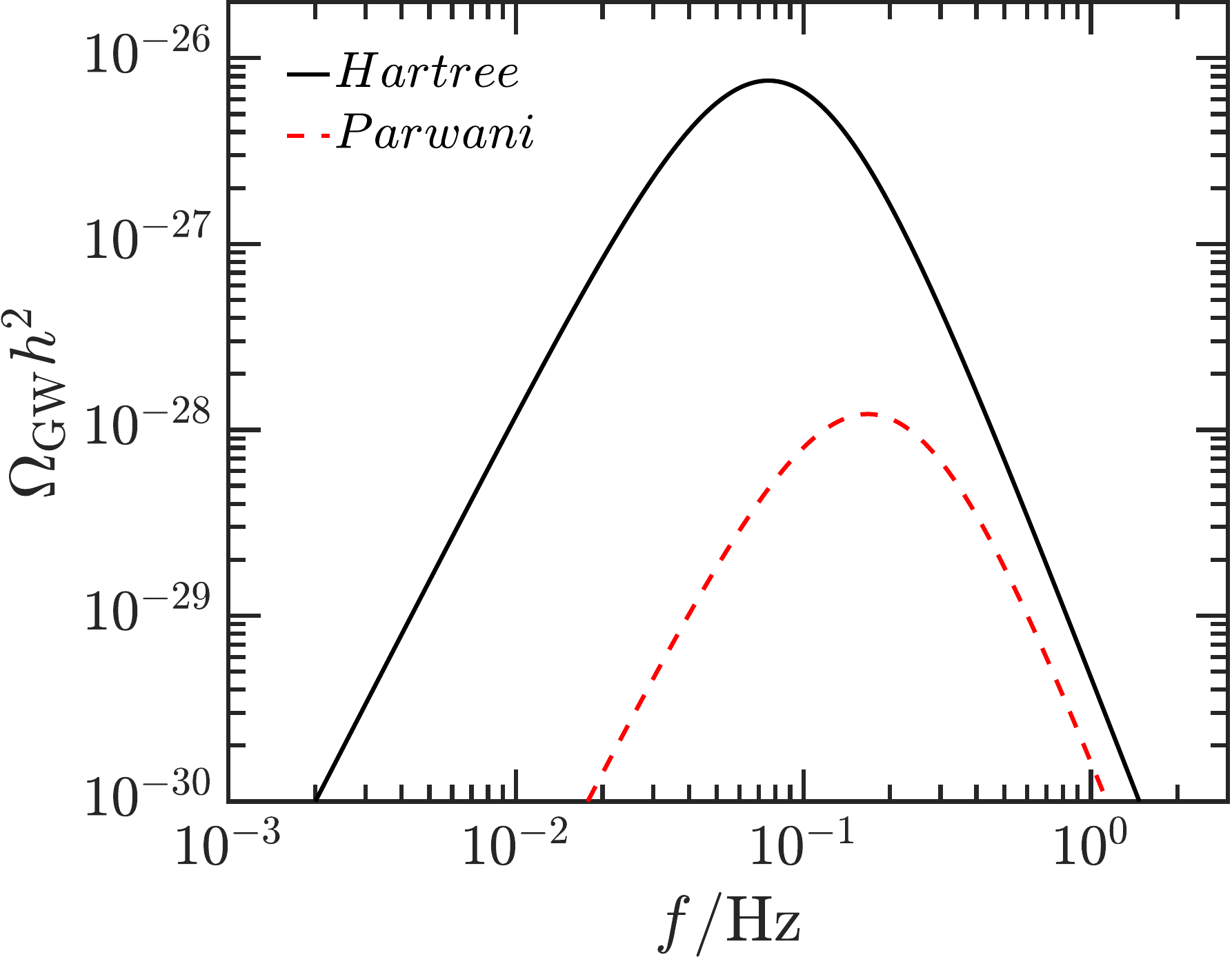}
    \caption{The gravitational wave spectrum computed for the one-step transition in Benchmark~1 \textit{(left)} and the two-step transition studied in Sec.~\ref{subsec:2stepPT} \textit{(right)} as obtained from the Hartree and Parwani effective potentials.
    }
    \label{fig:GWs}
\end{figure} 
%%%%%%%%%%%%%%%%%%%%%%%%%%%%%%%%%%%%%%%%%% 

In Fig.~\cref{fig:GWs}, we show the GW spectra predicted from each resummation scheme, corresponding to Benchmark Scenario 1 of Sec.~\cref{subsec:1stepPT} and the two-step transition analyzed in Sec.~\cref{subsec:2stepPT}, with the calculated PT parameters given in Table~\cref{tab:PT_params}. In these cases, the Hartree effective potential, on account of predicting a stronger transition, predicts a GW spectrum with an amplitude up to four orders of magnitude larger than the Parwani case. None of these spectra generated from PTs in this toy model we have investigated is within the reach of future GW observatories. However, we expect that our broad conclusions regarding the strength of the transition and the implications for the GW spectrum may carry over to more realistic models, such as extensions of the SM. It would be interesting to see how the predictions of the ring-resummed and Parwani approximations compare with the 2PI results when contributions from other fields, such as gauge bosons and fermions, are also incorporated. The calculation of the 2PI effective potential including gauge and fermion corrections is a separate, challenging task, which we plan to undertake in the future.

%%%%%%%%%%%%%%%%%%%%%%%%%%%%%%%%%%%%%%%%%%%%%%%%%%%%%%%%%%%%%%%%%%%%%%%%%%%%%%%
%%%%%%%%%%%%%%%%%%%%%%%%%%%%%%%%%%%%%%%%%%%%%%%%%%%%%%%%%%%%%%%%%%%%%%%%%%%%%%%
%
\section{Conclusions}
\label{sec:conclusion}
%
%%%%%%%%%%%%%%%%%%%%%%%%%%%%%%%%%%%%%%%%%%%%%%%%%%%%%%%%%%%%%%%%%%%%%%%%%%%%%%%
%%%%%%%%%%%%%%%%%%%%%%%%%%%%%%%%%%%%%%%%%%%%%%%%%%%%%%%%%%%%%%%%%%%%%%%%%%%%%%%

We have presented a consistent framework for the resummation of thermal masses based on 2PI effective action methods, applicable for studies of cosmological first-order phase transitions. Considering a model comprising two mixing real scalar fields, we first systematically renormalized the 2PI effective action in the Hartree approximation, paying particular attention to the equations of motion for the resummed propagator and the fields. As an outcome of the renormalization procedure, we showed that we were able to parameterize the effective action in terms of the $p^2 = 0$ masses and two sets of auxiliary couplings, one of which was scale invariant. We then derived the vacuum Hartree effective potential. Through field derivatives of this effective potential, we established connections between the auxiliary couplings and then related them to physical observables, such as the $p^2 = 0$ masses defined at the minimum of the potential and physical four-point scalar functions at zero external momentum. Although the calculations we have presented are specific to the Hartree approximation, the procedure to establish these relations is general, which we expect can be extended to higher truncations of the 2PI effective action.

We then studied thermal corrections to the Hartree effective potential, comparing it to the Parwani- and ring-approximated resummed effective potentials. In most scenarios, we found that the Hartree effective potential predicted {\em quantitatively} the strongest transitions.  We also found that the Parwani approximation predicted transitions {\em qualitatively} similar to the results of the Hartree effective potential, presumably because both potentials also augment the vacuum part of the potential with thermally corrected masses. However, in the Parwani approximation, the thermal masses are not self-consistently defined, arising instead from the high-temperature expansion approximation. In contrast, the masses in the Hartree approximation are well-defined solutions of the gap equations~\cref{eq:sqm_mixedbasis}, and hence consistently resummed to all orders. The ring approximation, in contrast, relies primarily on thermal corrections arising from the thermodynamic pressure and does not incorporate the thermal corrections to the vacuum, and typically exhibits the weakest transitions for this model. It would be interesting to study, once contributions from additional bosonic fields are included, whether such observations continue to persist. 

We also computed the dynamical nucleation rates in the Hartree case both including the non-trivial wave-function renormalization factors in the equations for the one-point functions, and by setting them to unity. This novel calculation of the wave-function renormalization factors was explicitly presented in Appendix~\cref{app:Z2_wave_function_ren_factors}. The numerical effect of wave-function renormalization factors on nucleation rates was found to be very small, by far inferior to the difference between the Hartree and one-loop potential approximations. 

Finally, we computed the gravitational wave spectrum arising from the benchmark scenarios for PTs we have considered, primarily as a demonstration of the impact on the phenomenology. The predictions for the GW spectra reflect our numerical results for the PTs: the Hartree approximation predicts the largest GW amplitude in most cases. We expect that these observations extend to other aspects of phenomenology associated with cosmological PTs, such as electroweak baryogenesis. Moreover, we envisage our computational techniques could also be extended to incorporate higher-order diagrams in the 2PI effective action, and a 2PI-based calculation of the bubble nucleation rate at finite temperature.

%%%%%%%%%%%%%%%%%%%%%%%%%%%%%%%%%%%%%%%%%%%%%%%%%%%%%%%%%%%%%%%%%%%%%%%%%%%%%%%
%%%%%%%%%%%%%%%%%%%%%%%%%%%%%%%%%%%%%%%%%%%%%%%%%%%%%%%%%%%%%%%%%%%%%%%%%%%%%%%
%
\acknowledgments
%
%%%%%%%%%%%%%%%%%%%%%%%%%%%%%%%%%%%%%%%%%%%%%%%%%%%%%%%%%%%%%%%%%%%%%%%%%%%%%%%
%%%%%%%%%%%%%%%%%%%%%%%%%%%%%%%%%%%%%%%%%%%%%%%%%%%%%%%%%%%%%%%%%%%%%%%%%%%%%%%

We gratefully acknowledge Werner Porod for his input and collaboration in the early stages of this project. AB is supported by the Basic Science Research Program through the National Research Foundation of Korea (NRF), funded by the Ministry of Education, under grant numbers [NRF-2021R1C1C1005076, NRF-2020R1I1A3068803, RS-2026-25484206]. AB acknowledges the discussion sessions during the ``Focus Workshop on Cosmological Phase Transitions'' at the Institute for Basic Sciences (IBS), Daejeon, Korea, where some preliminary results of this work were presented.

%%%%%%%%%%%%%%%%%%%%%%%%%%%%%%%%%%%%%%%%%%%%%%%%%%%%%%%%%%%%%%%%%%%%%%%%%%%%%%%
%%%%%%%%%%%%%%%%%%%%%%%%%%%%%%%%%%%%%%%%%%%%%%%%%%%%%%%%%%%%%%%%%%%%%%%%%%%%%%%
% APPENDIX
%
\appendix
%
%%%%%%%%%%%%%%%%%%%%%%%%%%%%%%%%%%%%%%%%%%%%%%%%%%%%%%%%%%%%%%%%%%%%%%%%%%%%%%%
%%%%%%%%%%%%%%%%%%%%%%%%%%%%%%%%%%%%%%%%%%%%%%%%%%%%%%%%%%%%%%%%%%%%%%%%%%%%%%%

%%%%%%%%%%%%%%%%%%%%%%%%%%%%%%%%%%%%%%%%%%%%%%%%%%%%%%%%%%%%%%%%%%%%%%%%%%%%%%%
%%%%%%%%%%%%%%%%%%%%%%%%%%%%%%%%%%%%%%%%%%%%%%%%%%%%%%%%%%%%%%%%%%%%%%%%%%%%%%%
%
\section{Wave-function renormalization factors for classical fields}
\label{app:Z2_wave_function_ren_factors}
%
%%%%%%%%%%%%%%%%%%%%%%%%%%%%%%%%%%%%%%%%%%%%%%%%%%%%%%%%%%%%%%%%%%%%%%%%%%%%%%%
%%%%%%%%%%%%%%%%%%%%%%%%%%%%%%%%%%%%%%%%%%%%%%%%%%%%%%%%%%%%%%%%%%%%%%%%%%%%%%%

In this appendix, we derive the wave-function renormalization factors $Z_\lin{\alpha,\!2}$ for the classical fields, which is a non-trivial problem in the 2PI formalism. In particular, the $Z_\lin{\alpha,\!2}$-factors differ from unity even in the Hartree approximation, as shown in the single-field case~\cite{Banik:2023nqm}. One practical point for our article is to show that these are scale independent, because only then may we equate the couplings $\lambda^\luin{2}_\lin{\alpha\beta}$ and $\lambda^\luin{0}_\lin{\alpha\beta}$, as we have done throughout the article. We note that the $Z_\lin{\alpha,\!2}$-factors appear in the classical equations for the fields, and hence they can affect physical predictions, such as for the GW spectrum, for example.

%=======================================================================================================
%=======================================================================================================
\begin{figure}[t!]
    \centering
    \includegraphics[width=0.8\columnwidth]{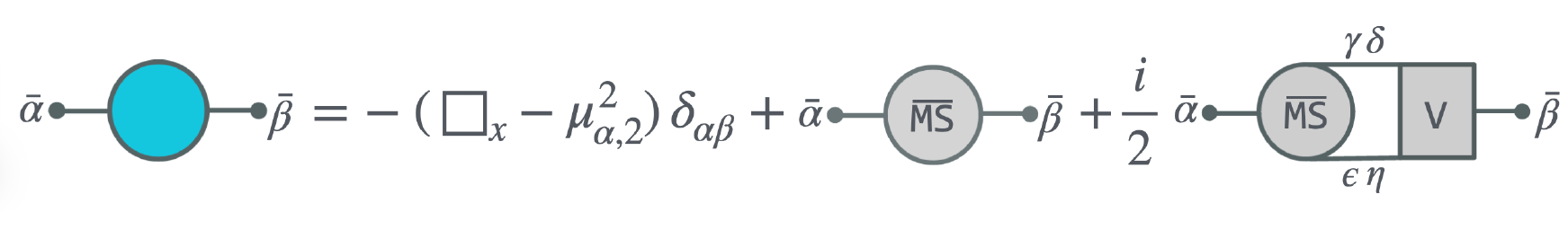}
    \vskip-0.3cm
    \caption{Graphical representation of the full inverse propagator (the 2-point function) as given in \eqref{eq:FullPropagator_2}. The blue circle represents the full 2PI action and the gray circle the ``interaction'' part $\Gamma_{\rm int}$. The lines ending in small black dots represent derivatives with respect to classical fields and the gray box denotes the vertex function $V_{\alpha\beta\bar\gamma}$.}
    \label{fig:EOM_propagator}
\end{figure} 
%=======================================================================================================
%=======================================================================================================

\begin{figure}[t!]
    \centering
    \includegraphics[width=0.6\columnwidth]{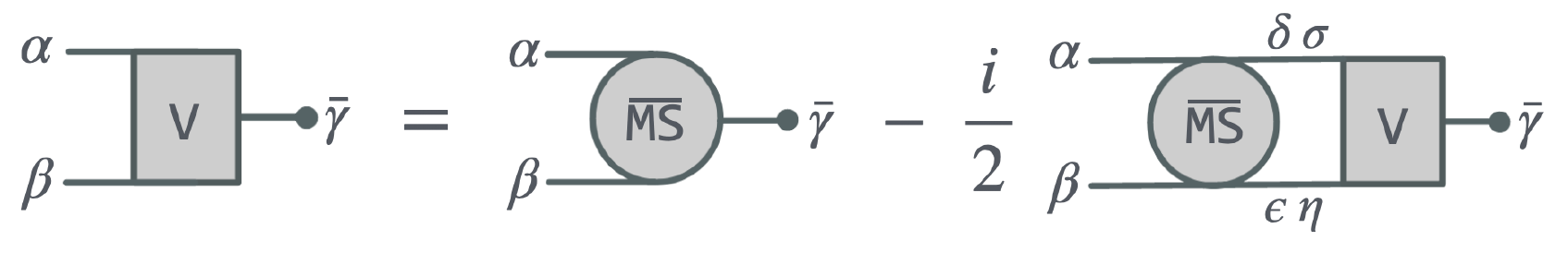}
    \vskip-0.3cm
    \caption{Graphical representation of the equation for the vertex function 
    $V_{\alpha\beta\bar\gamma}$ given in \eqref{eq:V-equation}.}
    \label{fig:EOM_vertex}
\end{figure}

The quantity of interest here is the full propagator of the theory, whose inverse is the full 2-point function computed from the 2PI effective action. We distinguish the full propagator from the classical two-point function $\Delta^{ab}_{\alpha\beta}$ by adding a bar over it. It is sufficient for us to concentrate only on the Feynman (11-) propagator at the renormalization point. Thus
\begin{equation}
i(\widebar\Delta^{11}_{\alpha\beta})^{-1} \equiv 
\frac{\bardelta^2\Gamma_{{\rm 2PI}}}{\bardelta\alpha\, \bardelta\beta} 
= - (\Box_x - \mu^2_\lin{\alpha,\!2})\delta_{\alpha\beta} 
  + \frac{\bardelta^2\Gamma_{{\rm int}}}{\bardelta\alpha\, \bardelta\beta}. 
\label{eq:FullPropagator_1}
\end{equation}
Here $\alpha,\beta \in \{\varphi,\chi\}$ and $\Gamma_{\rm int}$ extends $\Gamma_2$ to contain all terms in the 2PI action, except those in the quadratic classical action, which we treated explicitly in~\cref{eq:FullPropagator_1}. Moreover, $\bardelta$ denotes a full functional derivative, which also acts on the classical 2-point functions within the interaction term $\Gamma_{\rm int}[\varphi,\chi,\Delta[\varphi,\chi]]$. Using the chain rule for the functional differentiation in the last term, and then moving to momentum space, we can write Eq.~\cref{eq:FullPropagator_1} as 
\begin{equation}
i(\widebar\Delta^{11}_{\alpha\beta})^{-1} = (p^2 + \mu^2_\lin{\alpha,\!2})\delta_{\alpha\beta} 
  + C_{\bar\alpha\bar\beta} - \frac12 C_{\bar\alpha\gamma\epsilon} I^{\gamma\delta}_{\epsilon\eta} V_{\delta\eta\bar\beta}. 
\label{eq:FullPropagator_2}
\end{equation}
where the barred indices in the $C$-vertices indicate a differentiation of $\Gamma_{\rm int}$ {\em w.r.t.}~classical fields and ordinary indices {\em w.r.t.}~the two-point function:
\begin{equation}
C_{\bar\alpha\bar\beta} \equiv  \frac{\delta^2\Gamma_{{\rm int}}}{\delta\alpha\, \delta\beta}
\quad {\rm and} \quad
C_{\bar\alpha\gamma\eta} \equiv  -2\frac{\delta^2\Gamma_{{\rm int}}}{\delta\alpha\, \delta\Delta^\lin{\!11}_{\eta\gamma}}.
\end{equation}
\vskip-0.2cm \noindent
The loop integral $I^{\gamma\delta}_{\epsilon\eta}$ is an integral over the vacuum propagators:
\begin{align}
I^{\gamma\delta}_{\epsilon\eta} &\equiv - i \int \frac{{\rm d}^4p}{(2\pi )^4} 
           \Delta^\lin{11}_{\gamma\delta}(p)\Delta^\lin{11}_{\epsilon\eta}(p) 
           \nonumber\\
           &= \sum_{ij} \hat U_{\gamma i}\hat U_{\delta i}\hat U_{\epsilon j}\hat U_{\eta j}
           \big(\DeltaEps + iB_{0\widebar{\rm MS}}(p^2,\hat m_i^2,\hat m_j^2)\big)
           = \delta_{\gamma\delta}\delta_{\epsilon\eta} \DeltaEps + \bar I^{\gamma\delta}_{\epsilon\eta}.
\label{eq:master_eq_I}
\end{align}
\vskip-0.2cm \noindent
Here $\hat U_{\alpha n}$ are the elements of the orthogonal matrix defined in equation~\cref{eq:prop_diag} at renormalization point, $\hat m^2_i$ are the corresponding eigenvalues of the mass matrix $\hat m^2_{\alpha\beta}$~\cref{eq:mhatmatrix} and $iB_{0\widebar{\rm MS}}$ is the finite part of the standard Passarino-Veltman $B_0$-function~\cite{Denner:1991kt}. The finite part of the integral is denoted by $\bar I^{\gamma\delta}_{\epsilon\eta}$. Finally, it is easy to show that the vertex function $V_{\alpha\beta\bar\gamma} \equiv 
-2(\bardelta/\bardelta \gamma ) (\delta \Gamma_{\rm int} / \delta \Delta^\lin{\!11}_{\beta\alpha} )$, obeys an iterative equation 
\begin{equation}
V_{\alpha\beta\bar\gamma} = C_{\alpha\beta\bar\gamma} 
                   + \frac12C_{\alpha\beta\delta\epsilon} I^{\delta\sigma}_{\epsilon\eta} 
                            V_{\sigma\eta\bar\gamma}.
\label{eq:V-equation}
\end{equation}
where
\begin{equation}
C_{\alpha\beta\bar\gamma} \equiv -2\frac{\delta^2 \Gamma_{\rm int}}{\delta \gamma\, \Delta^\lin{\!11}_{\beta\alpha}}
\quad {\rm and} \quad
C_{\alpha\beta\delta\epsilon} \equiv  -4\frac{\delta^2\Gamma_{{\rm int}}}{\delta\Delta^\lin{11}_{\delta\alpha} \,\delta\Delta^\lin{\!11}_{\epsilon\beta}}
\label{eq:C3C4}
\end{equation}
Equations~\cref{eq:FullPropagator_2} and~\cref{eq:V-equation} are graphically represented in Figs.~\cref{fig:EOM_propagator} and~\cref{fig:EOM_vertex} respectively.

%=======================================================================================================
%=======================================================================================================
%=======================================================================================================
%=======================================================================================================

We start the evaluation of the terms not involving the vertex function $V_{\alpha\beta\bar\gamma}$ in~\cref{eq:FullPropagator_1}. The terms that make up the $\Gamma_{\rm int}$ are scattered through equations~\cref{eq:classical_counter_term_action,eq:inv_ct_prop,eq:ct_mass_matrix,eq:Gamma2_HT_renorm} and the quadratic part of the renormalized classical action. At any rate, one finds
\begin{align}
C_{\bar\varphi\bar\varphi} &= \delta^\luin{2}_\lin{\varphi\varphi}p^2
+ \delta^\duin{0}_{\mu\varphi} - \sfrac12\klpp{4}v^2 - \sfrac12\klpc{4}w^2 - \sfrac12\klpp{2} \Delta_\lin{\varphi\varphi}
\nonumber \\
C_{\bar\chi\bar\chi}       &= \delta^\luin{2}_\lin{\chi\chi} p^2
+ \delta^\duin{0}_{\mu\chi}    - \sfrac12\klcc{4}w^2 - \sfrac12\klpc{4}v^2 - \sfrac12\klcc{2} \Delta_\lin{\chi\chi}
\nonumber \\
C_{\bar\varphi\bar\chi}    &= \delta^\luin{2}_\lin{\varphi\chi}p^2
- \klpc{4}vw - \kblpc{2} \Delta_\lin{\varphi\chi}
\end{align}
From~\cref{eq:cancel_Deltapp_3,eq:coupled_eqs_1,eq:cancel_mu2}, we see\footnote{The order $\epsilon$ expansions would be sufficient for the purposes of this appendix. We expand the coupling kernels explicitly up to order $\epsilon^2$, however, because these expressions are needed in Section~\cref{sec:eff_pot} for the evaluation of the Hartree effective potential.} that all kernels are at least of the order $\epsilon$:
\begin{subequations}
\begin{align}
\mu^{2}_\lin{\alpha,\!0} + \delta^\duin{0}_{\mu\alpha} & = -B_\alpha/(A\DeltaEps) + {\cal O}(\epsilon^2),
\\
\blpcidx{0} + \dblpcidx{0} & = 1/\DeltaEps - 1/(\blpcidx{0}\DeltaEps^2) + {\cal O}(\epsilon^3),
\\
\lppidx{0}  + \dlppidx{0}  & = 2/\DeltaEps -2 \lccidx{0}/(A\DeltaEps^2) + {\cal O}(\epsilon^3),
\\
\lccidx{0}  + \dlccidx{0}  & = 2/\DeltaEps -2 \lppidx{0}/(A\DeltaEps^2) + {\cal O}(\epsilon^3),
\\
\lpcidx{0}  + \dlpcidx{0}  & = 2\lpcidx{0}/(A\DeltaEps^2) + {\cal O}(\epsilon^3),
\end{align}
\label{eq:limiting_kernels}%
\end{subequations}
and moreover, using also equations~\cref{eq:ct4-0_relations}, we obtain $\lambda^\luin{4}_\lin{\alpha\beta}+\delta^\duin{4}_\lin{\alpha\beta} =-2\lambda^\luin{4}_\lin{\alpha\beta} + {\cal O}(\epsilon)$. We can then write
\begin{equation}
(p^2 + \mu^2_\lin{\alpha\beta})\delta_{\alpha\beta} + C_{\bar\alpha\bar\beta} 
= Z^\luin{2}_\lin{\alpha\beta}p^2 - m^2_{\alpha\beta} 
+ \left\{ \begin{array}{l}
             \lppidx{4}v^2 + \lpcidx{4} w^2 \\[1mm]
             \lccidx{4}w^2 + \lpcidx{4} v^2 \\[1mm]
             2\lpcidx{4}vw 
          \end{array} 
  \right.
  \,.
\label{eq:first_part_of_full_propagator}
\end{equation}
%

%%%%%%%%%%%%%%%%%%%%%%%%%%%%%%%%%%%%%%%%%%%%%%%%%%%%%%%%%%%%%%%%%%%%%%%%%%%%%%%%%%%%%%%%%%%%%%%%%%%%%%%%%%%%%%%%
\paragraph{Three- and four-point $\widebar{\rm MS}$ vertices.}
%%%%%%%%%%%%%%%%%%%%%%%%%%%%%%%%%%%%%%%%%%%%%%%%%%%%%%%%%%%%%%%%%%%%%%%%%%%%%%%%%%%%%%%%%%%%%%%%%%%%%%%%%%%%%%%%

We next list the vertices $C_{\alpha\beta\bar\gamma}$ and $C_{\alpha\beta\delta\epsilon}$ defined in~\cref{eq:C3C4}, appearing in~\cref{eq:FullPropagator_2} and in the equation for the vertex function $V_{\alpha\beta\bar\gamma}$~\cref{eq:V-equation}. In the Hartree approximation, only a handful of vertices are non-vanishing:
\begin{equation}
C_{\alpha\alpha\bar\alpha} = (\laaidx{0} \!+\! \dlaaidx{0})\alpha\,,  \quad 
  C_{\beta\beta\bar\alpha} = (\lpcidx{0} \!+\! \dlpcidx{0})\alpha\,,  \quad
 C_{\alpha\beta\bar\alpha} = 2(\blpcidx{0} \!+\! \dblpcidx{0})\beta\,,
\label{eq:eq:C3}
\end{equation}
where $\alpha,\beta \in \{\varphi,\chi\}$ and it is assumed that $\alpha \neq \beta$. It should be obvious from the definition that the indices without bars can be switched: $C_{\beta\alpha\bar\alpha} = C_{\alpha\beta\bar\alpha}$. Moreover, it does not matter which order the differentiations were made, so the barred index can be rotated to be the first. Similarly, there are only three different types of non-vanishing $C_{\alpha\beta\delta\epsilon}$:
\begin{equation}
C_{\alpha\alpha\alpha\alpha} = \laaidx{0} \!+\! \dlaaidx{0}\,,  \quad 
C_{\alpha\alpha\beta\beta}   = \lpcidx{0} \!+\! \dlpcidx{0}\,, \quad
C_{\alpha\beta\alpha\beta}   = 2(\blpcidx{0} \!+\! \dblpcidx{0})\,,
\label{eq:eq:C4}
\end{equation}
where again $\alpha \neq \beta$ is assumed and $C_{\alpha\beta\beta\alpha}=C_{\alpha\beta\alpha\beta}$. Moreover, from~\cref{eq:limiting_kernels} we see that all non-vanishing $C$-vertices are at most of order $\epsilon$, and in particular $C_{\alpha\alpha\beta\beta} \sim \epsilon^2$.

%%%%%%%%%%%%%%%%%%%%%%%%%%%%%%%%%%%%%%%%%%%%%%%%%%%%%%%%%%%%%%%%%%%%%%%%%%%%%%%%%%%%%%%%%%%%%%%%%%%%%%%%%%%%%%%%
\paragraph{Integral functions.}
%%%%%%%%%%%%%%%%%%%%%%%%%%%%%%%%%%%%%%%%%%%%%%%%%%%%%%%%%%%%%%%%%%%%%%%%%%%%%%%%%%%%%%%%%%%%%%%%%%%%%%%%%%%%%%%%

Finally, we need to write down the integral functions $I^{\alpha\beta}_{\gamma\delta}$, and in particular resolve their divergence structure. Directly from~\cref{eq:master_eq_I} one finds that only the elements $I^{\alpha\alpha}_{\beta\beta}$ (where $\beta = \alpha$ is allowed) contain divergences: $I^{\alpha\alpha}_{\beta\beta}(Q) = \DeltaEps + \bar I^{\alpha\alpha}_{\beta\beta}(Q)$. All other integral elements are both finite and $Q$-independent. Explicitly, the finite parts of these functions are given by:
\begin{align}
\bar I^{\varphi\varphi}_{\varphi\varphi} &= c^4_\theta i\bar B_{11} + s^4_\theta i\bar B_{22} 
                                         + 2s^2_\theta c^2_\theta i\bar B_{12}\,,
\noindent\\
\bar I^{\varphi\varphi}_{\varphi\chi} &= c^3_\theta s_\theta (i\bar B_{12} - i\bar B_{11}) 
                                        -s^3_\theta c_\theta (i\bar B_{12} - i\bar B_{22})\,,
\noindent\\
\bar I^{\varphi\chi}_{\varphi\chi} &= s^2_\theta c^2_\theta ( i\bar B_{11} + i\bar B_{22} -2i\bar B_{12})\,.
\end{align}
where we used the shorthands $i\bar B_{ij} \equiv iB_{0\widebar{\rm MS}}(p^2,\hat m_i^2,\hat m_j^2)$ and $(c_\theta,s_\theta)\equiv(\cos \theta\,,\sin\theta)$.
Moreover, $\bar I^{\varphi\varphi}_{\chi\chi} = \bar I^{\varphi\chi}_{\varphi\chi} + i\bar B_{12}$ and
$\bar I^{\chi\chi}_{\chi\chi} = \bar I^{\varphi\varphi}_{\varphi\varphi}(s_\theta \leftrightarrow c_\theta)$ and
$\bar I^{\chi\chi}_{\varphi\chi} = \bar I^{\varphi\varphi}_{\varphi\chi}(s_\theta \leftrightarrow c_\theta)$. In addition, the indices within both upper and lower index pairs can be switched, and finally, the lower and upper index pairs can be swapped to no effect. 

%%%%%%%%%%%%%%%%%%%%%%%%%%%%%%%%%%%%%%%%%%%%%%%%%%%%%%%%%%%%%%%%%%%%%%%%%%%%%%%%%%%%%%%%%%%%%%%%%%%%%%%%%%%%%%%%
\paragraph{The vertex function $V_{\alpha\beta\bar\gamma}$.}
%%%%%%%%%%%%%%%%%%%%%%%%%%%%%%%%%%%%%%%%%%%%%%%%%%%%%%%%%%%%%%%%%%%%%%%%%%%%%%%%%%%%%%%%%%%%%%%%%%%%%%%%%%%%%%%%

We have now set up everything we need to solve the equation~\cref{eq:V-equation} for the vertex function $V_{\alpha\beta\bar\gamma}$. We consider as an explicit example the equation for $V_{\varphi\varphi\bar\varphi}$. Writing all non-vanishing terms subject to~\cref{eq:eq:C3,eq:eq:C4}, we find:
\begin{align}
V_{\varphi\varphi\bar\varphi} = C_{\varphi\varphi\bar\varphi} 
  &+ \sfrac12 \big( C_{\varphi\varphi\varphi\varphi} (\DeltaEps + I^{\varphi\varphi}_{\varphi\varphi}) 
                  + C_{\varphi\varphi\chi\chi} \bar I^{\varphi\chi}_{\varphi\chi} \big) V_{\varphi\varphi\bar\varphi}
\nonumber \\
  &+ \sfrac12 \big( C_{\varphi\varphi\varphi\varphi} \bar I^{\varphi\chi}_{\varphi\chi} 
                  + C_{\varphi\varphi\chi\chi} (\DeltaEps + \bar I^{\chi\chi}_{\chi\chi}) \big) V_{\chi\chi\bar\varphi}
\nonumber \\
  &+ \sfrac12 \big( C_{\varphi\varphi\varphi\varphi} \bar I^{\varphi\varphi}_{\varphi\chi} 
                  + C_{\varphi\varphi\chi\chi} \bar I^{\chi\varphi}_{\chi\chi} \big) V_{\varphi\chi\bar\varphi}
\label{eq:ppp_first}
\end{align}
Similar equations can be written for $V_{\chi\chi\bar\varphi}$ and $V_{\varphi\chi\bar\varphi}$. Using the explicit forms for the $C$-vertices and the $\bar I$-integrals, and expanding everything in $\epsilon$, one eventually finds a set of coupled equations that can be compactly written in the matrix form:
\begin{equation}
\begin{pmatrix}
    \frac{\lccidx{0}}{A} - \bar I^{\varphi\varphi}_{\varphi\varphi} & 
 - (\frac{\lpcidx{0}}{A} + \bar I^{\varphi\chi}_{\varphi\chi}) &
                        - 2\bar I^{\varphi\varphi}_{\varphi\chi} \\
 - (\frac{\lpcidx{0}}{A} + \bar I^{\varphi\chi}_{\varphi\chi}) & 
    \frac{\lccidx{0}}{A} - \bar I^{\chi\chi}_{\chi\chi} &
                        - 2\bar I^{\chi\chi}_{\varphi\chi} \\
   - \bar I^{\varphi\varphi}_{\varphi\chi} & 
   - \bar I^{\chi\chi}_{\varphi\chi} &
 \frac{1}{\bar\lambda^{(0)}_{\phi\chi}} 
 - \bar I^{\varphi\varphi}_{\chi\chi} - \bar I^{\varphi\chi}_{\varphi\chi} 
\end{pmatrix}
\left(
\begin{array}{c}
  V_{\varphi\varphi\bar\varphi} \\  
  V_{\chi\chi\bar\varphi} \\
  V_{\varphi\chi\bar\varphi} 
\end{array}
\right)
=
\left(
\begin{array}{c}
  2v\\
  0\\
  w
\end{array}
\right)\,.
\label{eq:Vphieq1}
\end{equation}
Let us set up some notation for brevity in what follows.  Denoting the $3\times3$ matrix in~\cref{eq:Vphieq1} by $G_{\bar\varphi}$, as well as $V_{\bar\varphi} \equiv (V_{\varphi\varphi\bar\varphi},V_{\chi\chi\bar\varphi},V_{\varphi\chi\bar\varphi} )^T$ and $S_{\bar \varphi} \equiv (2v, 0, w)^T$, we may write~\cref{eq:Vphieq1} as
\begin{equation}
G_{\bar\varphi} V_{\bar\varphi} = S_{\bar\varphi}\,.
\label{eq:Vphieq2}
\end{equation}
An analogous equation can be derived that relates the vertex functions to a classical $\chi$-field:
\begin{equation}
G_{\bar\chi} V_{\bar\chi} = S_{\bar\chi}\,.
\label{eq:Vchieq2}
\end{equation}
where $G_{\bar\chi} = G_{\bar\varphi}(\varphi \leftrightarrow \chi)$, $V_{\bar\chi} \equiv (V_{\chi\chi\bar\chi},V_{\varphi\varphi\bar\chi},V_{\chi\varphi\bar\chi} )^T$ and $S_{\bar\chi} \equiv (2w, 0, v)^T$.

It should be noted that the matrices $G_{\bar\varphi}$ and $G_{\bar\chi}$ are scale independent. For off-diagonal elements, this is clear, because there the $\bar I^{\alpha\beta}_{\gamma\delta}$-integrals appearing there are manifestly finite and scale-independent, and also the ratio $\lpcidx{0}/A$ is scale-independent according to~\cref{eq:running_couplings1}. For diagonal elements the scale-independence follows because the scale dependence of $1/\blpcidx{0}$ and $\laaidx{0}/A$, shown
in~\cref{eq:running_barcoupling0_soln,eq:running_couplings1}, exactly cancels the scale dependence coming from $\bar I^{\alpha\alpha}_{\beta\beta}$, given by
\begin{equation}
\bar I^{\alpha\alpha}_{\beta\beta}(Q) = \bar I^{\alpha\alpha}_{\beta\beta}(Q_0) + L_{QQ_0}\,,
\end{equation}
where $L_{QQ_0}$ was defined in~\cref{eq:running_barcoupling0_soln}. Because $G_{\bar\alpha}$ and clearly $S_{\bar\alpha}$ are thus $Q$-independent, so are also the six vertex functions contained in 3-vectors $V_{\bar\alpha}$:
\begin{equation}
V_{\bar\alpha}(p^2) = G^{-1}_{\bar\alpha}(p^2) S_{\bar\alpha}\,,
\label{eq:Vfinal}
\end{equation}
where the $p^2$-dependence comes from the finite integral functions $\bar I^{\alpha\beta}_{\gamma\delta}(p^2)$ within the matrices $G_{\bar\alpha}(p^2)$.

%%%%%%%%%%%%%%%%%%%%%%%%%%%%%%%%%%%%%%%%%%%%%%%%%%%%%%%%%%%%%%%%%%%%%%%%%%%%%%%%%%%%%%%%%%%%%%%%%%%%%%%%%%%%%%%%
\paragraph{The self-energy function.}
%%%%%%%%%%%%%%%%%%%%%%%%%%%%%%%%%%%%%%%%%%%%%%%%%%%%%%%%%%%%%%%%%%%%%%%%%%%%%%%%%%%%%%%%%%%%%%%%%%%%%%%%%%%%%%%%

We are now finally ready to evaluate the self-energy functions appearing in equation~\cref{eq:FullPropagator_2}:
\begin{equation}
\Pi_{\alpha\beta} \equiv \frac12 C_{\bar\alpha\gamma\epsilon} I^{\gamma\delta}_{\epsilon\eta} V_{\delta\eta\bar\beta}\,.
\label{eq:self-energy-correction}
\end{equation}
Because vertices $V_{\alpha\beta\bar\gamma}$ are finite order one quantities and all vertices $C_{\bar\alpha\gamma\epsilon}\sim \epsilon$, only the divergent parts from the integral functions contribute and~\cref{eq:self-energy-correction} immediately collapses to following three equations:
\begin{align}
\Pi_{\varphi\varphi} &= V_{\varphi\varphi\bar\varphi}v + V_{\varphi\chi\bar\varphi}w\,,
\nonumber \\
\Pi_{\chi\chi}       &= V_{\chi\chi\bar\chi}w + V_{\chi\varphi\bar\chi}v\,,
\nonumber \\
\Pi_{\varphi\chi}    &= V_{\varphi\varphi\bar\chi}v + V_{\chi\varphi\bar\chi}w\,.
\end{align}
Combining these solutions with~\cref{eq:first_part_of_full_propagator}, we can eventually write the renormalized full propagator as follows
\begin{align}
i(\widebar\Delta^\duin{11}_\lin{\varphi\varphi})^{-1} 
        &\;=\; Z^\lin{\,(2)}_\lin{\varphi\varphi} p^2 - \hat m^2_\lin{\varphi\varphi}
         + \big( \lppidx{0}v - V_{\varphi\varphi\bar\varphi} \big) v 
         + \big( \lpcidx{0}w - V_{\varphi\chi\bar\varphi} \big) w\,.
\noindent \nonumber \\
i(\widebar\Delta^\duin{11}_\lin{\chi\chi})^{-1} 
       &\;=\; Z^\lin{\,(2)}_\lin{\chi\chi} p^2 - \hat m^2_\lin{\varphi\varphi}
        + \big( \lccidx{0}w - V_{\chi\chi\bar\chi} \big) v 
        + \big( \lpcidx{0}w - V_{\chi\varphi\bar\chi}\big) w\,,
\noindent \nonumber \\
i(\widebar\Delta^\duin{11}_\lin{\varphi\chi})^{-1} 
      &\;=\; Z^\lin{\,(2)}_\lin{\varphi\chi} p^2 - \hat m^2_\lin{\varphi\chi}
       + \big( \lpcidx{0}w - V_{\varphi\varphi\bar\chi} \big) v 
       + \big( \lpcidx{0}v - V_{\chi\varphi\bar\chi}\big) w\,.
\label{eq:full_propagator_final}
\end{align}
We point out that for arbitrary constant fields the $V_{\alpha\beta\bar\gamma}$-functions by definition reduce to differentials of the mass function with respect to fields at zero momentum:
\begin{equation}
V_{\alpha\beta\bar\gamma}(p^2=0) = \frac{\partial m^2_{\alpha\beta}}{\partial \gamma}\,.
\label{eq:Valphazero}
\end{equation}
This implies that the terms in parentheses in the equations~\cref{eq:full_propagator_final} cancel at $p^2=0$, which implies the full propagator has the same $p^2=0$ masses as the classical propagator function $\Delta^\duin{11}_\lin{\alpha\beta}$.

%%%%%%%%%%%%%%%%%%%%%%%%%%%%%%%%%%%%%%%%%%%%%%%%%%%%%%%%%%%%%%%%%%%%%%%%%%%%%%%%%%%%%%%%%%%%%%%%%%%%%%%%%%%%%%%%
\paragraph{Wave-function renormalization factors.}
%%%%%%%%%%%%%%%%%%%%%%%%%%%%%%%%%%%%%%%%%%%%%%%%%%%%%%%%%%%%%%%%%%%%%%%%%%%%%%%%%%%%%%%%%%%%%%%%%%%%%%%%%%%%%%%%

The $Z^\lin{\,(2)}_{\alpha\beta}$ factors can now be read off directly from~\cref{{eq:full_propagator_final}}: 
\begin{align}
Z^\lin{\,(2)}_\lin{\varphi\varphi} -1 
   &\;=\;  v \frac{\partial V_{\varphi\varphi\bar\varphi}}{\partial p^2}\Big|_{p^2=0} 
        +  w \frac{\partial V_{\varphi\chi\bar\varphi}}{\partial p^2}\Big|_{p^2=0} \,.
\noindent \\
Z^\lin{\,(2)}_\lin{\chi\chi} -1
  &\;=\;   v \frac{\partial V_{\chi\chi\bar\chi}}{\partial p^2}\Big|_{p^2=0}  
         + w \frac{\partial V_{\chi\varphi\bar\chi}}{\partial p^2}\Big|_{p^2=0} \,,
\noindent \\
Z^\lin{\,(2)}_\lin{\varphi\chi} -1
 &\;=\;   v \frac{\partial V_{\varphi\varphi\bar\chi} }{\partial p^2}\Big|_{p^2=0} 
        + w \frac{\partial V_{\chi\varphi\bar\chi}}{\partial p^2}\Big|_{p^2=0}\,.
\label{eq:wfr_2}
\end{align}
The $p^2$-derivatives appearing in~\cref{eq:wfr_2} can be evaluated from~\cref{eq:Vfinal}. The result can be compactly given in the vector form:
\begin{equation}
\frac{\partial V_{\bar\alpha}}{\partial p^2}\Big|_{p^2=0} 
= G^{-1}_\alpha(0) \frac{\partial G_\alpha}{\partial p^2}\Big|_{p^2=0} G^{-1}_\alpha(0) \; S_\alpha \,.
\end{equation}

For completeness, we finally give explicit expressions for the $iB_{0\widebar{\rm MS}}$-function and its derivative at $p^2=0$, which are needed to evaluate the various $\bar I^{\alpha\beta}_{\gamma\delta}$ and $\partial_{p^2}\bar I^{\alpha\beta}_{\gamma\delta}$ factors that appear in the $G_{\bar\alpha}$-matrices and their derivatives at zero momentum:
\begin{equation}
iB_{0\widebar{\rm MS}}(0,m_i^2,m_j^2) 
= \frac{1}{16\pi^2}\Bigg\{ \frac{1}{m_i^2-m_j^2}
           \Big[ m_i^2 \ln\Big(\frac{m_i^2}{Q^2}\Big) - m_j^2 \ln\Big(\frac{m_j^2}{Q^2}\Big)\Big]  - 1\Bigg\}\,,
\end{equation}
\begin{equation}
\frac{\partial iB_{0\widebar{\rm MS}}(p^2,m_i^2,m_j^2)}{\partial p^2}\Big|_{p^2=0} 
= \frac{1}{32\pi^2}\Bigg\{\frac{m_i^2+m_j^2}{(m_i^2-m_j^2)^2} - \frac{2m_i^2m_j^2}{(m_i^2-m_j^2)^3}\ln\Big(\frac{m_i^2}{m_j^2}\Big) \Bigg\}\,.
\label{eq:appdB0dp2}
\end{equation}
This completes the renormalization of the full propagator function and our evaluation of the wave-function renormalization factors.

%%%%%%%%%%%%%%%%%%%%%%%%%%%%%%%%%%%%%%%%%%%%%%%%%%%%%%%%%%%%%%%%%%%%%%%%%%%%%%%
\paragraph{Field- and $T$-dependent wave-function renormalization factors.}
%%%%%%%%%%%%%%%%%%%%%%%%%%%%%%%%%%%%%%%%%%%%%%%%%%%%%%%%%%%%%%%%%%%%%%%%%%%%%%%

We close this appendix with some speculation. The entire derivation above can be easily redone by replacing $v$ and $w$ with arbitrary constant fields $\varphi$ and $\chi$, and extending the definition of the gap masses and possibly $I^\lin{\alpha\beta}_{\gamma\delta}$ functions to finite temperature. One could then extend the definition of $Z_\lin{\alpha,\!2}$'s from constants to functions of fields and of temperature, defined to ensure that the full propagator $\widebar \Delta_{\alpha\beta}$ has a residue equal to one at $p^2=0$ for any $\varphi$, $\chi$ and $T$. This definition would imply that the masses given by the gap equation would be the true physical masses of excitations around any field configuration considered, and it would be somewhat similar to going to the quasi-particle basis in usual thermal field theory. Still restricting to the adiabatic limit, this would give a generalized effective action, not with constant $Z_\lin{\alpha,\!2}$'s as in~\cref{eq:S3}, but with {\em functional} ones:
\begin{align}
    S_3(T) = 4\pi\int_0^\infty dr \, r^2 
    \left[ \frac12 \sum_{\alpha =\varphi,\chi}Z_\lin{\alpha,\!2}(\varphi,\chi,T)\Big(\frac{d\alpha}{dr}\Big)^2
           + V(\varphi,\chi;T)\right].
\label{eq:S3app}
\end{align}
This action would lead to somewhat larger corrections away from the simple $Z_\lin{\alpha,\!2}\equiv 1$ limit in nucleation calculations. However, the functional wave-function factors display singularities at the points where the mass eigenvalues vanish, due to the function~\cref{eq:appdB0dp2}, making the functional $Z_\lin{\alpha,\!2}$-factors more challenging to implement numerically. We therefore leave the study of this possibility to future work.

%%%%%%%%%%%%%%%%%%%%%%%%%%%%%%%%%%%%%%%%%%%%%%%%%%%%%%%%%%%%%%%%%%%%%%%%%%%%%%%
%%%%%%%%%%%%%%%%%%%%%%%%%%%%%%%%%%%%%%%%%%%%%%%%%%%%%%%%%%%%%%%%%%%%%%%%%%%%%%%
%
\section{Scale invariance of the Hartree potential}
\label{app:scale_invariance}
%
%%%%%%%%%%%%%%%%%%%%%%%%%%%%%%%%%%%%%%%%%%%%%%%%%%%%%%%%%%%%%%%%%%%%%%%%%%%%%%%
%%%%%%%%%%%%%%%%%%%%%%%%%%%%%%%%%%%%%%%%%%%%%%%%%%%%%%%%%%%%%%%%%%%%%%%%%%%%%%%

In this appendix, we explicitly show that the Hartree effective potential in Eq.~\cref{eq:Vht_final} is independent of the renormalization scale $Q$. To this end, we first show that the masses are scale-invariant. It is convenient to rewrite the diagonal elements in Eq.~\cref{eq:sqm_mixedbasis} into the following matrix form:
\begin{equation}
    \begin{pmatrix}
        \sqmpp \\
        \sqmcc
    \end{pmatrix}
    =
    \begin{pmatrix}
        m^2_{0\varphi\varphi}  \\
        m^2_{0\chi\chi} 
    \end{pmatrix}
    + 
    \frac12
    \Lambda
    \begin{pmatrix}
        \DeltaF_\lin{\varphi\varphi} \\
        \DeltaF_\lin{\chi\chi}
    \end{pmatrix}\,,
    \label{eq:mass_mateqn}
\end{equation}
where $m^2_{0\varphi\varphi}$ and $m^2_{0\chi\chi}$ were defined in \eqref{eq:mass0matrix}, and we now define the matrix of couplings:
\begin{equation}
    \Lambda \equiv
    \begin{pmatrix}
        \lppidx{0} &  \lpcidx{0} \\
        \lpcidx{0} &  \lccidx{0}
    \end{pmatrix}\,.
    \label{eq:coup_mat}
\end{equation}
We need the following intermediate result:
\begin{equation}
    Q\,\partial_Q\left[\DeltaFO(m^2_i)\right] = Q\left(\partial_Q m^2_i\right)L_i - \frac{m_i^2}{8\pi^2}\,,
    \label{eq:delQDelta}
\end{equation}
where $L_i \equiv (16\pi^2)^{-1}\,\ln(m_i^2/Q^2)$. Also, we need the following formula for the mixing angle:
\begin{equation}
    \sin 2\theta = \frac{2\blpcidx{0}\varphi\chi}{\delta m^2 - \lpcidx{0}\delta \Delta}\,,
\label{eq:mixing_angle_sin}
\end{equation}
whose $Q$-derivative becomes:
\begin{align}
    \partial_Q \theta &= K_1 \partial_Q m^2_1 - K_2 \partial_Q m^2_2\,, 
    \quad {\rm where} \quad 
    K_i \equiv \frac{s^2_{2\theta}}{4\,c_{2\theta}}
               \left(\frac{1-\blpcidx{0}L_i}{\blpcidx{2}\,\varphi\chi}\right)\,.
    \label{eq:delQtheta}
\end{align}
Now, taking a derivative of~\cref{eq:mass_mateqn} {\em w.r.t.}~$Q$, and invoking the running equations~\cref{eq:running_sqmu} and~\cref{eq:running_couplings0}, we obtain
\begin{align}
    \begin{pmatrix}
    \partial_Q\sqmpp \\
    \partial_Q\sqmcc
    \end{pmatrix}
    =
    \frac12
    \Lambda
    \begin{pmatrix}
        \partial_Q\DeltaF_\lin{\varphi\varphi} \\
        \partial_Q\DeltaF_\lin{\chi\chi}
    \end{pmatrix}\,.
\end{align}
Converting to the diagonal basis and using~\cref{eq:delQDelta}, we first have,
\begin{align}
    (\mathbb{I}_2- {\cal B})\begin{pmatrix}
        \partial_Q m^2_1 \\
        \partial_Q m^2_2
    \end{pmatrix}
    = {\cal C} \partial_Q\theta\,,
    \label{eq:delQsqm}
\end{align}
where the $2\times 2$ matrix ${\cal B}$ and the 2-vector ${\cal C}$ are 
\begin{subequations}
\begin{equation}
    {\cal B} =
    \frac{1}{2c_{2\theta}} 
    \begin{pmatrix}
    \phantom{-}c^2_\theta  & -s^2_\theta\\
    -s^2_\theta & \phantom{-}c^2_\theta
    \end{pmatrix}
    \Lambda
    \begin{pmatrix}
    c^2_\theta  & s^2_\theta\\
    s^2_\theta & c^2_\theta
    \end{pmatrix}
    \begin{pmatrix}
    L_1  & 0\\
    0 & L_2
    \end{pmatrix}\,,
    \label{eq:mat_B}
\end{equation}
\begin{equation}
    {\cal C} =
    \tan 2\theta
    \begin{pmatrix}
    \phantom{-}c^2_\theta  & -s^2_\theta\\
    -s^2_\theta & \phantom{-}c^2_\theta
    \end{pmatrix}
    \begin{pmatrix}
    -\delta m^2  + \frac{1}{2}(\lppidx{0}-\lpcidx{0})\delta \Delta \\
    \phantom{-}\delta m^2  - \frac{1}{2}(\lccidx{0}-\lpcidx{0})\delta \Delta
    \end{pmatrix} 
    \equiv 
    \begin{pmatrix}
    C_1 \\
    C_2
    \end{pmatrix}\,,
    \label{eq:mat_C}
\end{equation}
\end{subequations}
where $\delta m^2 = m^2_2- m^2_1$, $\delta \Delta = \DeltaFO(m^2_2)- \DeltaFO(m^2_1)$. Then using~\cref{eq:delQtheta}, one arrives at the matrix equation:
\begin{equation}
    (\mathbb{I}_2 - {\cal B} - \tilde{{\cal C}})
    \begin{pmatrix}
        \partial_Q m^2_1 \\
        \partial_Q m^2_2
    \end{pmatrix}
    = 0\,,\quad {\rm where} \quad {\tilde {\cal C}} = 
    \begin{pmatrix}
    K_1C_1 & -K_2C_1 \\
    K_1C_2 & -K_2C_2
    \end{pmatrix}\,.
    \label{eq:dQmassmatdiag}
\end{equation}
In general, $(\mathbb{I}_2 - {\cal B} - \tilde{{\cal C}})$ is non-singular and therefore one can conclude $\partial_Q m^2_i = 0$. As a corollary, this implies from~\cref{eq:delQsqm} that $\partial_Q\theta = 0$, and thus $\partial_Q m^2_{\alpha\beta} = 0$. Finally, taking a derivative of~\cref{eq:Vht_final} {\em w.r.t.}~$Q$,
\begin{align}
    \partial_QV_{\rm HT} = 
    &\frac12 \Big[\partial_Q\Big(\frac{\lppidx{0}}A\Big)m^4_{\varphi\varphi}
                 +\partial_Q\Big(\frac{\lccidx{0}}A\Big)m^4_{\chi\chi}
                -2\partial_Q\Big(\frac{\lpcidx{0}}A\Big)\sqmpp\sqmcc \Big] 
\nonumber \\
    &-\frac{1}{2(\blpcidx{0})^2}\big(\partial_Q\blpcidx{0}\big) m^4_{\varphi\chi}\,.
\end{align}
Invoking the running of the couplings given in \cref{eq:running_barcoupling0} and~\cref{eq:running_couplings0}, then converting to the diagonal basis, we find that indeed $\partial_QV_{\rm HT} = 0$, as desired.

%%%%%%%%%%%%%%%%%%%%%%%%%%%%%%%%%%%%%%%%%%%%%%%%%%%%%%%%%%%%%%%%%%%%%%%%%%%%%%%
%%%%%%%%%%%%%%%%%%%%%%%%%%%%%%%%%%%%%%%%%%%%%%%%%%%%%%%%%%%%%%%%%%%%%%%%%%%%%%%
%
\section{Derivatives of the mass matrix elements}
\label{app:mass-derivatives}
%
%%%%%%%%%%%%%%%%%%%%%%%%%%%%%%%%%%%%%%%%%%%%%%%%%%%%%%%%%%%%%%%%%%%%%%%%%%%%%%%
%%%%%%%%%%%%%%%%%%%%%%%%%%%%%%%%%%%%%%%%%%%%%%%%%%%%%%%%%%%%%%%%%%%%%%%%%%%%%%%

In the course of connecting the auxiliary couplings to the physical parameters, as described in Sec.~\ref{subsec:phys_params}, we require the first field derivatives of the resummed masses, cf. \cref{eq:lambda_connection}. Let us first convert \cref{eq:mass_mateqn} to the diagonal basis:
\begin{equation}
    \begin{pmatrix}
        m^2_1 \\
        m^2_2
    \end{pmatrix}
    =
    \frac{1}{c_{2\theta}}
    \begin{pmatrix}
        \phantom{-}c^2_\theta & -s^2_\theta  \\
        -s^2_\theta & \phantom{-}c^2_\theta 
    \end{pmatrix}
    \Bigg\{
    \begin{pmatrix}
        m^2_{0\varphi\varphi}  \\
        m^2_{0\chi\chi} 
    \end{pmatrix}
    + 
    \frac12
    \Lambda
    \begin{pmatrix}
        c^2_\theta & \,s^2_\theta  \\
        s^2_\theta & \,c^2_\theta 
    \end{pmatrix}
    \begin{pmatrix}
        \DeltaFO(m^2_1)\\
        \DeltaFO(m^2_1)
    \end{pmatrix}
    \Bigg\}
    \,.
    \label{eq:massdiag_mateqn}
\end{equation}
Here, $\Lambda$ is the same matrix of couplings in \cref{eq:coup_mat}. We now take a field derivative $\partial_\eta$, where $\eta \in \{\varphi,\chi\}$, of the above expression to arrive at
\begin{equation}
    (1-{\cal B})
    \begin{pmatrix}
        \partial_\eta m^2_1 \\
        \partial_\eta m^2_2
    \end{pmatrix}
    =
    {\cal C}\partial_\eta \theta 
    + 
    \frac{1}{c_{2\theta}}
    \begin{pmatrix}
        \phantom{-} c^2_\theta & -s^2_\theta  \\
        -s^2_\theta & \phantom{-} c^2_\theta 
    \end{pmatrix}
    \begin{pmatrix}
        \partial_\eta m^2_{0\varphi\varphi}  \\
        \partial_\eta m^2_{0\chi\chi} 
    \end{pmatrix}\,,
\end{equation}
where the matrices $\cal B$ and $\cal C$ are the same as in \cref{eq:mat_B} and \cref{eq:mat_C}. The field derivative of the mixing angle~\cref{eq:mixing_angle_sin} is given by:
\begin{equation}
    \partial_\eta \theta = \frac{1}{2\eta}\tan 2\theta + K_1 \partial_\eta m^2_1 - K_2\partial_\eta m^2_2 \,,
    \label{eq:dthetafield}
\end{equation}
where the $K_i$ have the same expressions as in \cref{eq:delQtheta}. Together, we finally obtain,
\begin{equation}
    (\mathbb{I}_2 - {\cal B} - \tilde{{\cal C}})
    \begin{pmatrix}
        \partial_\eta m^2_1 \\
        \partial_\eta m^2_2
    \end{pmatrix}
    = 
    \frac{1}{2\eta}\tan2\theta\,{\cal C}
    + 
    \frac{1}{c_{2\theta}}
    \begin{pmatrix}
        \phantom{-} c^2_\theta & -s^2_\theta  \\
        -s^2_\theta & \phantom{-} c^2_\theta 
    \end{pmatrix}
    \begin{pmatrix}
        \partial_\eta m^2_{0\varphi\varphi}  \\
        \partial_\eta m^2_{0\chi\chi} 
    \end{pmatrix}\,,
    \label{eq:detamassmatdiag}
\end{equation}
where $\tilde{C}$ is defined in \cref{eq:dQmassmatdiag}. Equation~\cref{eq:detamassmatdiag} is a set of linear equations using which one can solve for the quantities $\partial_\eta m^2_i$. Having found these solutions (practically for a choice of minimum), one obtains first the derivative of the angle in \cref{eq:dthetafield} and then finally the derivatives of the masses in the field basis through:
\begin{subequations}
    \begin{align}
        \partial_\eta \sqmpp &= \phantom{-} s_{2\theta}\,\delta m^2(\partial_\eta \theta) + c^2_\theta\partial_\eta m^2_1 + s^2_\theta\partial_\eta m^2_2\,,
    \\[0.5ex]
        \partial_\eta \sqmcc &= -s_{2\theta}\,\delta m^2(\partial_\eta \theta) + s^2_\theta\partial_\eta m^2_1 + c^2_\theta\partial_\eta m^2_2\,,
    \\[0.5ex]
        \partial_\eta \sqmpc &= \phantom{-} c_{2\theta}\,\delta m^2(\partial_\eta \theta) 
        + \sfrac12 s_{2\theta}\big(\partial_\eta m^2_2 - \partial_\eta m^2_1\big)\,.
    \end{align}
\end{subequations}
where again $\delta m^2 = m_2^2-m_1^2$.

%%%%%%%%%%%%%%%%%%%%%%%%%%%%%%%%%%%%%%%%%%%%%%%%%%%%%%%%%%%%%%%%%%%%%%%%%%%%%%%
%%%%%%%%%%%%%%%%%%%%%%%%%%%%%%%%%%%%%%%%%%%%%%%%%%%%%%%%%%%%%%%%%%%%%%%%%%%%%%%
%
\bibliographystyle{JHEP}
\bibliography{lit}
%
%%%%%%%%%%%%%%%%%%%%%%%%%%%%%%%%%%%%%%%%%%%%%%%%%%%%%%%%%%%%%%%%%%%%%%%%%%%%%%%
%%%%%%%%%%%%%%%%%%%%%%%%%%%%%%%%%%%%%%%%%%%%%%%%%%%%%%%%%%%%%%%%%%%%%%%%%%%%%%%

\end{document}